\documentclass{article}
\usepackage[a4paper,
            bindingoffset=0.2in,
            left=0.6in,
            right=0.6in,
            top=1in,
            bottom=1in,
            footskip=.25in]{geometry}
\usepackage[utf8]{inputenc}
\usepackage{jcappub}
\usepackage{natbib}
\setcitestyle{square,numbers}
\usepackage{graphicx}
\usepackage{xcolor}
\usepackage{float}
\usepackage{multirow,array,longtable}
\usepackage{booktabs}
\usepackage{arydshln} 
\usepackage[makeroom]{cancel}
\usepackage{amsfonts}
\usepackage{slashed}

\renewcommand\eqref[1]{Eq.~(\ref{#1})}

\newcommand{\be}{\begin{equation}}
\newcommand{\ee}{\end{equation}}
\newcommand{\bear}{\begin{eqnarray}}
\newcommand{\eear}{\end{eqnarray}}
\newcommand{\nn}{\nonumber}

\def\1loop{one-loop}

\newcommand{\mL}{\mathcal{L}}

\def\cw{c_{\rm w}}

\def\mh{m_{h}}
\def\mw{m_{W}}
\def\mz{m_{Z}}

\title{Exploring correlations between  HEFT Higgs couplings $\kappa_V$ and $\kappa_{2V}$ via HH  production at  $e^+e^-$ colliders}
\author[a]{J.M. D\'avila,}
\author[a]{D. Domenech,}
\author[a]{M. J.  Herrero,}
\author[b]{and R. A.  Morales}
\affiliation[a]{Departamento de F\'{\i}sica Te\'orica and Instituto de F\'{\i}sica Te\'orica, IFT-UAM/CSIC,\\
Universidad Aut\'onoma de Madrid, Cantoblanco, 28049 Madrid, Spain}
\affiliation[b]{IFLP, CONICET - Dpto. de F\'{\i}sica, Universidad Nacional de La Plata, \\ 
C.C. 67, 1900 La Plata, Argentina}
\emailAdd{juanm.davila@estudiante.uam.es}
\emailAdd{jose.domenech@estudiante.uam.es}
\emailAdd{maria.herrero@uam.es}
\emailAdd{roberto.morales@fisica.unlp.edu.ar}

\abstract{In this work we explore the phenomenological implications at future $e^+e^-$  colliders of assuming anomalous couplings of the Higgs boson to gauge bosons $HVV$ and $HHVV$ ($V=W,Z$)   given by the $\kappa$-modifiers with respect to the Standard Model couplings,  $\kappa_V$ and $\kappa_{2V}$, respectively.  For this study we use the Higgs Effective Field Theory (HEFT) where these two $\kappa$ parameters are identified with the two most relevant  effective couplings at leading order, concretely $a=\kappa_V$ and $b=\kappa_{2V}$.  
Our focus is put on these two couplings and their potential correlations which we believe carry interesting information on the underlying ultraviolet theory.  The particular studied process is 
$e^+e^- \to HH \nu \bar \nu$,  where the vector boson scattering subprocess $WW \to HH$ plays a central role,  specially at the largest planned energy colliders.  Our detailed study of this process as a function of the energy and the angular variables indicates that the produced Higgs bosons in the BSM scenarios will have in general a high transversality as compared to the SM case if $\kappa_V^2 \neq \kappa_{2V}$.   In order to enhance the sensitivity to these HEFT parameters $\kappa_V$ and $\kappa_{2V}$ and their potential correlations we propose here some selected differential cross sections for the $e^+e^- \to HH \nu \bar\nu$ process, where different kinematic properties of the BSM case with respect to the SM are manifested.  Finally,  we will focus on the dominant Higgs decays to $b \bar b$ pairs leading to final events with 4 $b$-jets and missing transverse energy from the undetected neutrinos and will provide the expected accessibility to the $(\kappa_V,\kappa_{2V})$ effective couplings and their potential correlations.  In our study we will consider the three projected energies for $e^+e^-$ colliders of 500 GeV, 1000 GeV and 3000 GeV. }

\begin{document}
\begin{flushright}
	IFT-UAM/CSIC-23-152 
\end{flushright}
\maketitle
\section{Introduction}
\label{intro}

After the Higgs particle discovery ~\cite{ATLAS:2012yve, CMS:2012qbp} with a mass of 125 GeV a big effort has been done in exploring its properties at colliders within the context of both the Standard Model of Particle Physics (SM) and also in scenarios beyond the Standard Model (BSM).  One of the most interesting avenues to test possible signals of BSM Higgs physics at colliders is pursued via the so-called anomalous Higgs couplings to SM particles which are usually defined with respect to the SM Higgs couplings by means of the so-called $\kappa$ modifiers (for a review,  see \cite{Workman:2022ynf}).
  We will focus here in the anomalous couplings belonging to the bosonic sector,  and more specifically in those describing the effective interactions of the Higgs particle to the electroweak (EW) gauge bosons,  $HVV$ and $HHVV$ (with $V=W, Z$) which are given by the $\kappa$ modifiers,  $\kappa_V$ and $\kappa_{2V}$ respectively.  The $\kappa$ parameters have been explored by the Large Hadron Collider (LHC) collaborations, ATLAS  and CMS,  and several  constraints on their values have been set by means of the various Higgs production and decay channels.  Whereas $\kappa_V$ is mostly constrained by single Higgs production at the LHC \cite{Workman:2022ynf},  
$\kappa_{2V}$ is mostly constrained by double Higgs production at the LHC~\cite{ATLAS:2023qzf,ATLAS:2015sxd,CMS:2017hea, CMS:2020tkr, ATLAS:2021ifb, CMS:2022hgz, ATLAS:2022xzm}.  The $HH$ production via the $WW$ vector boson fusion (WBF) mechanism is indeed the most sensitive test of both parameters $\kappa_{V}$ and $\kappa_{2V}$ together since they both appear in the same relevant subprocess taking place which is the $WW \to HH$ scattering where the two $W$ gauge bosons are radiated from the initial quarks which are inside the protons.  The present best experimental constraints, at 95\%CL,  on these two $\kappa$ parameters can be summarized as,   $\kappa_V \in (0.97,1.13)$~\cite{ATLAS:2019nkf} and 
$\kappa_{2V}\in (0.0,2.1)$~\cite{CMS:2022hgz}.

From the theoretical side,  the most proper approach to explore these anomalous couplings and their phenomenological consequences at colliders is provided by the tool of Effective Field Theories.  The two most popular ones to study the Higgs effective couplings are the the Standard Model Effective Field Theory (SMEFT) and the Higgs Effective Field Theory (HEFT) (for reviews,  see for instance,  \cite{Brivio:2017vri, Dobado:2019fxe}).  We choose here the second one because the Higgs effective couplings of our interest,  $HVV$ and $HHVV$,  are given in the HEFT by the leading order (LO) effective Lagrangian parameters $a$ and $b$ (also called Wilson coefficients) which are independent parameters and directly identifiable with the previously commented $\kappa$'s,  specifically by $a=\kappa_V$ and $b=\kappa_{2V}$.  With this LO-HEFT Lagrangian one can also recover the SM predictions by simply setting $a=b=1$ (in addition to also setting the triple and quartic Higgs self-couplings modifiers to $\kappa_3=\kappa_4=1$).  The  LO-HEFT then provides the proper description of the BSM Higgs anomalous couplings to EW gauge bosons for $a \neq 1$ and $b \neq 1$, and these are usually parametrized with respect to their SM values by $a=1- \Delta a$ and $b=1-\Delta b$,  with the deviations given by $\Delta a \neq 0$ and $\Delta b \neq 0$.  Another important and distinctive feature of the HEFT is that the  $SU(2)_L \times U(1)_Y$ gauge invariant effective operators containing the $HVV$ and $HHVV$ effective interactions are directly written in terms of the Higgs field $H$, which is a singlet.  
Then the corresponding  low-energy parameters $a$ and $b$ in front of the effective operators  in this HEFT framework appear generically as uncorrelated parameters, in contrast to the SMEFT case where the $HVV$ and $HHVV$ effective interactions appear correlated due to the fact that the Higgs field in the SMEFT  is not a singlet but a component of the usual scalar doublet $\Phi$.  Thus we choose here the HEFT that provides the most general gauge invariant framework where there are not a priori correlations between 
$\kappa_V$ and $\kappa_{2V}$, therefore allowing us to treat them as independent parameters for the present study. 
It is also important to mention that we will work exclusively within the bosonic sector of the HEFT. Correspondingly, we will assume here that the new physics is in the bosonic-HEFT and all the interactions in the fermionic sector are the same as in the SM.

The search of potential correlations between  $\kappa_V$ and $\kappa_{2V}$ is by itself an interesting subject,  both from the theory and the experimental sides.  On the one hand,  from the EFT point of view,  one expects that once a particular underlying ultraviolet (UV) theory is assumed to be the generator of such a low-energy theory, the corresponding Wilson coefficients in front of the effective operators defining this EFT can be derived by some matching procedure at low energies between the EFT and the underlying UV theory after the BSM  heavy modes have been integrated out.  This matching not only provides the values of the coefficients in terms of the fundamental parameters of the UV theory,  but also could lead to correlations among these coefficients which are even more interesting.  For instance,  assuming Minimal Composite Higgs Models ~\cite{Agashe:2004rs, Contino:2006qr}, or in the Strongly Interacting Light Higgs Model ~\cite{Giudice:2007fh},  or in the SMEFT~\cite{RoberMariaDaniMJ},  where the Higgs field is a component inside a $SU(2)$ doublet, one finds the following common correlation between the deviations of the $a$ and $b$ parameters  
$4\Delta a= \Delta b$.   In contrast, assuming dilaton models~\cite{Goldberger:2007zk} or in models with iso-singlet mixing~\cite{Englert:2011yb,Englert:2023uug} (see also references therein), one finds another correlation given by $2\Delta a= \Delta b$ which is the consequence of having the particular relation between $a$ and $b$ given by $a^2=b$ in those models.  Recently, the case of 2HDM has been studied in \cite{Arco:2023sac, Dawson:2023ebe} and a new correlation has been obtained.  In that case the observed Higgs particle is identified with the lightest 2HDM  boson and the other four Higgs bosons of the 2HDM are considered very heavy and integrated out.  The correlation found in this matching of the 2HDM to the HEFT \cite{Arco:2023sac} is given by  $2\Delta a= -\Delta b$,  in the region close to the alignment condition defined by $\cos(\alpha-\beta)\ll 1$. 
On the other hand,  from the experimental side,  the colliders will be able to provide a direct test of which of those correlations are better favoured or disfavoured  by data,  providing an even more efficient test when these two parameters participate in the same process.  This is clearly the case of $HH$ production via WBF at colliders where, as we have said,  the two $\kappa_V$ and $\kappa_{2V}$ participate in the same subprocess $WW \to HH$.  Thus,  one hopes that exploring specific observables where this subprocess is involved in a relevant way,  like,  for instance,  some particular differential cross sections,  one could find an efficient way to
access with the highest sensitivity to these anomalous couplings in the ($\kappa_V$, $\kappa_{2V}$)=($a$, $b$) plane.  One clear example capturing a good sensitivity to the combination 
$(\kappa_V^2-\kappa_{2V})=(a^2-b)$ is the differential cross section with respect to the $HH$ invariant mass, which has been discussed at both colliders,  $e^+e^-$ \cite{Contino:2013gna, Gonzalez-Lopez:2020lpd} and $pp$ \cite{Contino:2010mh}.  Here we will re-analize this differential cross section for $e^+e^-$ colliders and will propose other differential cross sections with respect to other kinematical variables which we have found to be even more efficient to improve the sensitivity to BSM departures in terms of the mentioned combination $(\kappa_V^2-\kappa_{2V})$. 
The particular process of our interest here is $e^+e^- \to HH \nu \bar \nu$, which contains $WW \to HH$ as the most relevant subprocess.  Other competing subprocesses,  like those mediated by $Z$ gauge bosons, are subdominant in the case of collider energies in the TeV range.  Therefore,   the combination  $(\kappa_V^2-\kappa_{2V})$ being present in $WW \to HH$ is expected to be accessible singularly by means of $e^+e^- \to HH \nu \bar \nu$, which will be our main focus in this work. 

The accessibility  to  $a=\kappa_V$ and $b=\kappa_{2V}$ within the HEFT via the total cross section of double Higgs production from WBF has been studied previously in the literature for both types of colliders,  future $e^+e^-$ colliders  
\cite{Contino:2013gna, Gonzalez-Lopez:2020lpd,RoberMariaDaniMJ, Englert:2023uug} like the International Linear Collider (ILC)  and the Compact Linear Collider (CLIC) and also at $pp$ colliders like the LHC and its future upgrades in luminosity and energy \cite{Contino:2010mh,Anisha:2022ctm, Englert:2023uug}.  
The reason to choose $e^+e^-$ future planned colliders in the present paper is that these colliders are well known to offer a cleaner environment (i.e. with less background) to study the phenomenological implications of BSM couplings as compared to hadronic colliders like LHC.  In particular,  studying the role of correlations among $\kappa_V$ and $\kappa_{2V}$ at colliders, which is our main purpose in this work,  seems to be easier at $e^+e^-$ than at LHC.
We will consider here two cases of future $e^+e^-$ colliders: 1)  ILC \cite{Bambade:2019fyw} with two options for total energy and integrated luminosity 
$(\sqrt{s}, \mL)$ of $(500 \,{\rm GeV}, 4 \,{\rm ab}^{-1} )$ and  $(1000\, {\rm GeV}, 8\, {\rm ab}^{-1} )$,  and  2) CLIC \cite{CLICdp:2018cto} with the highest expected energy and luminosity of $(3000 \, { \rm GeV},  5\,{\rm ab}^{-1})$.  We will devote our study mainly to the accessibility to test the potential correlations between the effective parameters $a$ and $b$ in the ($\kappa_V$, $\kappa_{2V}$) plane, computing both the total and the differential cross sections.  We also wish to explore and determine here which kind of differential cross sections and specific final states will provide the highest sensitivity to the different hypotheses for these correlations at $e^+e^-$ colliders.  

The paper is organized as follows:  in section \ref{effHcoup} we review the relevant effective Higgs couplings $HVV$ and $HHVV$ within the HEFT in terms of the low-energy parameters $a$ and $b$ and explain their relation with the $\kappa_V$ and $\kappa_{2V}$ parameters, i.e.  we derive $a=\kappa_V$ and $b=\kappa_{2V}$.   In section \ref{WBF-HH} we review and discuss the main features of the relevant subprocess $WW \to HH$,  differentiating the two cases of our interest: $\kappa_V^2 \neq \kappa_{2V}$ and $\kappa_V^2 = \kappa_{2V}$.  Our findings of the high transversality of the final $H$'s for $\kappa_V^2 \neq \kappa_{2V}$ are first pointed out in this section.  In section \ref{xsectionee}
we explore the consequences of assuming different correlations between  $\kappa_V$ and $\kappa_{2V}$ both in total and differential cross sections for the  $e^+e^- \to HH \nu \bar \nu$ process. In particular,  we propose three differential cross sections that are sensitive to these correlations.  In section \ref{correlations} we consider the full process $e^+e^- \to HH \nu \bar \nu \to b \bar b b \bar b \nu \bar\nu$ where the Higgs particles decay into $b \bar b$ pairs and study the accessibility to $\kappa_V$ and $\kappa_{2V}$ and their possible correlations by analizing the final events with 4 $b$-jets and missing transverse energy.  In section \ref{conclu} we finally summarize our conclusions.

\section{Effective Higgs couplings to gauge bosons within HEFT}
\label{effHcoup}

As we have said in the Introduction,  the HEFT is the proper EFT to describe the effective interactions of our interest here,  given by the anomalous couplings of one Higgs boson to two $W$ gauge bosons, $HWW$,  and of two Higgs bosons to two $W$ gauge bosons, $HHWW$.  For the present work,  we only need to specify the EW bosonic part of the HEFT.  The bosonic Lagrangian of 
the HEFT is written in terms of effective operators  which are built with the relevant bosonic fields,  namely,  the Higgs field $H$;   the  $SU(2)_L \times U(1)_Y$ EW gauge bosons,  $W^i$,  $B$;  and the would-be-Goldstone bosons $w^i$ which are introduced in a non-linear representation of the EW symmetry group.  The principle guide to build this Lagrangian is,  as usual,  the requirement of $SU(2)_L \times U(1)_Y$ gauge invariance.  The series of operators contributing in the HEFT are organized in terms of the chiral dimension (counting derivatives and  masses) instead of the other more usual counting in terms of canonical dimension.  Thus,  the  leading order (LO) corresponds to chiral dimension 2,  the next to leading order (NLO) corresponds to chiral dimension 4,  and so on.  The complete set of effective operators up to NLO can be found in \cite{Alonso:2012px, Brivio:2013pma, Buchalla:2013rka,Sun:2022ssa}.  The HEFT is a renormalizable quantum field theory in the EFT sense,  meaning that all the divergences generated to one loop by the LO-HEFT Lagrangian can be absorbed  by redefinitions of the LO and NLO HEFT parameters and fields.  The full renormalization program for the bosonic sector of the HEFT to one-loop has been achieved in a general covariant $R_\xi$ gauge in refs. \cite{Herrero:2020dtv, Herrero:2021iqt,Herrero:2022krh},  where the explicit running equations for all the effective couplings involved in the bosonic sector,  LO and NLO,  can also be found,  including the ones related to the parameters $\kappa_V$ and $\kappa_{2V}$ whose phenomenological consequences we are interested in here.  Other alternative renormalization programs within the HEFT have also been considered in the literature,  see for instance \cite{Delgado:2013hxa, Espriu:2013fia, Gavela:2014uta, Guo:2015isa, Buchalla:2017jlu,  Asiain:2021lch}.
 
The relevant effective bosonic operators for the present work are contained in the LO-HEFT Lagrangian. Therefore,  here we restrict ourselves to this LO-HEFT Lagrangian,  which for an arbitrary covariant $R_\xi$ gauge is given by: 
    
       \begin{eqnarray}
        {\mL}_{\rm LO}^{\rm HEFT} \, = \, \frac{v^2}{4}  \left(1 + 2 a \frac{H}{v} + b \frac{H^2}{v^2}+\dots \right)
        \text{Tr}[D_{\mu} U^{\dagger} D^{\mu} U] + \frac12 \partial_{\mu} H \partial^{\mu} H - V(H) \nn \\
        - \frac{1}{2 g^2} \text{Tr}[\hat{W}_{\mu \nu} \hat{W}^{\mu \nu}] - \frac{1}{2 g'^2} \text{Tr}[\hat{B}_{\mu \nu} \hat{B}^{\mu \nu}] + {\mL}_{GF} + {\mL}_{FP},
        \label{eqn: leading}
    \end{eqnarray}
The relevant fields and quantities appearing in this Lagrangian are:  
$H$ is the Higgs field which is introduced in the HEFT as a singlet field, in contrast to the SM or the SMEFT where it is introduced as a component of the usual doublet $\Phi$.  The $U$ field is a $2\times2$ matrix
\begin{equation}
        U \, = \, \exp \left( i \frac{{\omega_i} {\tau_i}}{v} \right),
    \end{equation}  
   that contains the three GB fields ${\omega}_i$ ($i=1,2,3$) in a non-linear representation of the $SU(2)$ symmetry group and the three Pauli matrices $\tau_i$.  The EW covariant derivative of this $U$ field 
   is defined as:
   \begin{equation}
        D_{\mu} U \, = \, \partial_{\mu} U + i \hat{W}_{\mu} U - i U \hat{B}_{\mu},
    \end{equation} 
    that contains the EW gauge fields,  $\hat{W}_{\mu} = \frac{g}{2} W^i_{\mu} \tau^i$ and $\hat{B}_{\mu} = \frac{g'}{2} B_{\mu} \tau^3$ and the EW gauge couplings $g$ and $g'$.  The corresponding EW field strength tensors are given by:
       \begin{equation}
        \hat{W}_{\mu \nu} \, = \, \partial_{\mu} \hat{W}_{\nu} - \partial_{\nu} \hat{W}_{\mu} + i [\hat{W}_{\mu}, \hat{W}_{\nu}], \hspace{8mm} \hat{B}_{\mu \nu} \, = \, \partial_{\mu} \hat{B}_{\nu} - \partial_{\nu} \hat{B}_{\mu}.
    \end{equation}
    The physical gauge fields are then given,  as usual,  by:
\be
W_{\mu}^\pm = \frac{1}{\sqrt{2}}(W_{\mu}^1 \mp i W_{\mu}^2) \,,\quad
Z_{\mu} = c_W W_{\mu}^3 - s_W B_{\mu} \,,\quad
A_{\mu} = s_W W_{\mu}^3 + c_W B_{\mu} \,,
\label{eq-gaugetophys}
\ee
where we use the short notation $s_W=\sin \theta_W$ and $c_W=\cos \theta_W$,  with $\theta_W$ the weak angle.
$V(H)$ is the Higgs potential within the HEFT, which includes the triple and quartic  Higgs self-interactions:
    \begin{equation}
        V(H) \, = \, \frac12 m_H^2 H^2 + \kappa_3 \lambda v H^3 + \kappa_4 \frac{\lambda}{4} H^4.
    \end{equation}
It should be noticed that the previous bosonic HEFT Lagrangian is built,  by definition,  on top of the asymmetric EW vacuum,  following the same spirit as the chiral Lagrangian and ChPT of QCD being built on top of the $SU(2)_L \times SU(2)_R$  asymmetric vacuum of low energy QCD.   The chiral counting dimensions in the bosonic HEFT then follows closely the usual chiral dimension counting of ChPT.  Both chiral Lagrangians are written as the sum of two contributions: $\mL_2$ (LO Lagrangian), with chiral dimension 2,  and $\mL_4$ (NLO Lagrangian) with chiral dimension 4.  In ChPT both the pion mass and the derivatives of the $U$ field (which includes the pions in a non-linear representation) count as ${ \cal O} (p)$ in the momentum expansion.  In the case of the bosonic HEFT,  both the boson masses $m_H$,  $m_W$,  $m_Z$ and the derivatives of the $U$ field (which includes the EW GBs in a non-linear representation) count as $ {\cal O}(p)$ in the momentum expansion.  This counting is fixed in the HEFT at the very beginning, without any reference to the particular UV theory, which should be the final responsible for generating the low energy effective Lagrangian.  Similarly to the case of ChPT,  the integration out of the heavy modes from the particular UV theory in the EW case will provide a particular prediction for the values of low energy couplings ($a$, $b$, etc.)  that are present in the HEFT.  However this will not change the chiral counting nor the ordering of the LO and NLO effective operators that define the HEFT Lagrangian terms $\mL_2$ ($\equiv {\mL}_{\rm LO}^{\rm HEFT}$ in Eq. \ref{eqn: leading})  and $\mL_4$ (not given explicitly here), respectively.
    
 Notice also that the terms in $V(H)$ do not contain derivatives but they also belong to the LO Lagrangian with chiral dimension 2 because they can be written with  squared masses in front.  In particular, in the EW bosonic sector,  the relevant  masses of the low energy modes are the Higgs boson mass $\mh$, the $W$ boson mass $\mw$ and the $Z$ boson mass $\mz$ and all of them are input parameters within the HEFT.  At LO, these input masses can be written  in terms of the SM couplings and the vacuum expectation value $v=246$ GeV as usual,  i.e.  $\mh^2  = 2 \lambda v^2$,  $\mw=g v /2$ and $\mz=\mw/\cw$.  For the rest of this work, the Higgs self-couplings within the HEFT will be set to their SM values,  i.e.  we will assume here $\kappa_3=1$ and $\kappa_4=1$.   Finally,   ${\mL}_{GF}$ and ${\mL}_{FP}$ are the gauge fixing  and Faddeev-Popov ${\mL}_{FP}$ terms,  respectively, whose explicit expressions in the general $R_\xi$ covariant gauges can be found in 
\cite{Herrero:2020dtv, Herrero:2021iqt,Herrero:2022krh},  together with more definitions and specifications within the HEFT.  Last,  but not least,  regarding the counting rules for the ordering of the effective operators in the HEFT,  it is fair to say that there has been some debate in the literature where the counting by chiral dimensions,  the counting  by inverse powers of the UV cut-off $\Lambda$ and the counting by the loop suppression factor $1/(4 \pi v)$  have been compared and discussed (see,  for instance~\cite{Brivio:2017vri,Buchalla:2013eza,Gavela:2016bzc}).   Indeed,  some authors have questioned the counting by chiral dimensions~\cite{Gavela:2016bzc}.   Since our study here is with the LO Bosonic-HEFT and it will be devoted exclusively to the computation of tree level scattering amplitudes,  these counting rules issues are not relevant here and do not affect the phenomenological study of the present work.

\begin{figure}[!t]
    \centering
        \includegraphics[scale=0.3]{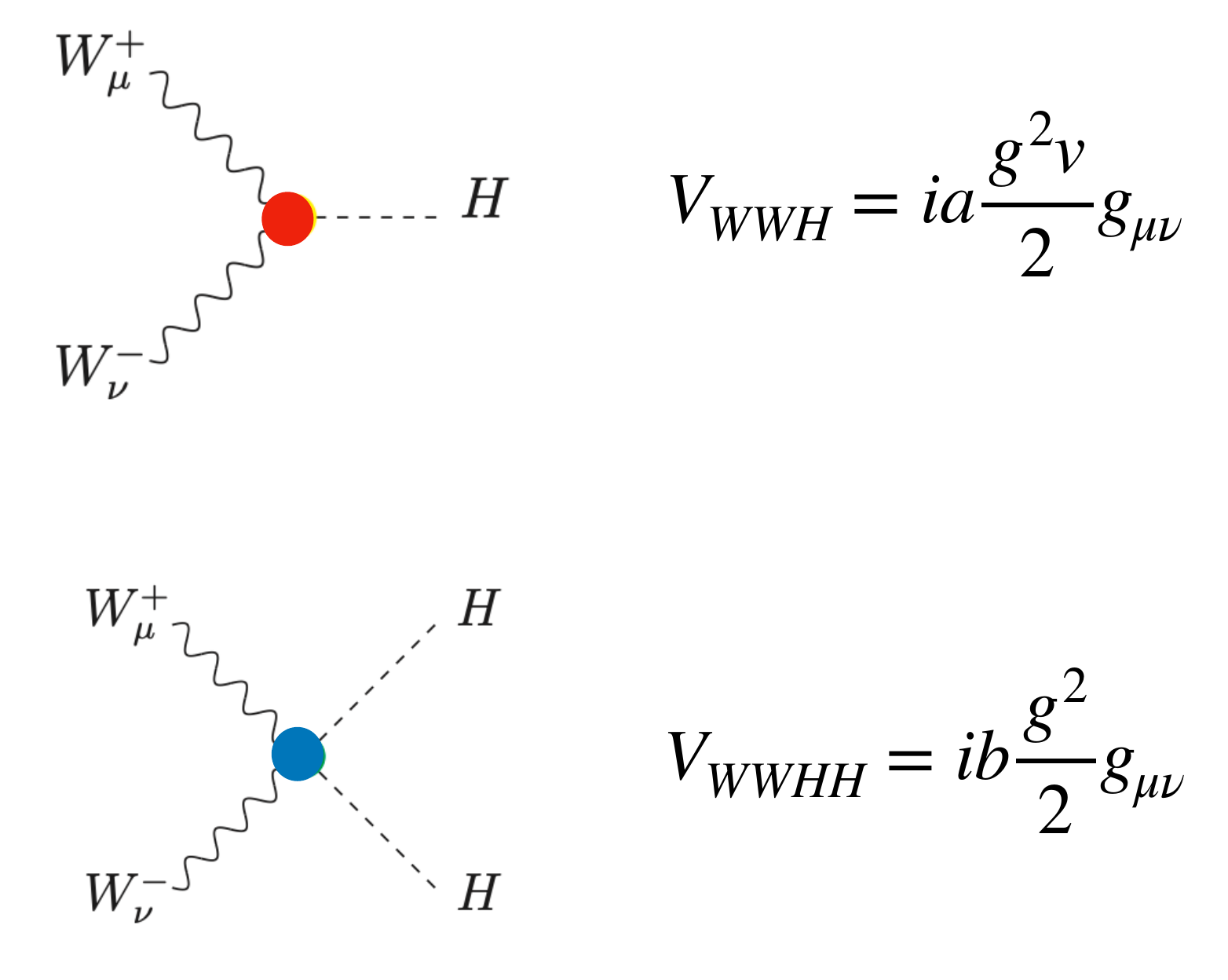}
    \caption{Feynman rules of the $WWH$ and $WWHH$ effective interactions within HEFT. The corresponding Feynman rules for the $ZZH$ and $ZZHH$ effective interactions within HEFT (not displayed here) are 
    $V_{ZZH}= i a g^2 v /(2 c_W^2) g_{\mu \nu}$ and $V_{ZZHH}=i b g^2/(2 c_W^2) g_{\mu \nu}$. }
    \label{FR}
    \end{figure}

For the forthcoming computation we first need to identify in the LO-HEFT Lagrangian of Eq. \ref{eqn: leading} the relevant effective interactions of one Higgs boson to two $W$ bosons and of two Higgs bosons to two $W$ bosons. These are contained in the effective operators with two covariant derivatives that go with the low-energy coefficients $a$ and $b$ respectively.  The corresponding Feynman rules for these effective interaction vertices are given in Fig. \ref{FR},  which are just like in the SM but with modifying  factors $a$ and $b$ in front.  Therefore,  one can immediately identify them with the corresponding $\kappa$ modifiers:
\begin{equation}
\label{ab-kappas}
a=\kappa_V \,\,\,, \, \, \, b=\kappa_{2V} \,.
\end{equation}
In the following we will then refer to these two sets of parameters,  ($a,b$) and ($\kappa_V$, $\kappa_{2V}$) equally.  We will discuss our results in terms of these parameters and also in terms of their deviations respect to the SM value,  i.e.  in terms of the corresponding $\Delta$'s,  defined as:
\begin{equation}
a= 1-\Delta a \, \, , \, \, b=1-\Delta b \,.
\label{deltas}
\end{equation}
Notice that when assuming a given correlation between $a$ and $b$ it implies trivially a correlation between $\Delta a$ and $\Delta b$.  For instance,  assuming $a^2=b$ implies (in the linear approximation for the deviations) $2 \Delta a=  \Delta b$,  etc.  In the following sections we will explore some interesting  phenomenological consequences for colliders by assuming different correlations between the HEFT parameters $a$ and $b$. 

\section{The role of $(\kappa_V^2-\kappa_{2V})$  in $HH$ production via WBF}
\label{WBF-HH}
\begin{figure}[!t]
    \centering
        \includegraphics[scale=0.4]{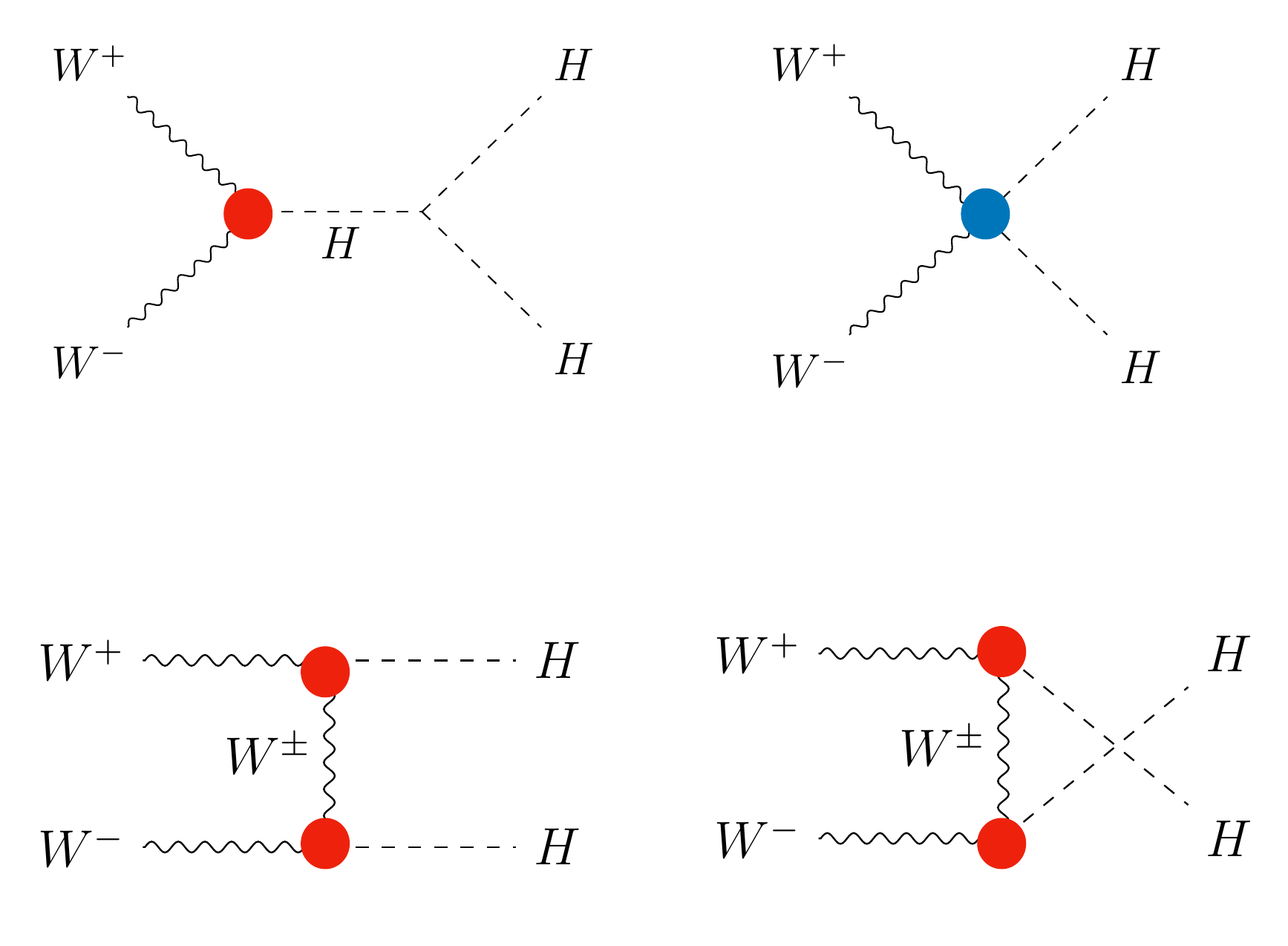}
    \caption{Feynman diagrams within HEFT contributing to $WW \to HH$ in the unitary gauge: $s$-channel (upper left),  $c$-channel (upper right),  $t$-channel (lower left), $u$-channel (lower right).  The coloured dots represent the effective interactions depicted in Fig. \ref{FR}. }
    \label{FD}
    \end{figure}

In this section, we review the main aspects of the relevant subprocess of our interest $WW \to HH$ where both low-energy effective parameters $a$ and $b$ participate.  The scattering amplitude for this process is a gauge invariant quantity and therefore it can be computed in different gauges leading to the same result.  Within the HEFT it has been computed both in the $R_\xi$ covariant gauges, including the GB fields in the internal lines of the diagrams;  and in the unitary gauge, which does not include the GB's.  Both results obviously coincide (an explicit demonstration of the gauge invariance of this HEFT amplitude can be found in \cite{Herrero:2022krh}).   In this section  we use the unitary gauge, where the contributing diagrams within the LO-HEFT are summarized in Fig. \ref{FD}. 
At the tree level within the HEFT there are just four diagrams contributing in the unitary gauge,  corresponding to the contact $c$-channel, the $s$-channel,  the $t$-channel and the $u$-channel.  The total scattering amplitude is obtained by adding these four contributions:
    \begin{equation}
        \mathcal{A} = \mathcal{A}_c + \mathcal{A}_s + \mathcal{A}_t + \mathcal{A}_u,
        \label{eqn: total_amp}
    \end{equation} 
    where each contribution is given respectively by:
    \begin{eqnarray}
         \label{eqn: amplitudes}
        \mathcal{A}_c & = & \frac{g^2 b}{2} \epsilon_+ \cdot \epsilon_- \,, \nonumber \\
        \mathcal{A}_s & = & \frac{3 g^2 a}{2} \frac{m_H^2}{s - m_H^2} \epsilon_+ \cdot \epsilon_- \,, \nonumber \\
        \mathcal{A}_t & = & g^2 a^2 \frac{m_W^2 (\epsilon_+ \cdot \epsilon_-) + (\epsilon_+ \cdot k_1)(\epsilon_- \cdot k_2)}{t - m_W^2} \,, \nonumber \\
        \mathcal{A}_u & = & g^2 a^2 \frac{m_W^2 (\epsilon_+ \cdot \epsilon_-) + (\epsilon_+ \cdot k_2)(\epsilon_- \cdot k_1)}{u - m_W^2}.
    \end{eqnarray}
    Here,  $s$, $t$ and $u$ are the Mandelstam variables of this subprocess, $\epsilon_+$ and $\epsilon_-$ are the polarizations of the incoming $W^+$ and $W^-$ bosons,  with momenta $p_+$ and $p_-$ respectively, $k_1$ and $k_2$ are the momenta of the outgoing Higgs bosons and $g$ is the weak coupling.  Notice that the SM prediction is recovered for $a=b=1$,  as expected. 
    In the center of mass frame (CM),   the momenta of the Higgs bosons  can be written as: \par
    \begin{equation}
        k_{1, 2} \, = \, \Big( \sqrt{s}/2, \pm \sin \theta \sqrt{s-4m_H^2}/2, 0, \pm \cos \theta \sqrt{s-4m_H^2}/2 \Big),
        \label{eqn: momenta}
    \end{equation}
    where $\theta$ is the angle between the momentum $p_+$  of the incoming $W^+$ boson and $k_1$.  The values of the polarization vectors, $\epsilon_{\pm}$,  are different for the longitudinal $W_L$ (with helicity 0) and the transverse $W_T$ (with helicities $\pm 1$).    They can be written as:
\begin{eqnarray}
        \epsilon_{\pm}^L &= & \Big( \sqrt{s-4m_W^2}/(2 m_W), 0, 0, \pm \sqrt{s}/(2 m_W) \Big) \,,  \nonumber \\
            \epsilon^{T_+}_\pm &=& \left(0,1/\sqrt{2},\pm i/\sqrt{2},0\right) \,,  \nonumber\\
        \epsilon^{T_-}_\pm &=& \left(0,1/\sqrt{2},\mp i/\sqrt{2},0\right) \,.
        \label{eqn: pol_vec}
   \end{eqnarray}
   As can be seen in the Eq. \ref{eqn: amplitudes}, the parameter  $a$ enters quadratically in the $t$ and $u$ channels,   and linearly in the $s$-channel.  In contrast,  the parameter $b$ enters only in the $c$ channel.  This may seem a trivial statement,  however, this dependence turns out to be crucial in the final behaviour of the amplitude  at high energies.  For instance, in the SM case with $a=b=1$, there is a well known strong cancellation of the fast growing behaviour with $\sqrt{s}$ of the separate contributions from the various channels,  and more concretely among the $t$, $u$ and $c$ channels, that leads to a flat behaviour with $\sqrt{s}$ at large energy values.  This can be understood in very simple terms by analysing  the dominant polarization channel which is the one with longitudinally polarized $W$ gauge bosons in the high energy region.  By using  the polarization vector for longitudinal $W$'s in Eq. \ref{eqn: pol_vec}
and making an expansion of the amplitude $ \mathcal{A}^L= \mathcal{A}(W^+_L W^-_L \to HH)$ in powers of the total energy ($ \mathcal{A}^L \simeq {\cal O}(s)+ {\cal O}(s^0)+{\cal O}{(s^{-1})}+\dots$) that is valid at high energies,  $\sqrt{s} \gg m_H, m_W$,   we get the following results for the dominant terms in this expansion for the various channels:
    \begin{eqnarray}
        \mathcal{A}_c^L & = & b \frac{g^2}{4 m_W^2} s + \mathcal{O}(s^0) \,, \nonumber \\
        \mathcal{A}_s^L & = & 0+\mathcal{O}(s^0) \,, \nonumber \\
        \mathcal{A}_t^L & = & a^2 \frac{g^2}{8 m_W^2} (\cos \theta - 1) s + \mathcal{O}(s^0) \,, \nonumber \\
        \mathcal{A}_u^L & = & - a^2 \frac{g^2}{8 m_W^2} (\cos \theta + 1) s + \mathcal{O}(s^0)\,.
        \label{eqn: amplitudes_long}
    \end{eqnarray}
    Thus,  the total amplitude for longitudinally polarized $W$'s  in the high energy region is given by:
    \begin{equation}
        \mathcal{A}^L \, = \, -(a^2-b) \frac{g^2}{4 m_W^2} s + \mathcal{O}(s^0)\,.
        \label{eqn: total_amp_long}
    \end{equation}
 From this expression, we learn several important features.  First,  as expected, we get the same result for the physical amplitudes as using the Equivalence Theorem (ET) \cite{Cornwall:1974km, Vayonakis:1976vz, Lee:1977eg} 
 where one replaces the external $W_L^{\pm}$ by the corresponding GB $\omega^{\pm}$.  Compare, for instance,  our results above with the results in \cite{Delgado:2013hxa} where they use the ET.    Second, the  growing with the subprocess squared energy $s$ of the separate contributions from the $c$, $t$ and $u$ channels  cancel in the SM case when they are summed.  We also see that for the BSM prediction this cancellation does no happen for generic $a$ and $b$ values with  $b \not= a^2$.  It is only when the particular correlation given by $a^2=b$ is assumed that the cancellation of the dominant ${\cal O}(s)$ contribution happens.  We also learn from this analytical result that this dominant contribution proportional to $(a^2-b)$ does not depend on the angular variable $\theta$ and this has important phenomenological implications for the BSM signals at colliders as we will see in the next sections.  For a complete analysis of the scattering amplitude in all the polarization channels $ \mathcal{A}(W^+_X W^-_Y \to HH)$ with  $XY=LL, TT,LT,TL$ including  also the next to leading order effective operators within  the HEFT we address the reader to \cite{RoberMariaDaniMJ}.  In this reference a similar large energy expansion in powers of $s$ has been done for all channels and the role of the combination $(a^2-b)$ in the dominant $LL$ channel has been confirmed. 

\begin{figure}[!t]
    \centering   
        \includegraphics[scale=0.24]{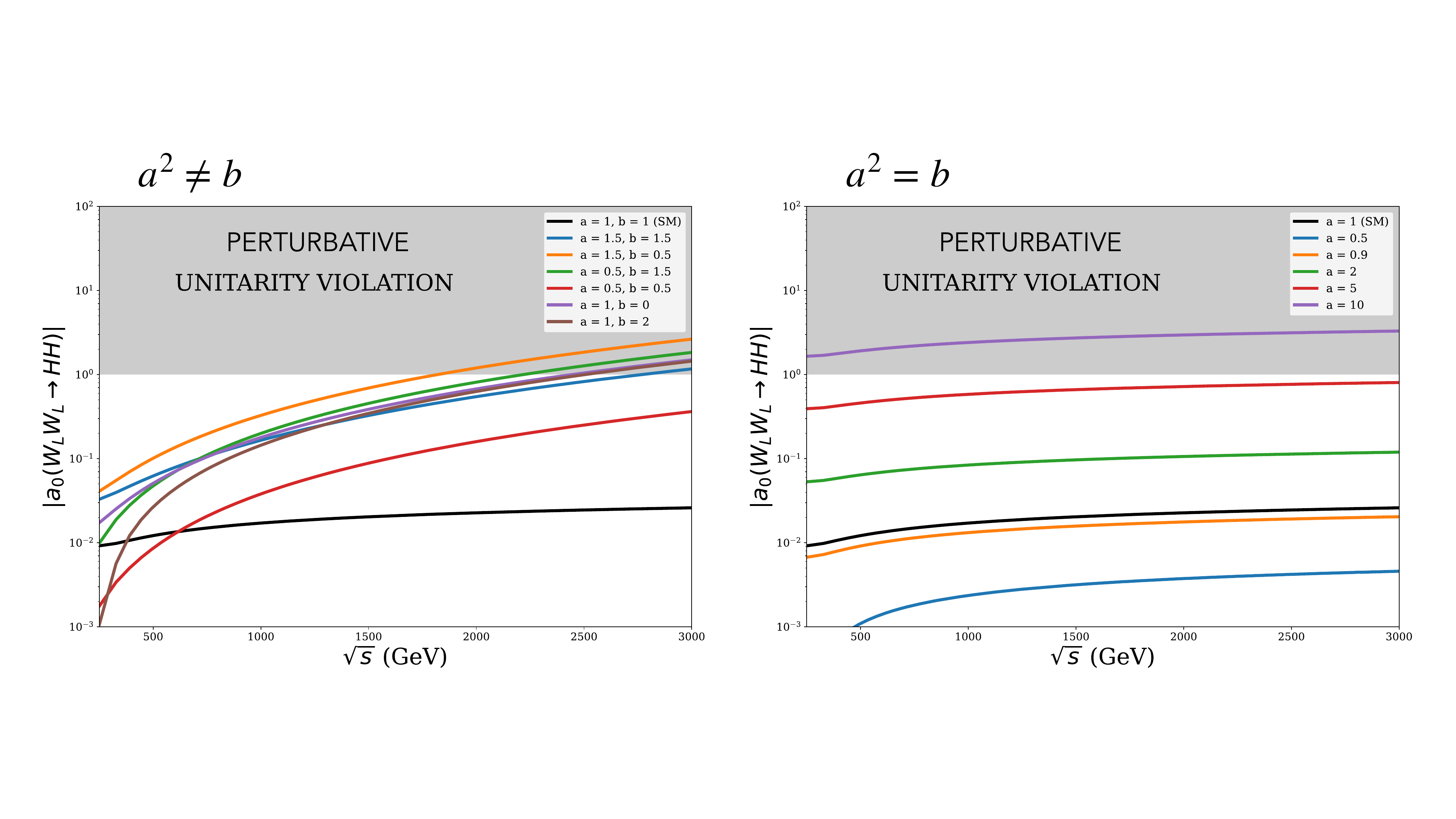}        
    \caption{Predictions of the $J=0$ partial wave amplitude $|a_0(W^+_L W^-_L \to HH)|$ within  the HEFT as a function of the CM energy $\sqrt{s}$ for various choices of the LO-HEFT parameters $a$ and $b$.  Left plot assumes $a^2 \neq b$.  Right plot assumes $a^2=b$.  The shaded region denotes the region of perturbative unitarity violation where $|a_0|>1$. }
    \label{unitarity}
    \end{figure}

The important role of the value for $(a^2-b)$ in the above commented cancellation at high energies was already noticed previously in the literature, and also its implications for collider physics have been explored.   For instance,  this was explored in \cite{Contino:2010mh} within the context of LHC and in \cite{Gonzalez-Lopez:2020lpd} within the context of $e^+e^-$ colliders.  One of the most relevant implications of assuming $(a^2-b)\neq 0$ is the potential violation of perturbative unitarity in the cross section predictions.  This can be seen in Fig. \ref{unitarity} where  the dominant partial wave amplitude $a_0$ (corresponding to angular momentum $J=0$) is shown for this $W_LW_L \to HH$ scattering as a function of the CM energy of the subprocess,  $\sqrt{s}=M_{HH}$,  for various values of $a$ and $b$.  Then,  the simple perturbative unitarity condition given by $|a_0|<1$ sets constraints on the maximum allowed size for $a$ and $b$.  

In the left plot of Fig. \ref{unitarity}, we see that for $(a^2-b) \neq 0$ the prediction for $|a_0|$ grows rapidly  with the energy and enters into the perturbative unitarity violation region at the ${\cal O}({\rm TeV})$ energy scales.  Concretely,  in this $b \neq a^2$ case and for the particular parameter values  chosen in Fig. \ref{unitarity},  this crossing into the perturbative unitarity violation region lays approximately in the interval $(1800,3000)$ GeV.   Therefore,  for the parameter values that will be studied in the following sections with $a \in (0.5, 1.5)$ and $b \in (0.5, 1.5)$ the perturbative unitarity is preserved for the $e^+e^-$ colliders with CM energies of 500 GeV and 1000 GeV.  For the $e^+e^-$ collider case with the highest energy of 3000 GeV,  there are yet some coefficient values that lead to crossing the unitarity line at subprocess energies $M_{HH}$ below 3000 GeV.  For instance,  for the choices $(a,b)=(1,0)$ and $(a,b)=(1,2)$ which correspond to the experimentally allowed extreme $\kappa_{2V}$ values,  the unitarity line is crossed at the invariant mass of the Higgs pair  around $M_{HH}=$ 2500 GeV.  So in these two cases,  with $b=0$ and $b=2$,  the collider predictions corresponding to the high $M_{HH}$ region in the interval $M_{HH} \in (2500,3000)$ GeV could in principle violate perturbative unitarity.  However,   in practical computations,  these problematic large $M_{HH}$ regions turn out not to have any impact on the final predicted rates at  colliders,  because at those large values the contribution to the cross section is already very small  and can be safely neglected.   Other choices of $a$ and $b$ in the range 0.5-1.5,  that will be the ones explored in the following sections devoted to $e^+e^-$ colliders,  lead also to this same conclusion.  For instance,  for 
$(a,b)=(1.4, 0.8)$  the crossing point  is at 2200 GeV and we have found that for CLIC(3000 GeV) just a $1\%$ of the total cross section comes from the problematic interval $M_{HH} \in (2200,3000)$ GeV.  We also get similar conclusions  for the values that lead to the highest cross sections.  Concretely for $(a,b)=(1.5, 0.5)$ we find the crossing at 1800 GeV,  and our computation gives that only a $5\%$ of the total cross section comes from the problematic interval $M_{HH} \in (1800, 3000)$ GeV. 

In contrast,  for the $b=a^2$ case,   the situation regarding perturbative unitarity violation is very different.  As can be seen in the right plot in Fig. \ref{unitarity},  for  $b=a^2$, the behavior with energy changes  leading to a flat plateau at large energies that is below the unitarity crossing line. The only exception is for extremely large values of the low-energy effective parameters,  like $a=10$,  which are unrealistic nowadays since they are not allowed by present LHC constraints.  Therefore, one can conclude from this behaviour that assuming the particular correlation $b=a^2$ ensures that the predictions for $HH$ production via WBF preserve perturbative unitarity at the energies presently explored at the LHC and also at future planned $e^+e^-$ colliders. 

\begin{figure}[!t]
    \centering
        \includegraphics[scale=0.21]{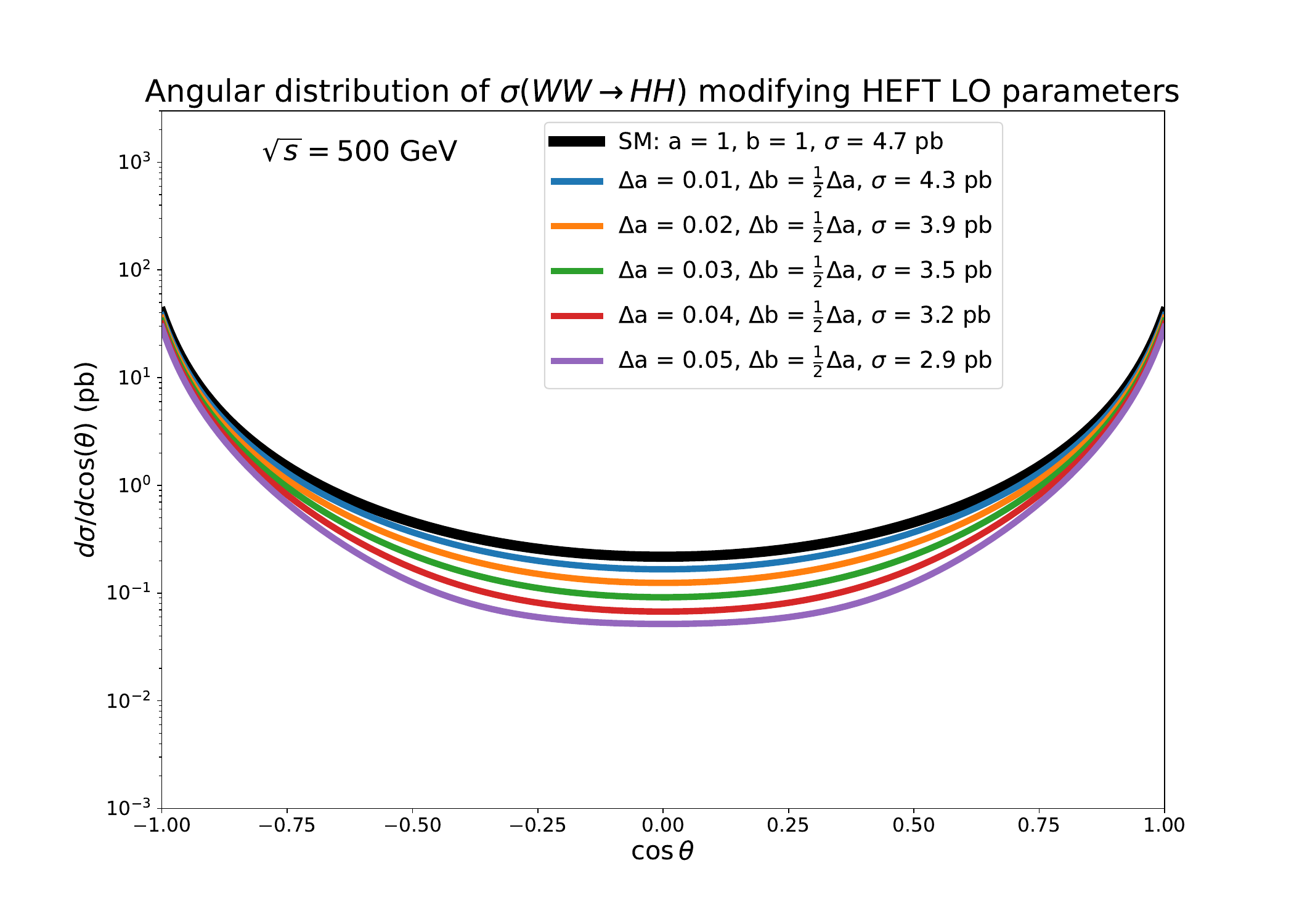}
        \includegraphics[scale=0.21]{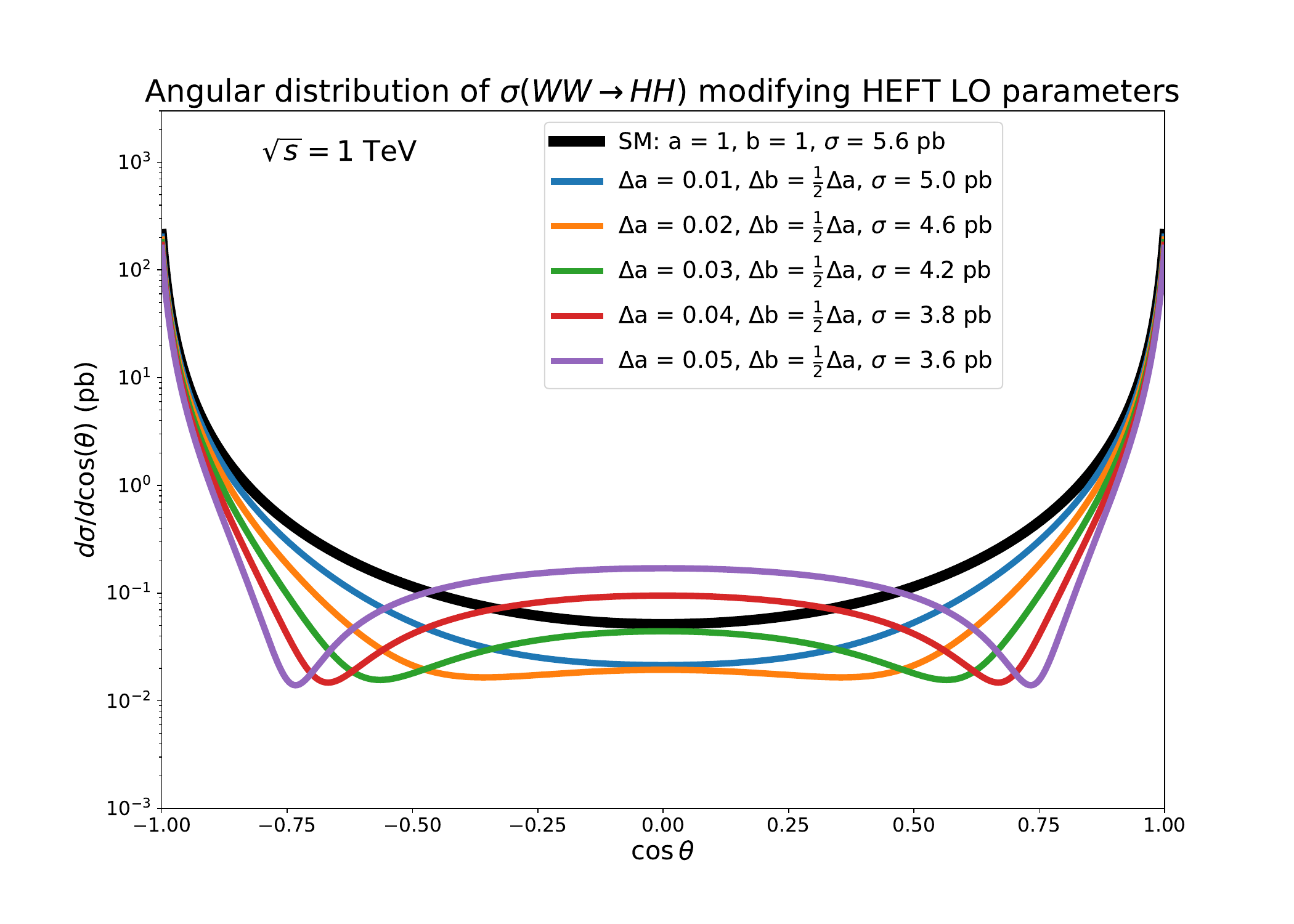}
        \\\hspace{0.1cm}
        \includegraphics[scale=0.4]{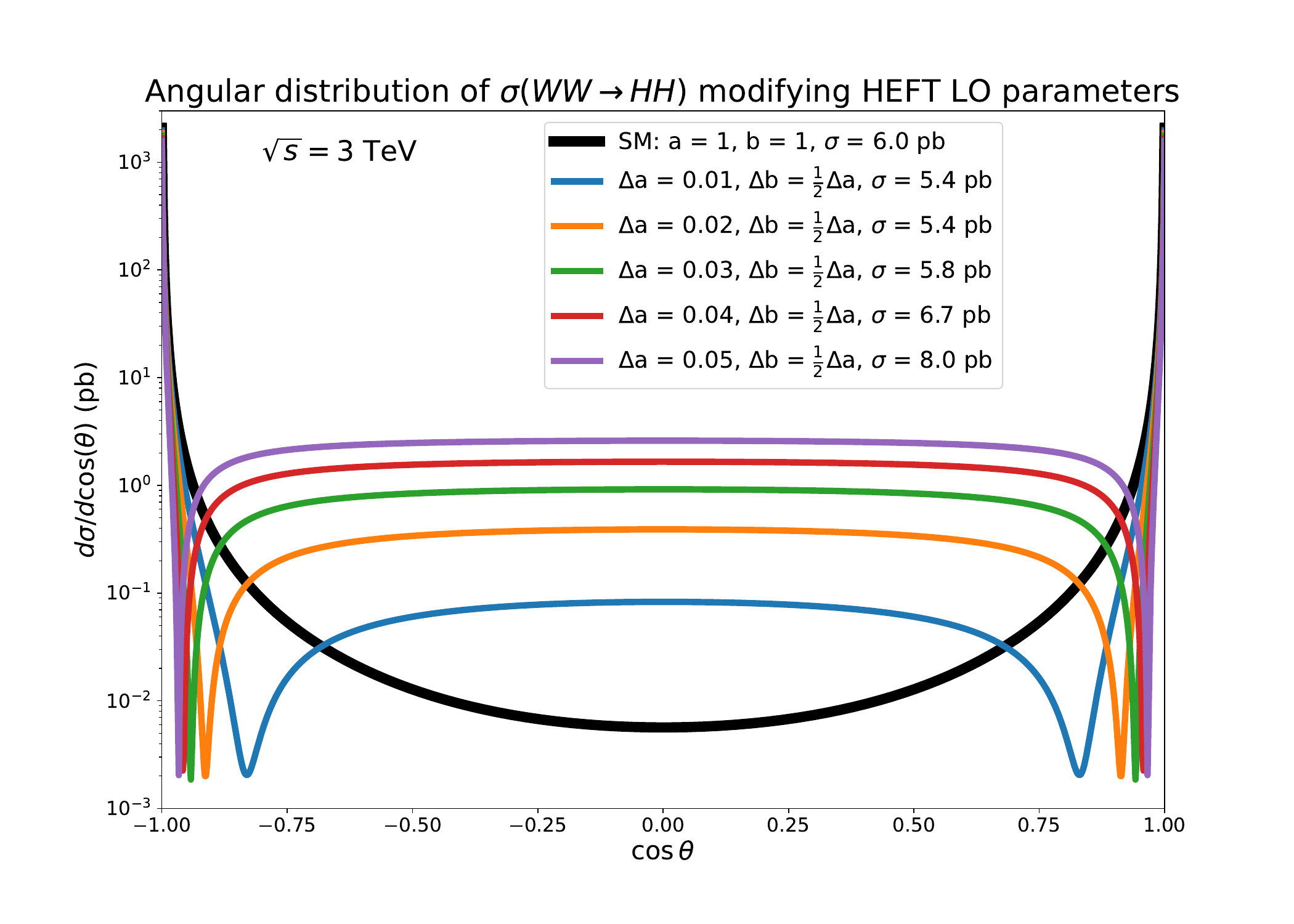}
    \caption{Predictions from the HEFT for the differential cross section of the process $W^+W^-\rightarrow HH$ with respect to $\cos \theta$, at center-of-mass energies of 500 GeV (upper left plot), 1 TeV (upper right plot) and 3 TeV (lower plot).  Several values of $a$ and $b$ are explored.  The corresponding $\Delta a$ and $\Delta b$ are assumed here to be correlated by  $\Delta b = \frac12 \Delta a$.}
    \label{angular_plots}
    \end{figure}

Other sensitive quantities where the combination of parameters $(a^2-b)$ also plays an important role are the differential cross sections with respect to angular variables  of the final state.  In particular,  for $W^+W^- \to HH$ scattering,  the differential cross section with respect to $\cos \theta$ already presents some interesting features. This can be easily computed from the previous amplitude in Eqs. \ref{eqn: total_amp} and \ref{eqn: amplitudes}.
In the CM frame this is given by: 
   \begin{equation}
        \frac{\text{d} \sigma}{\text{d} \cos\theta} \, = \,  \frac{1}{64 \pi s} \frac{\sqrt{s - 4m_H^2}}{\sqrt{s - 4m_W^2}}|\bar{\mathcal{A}}|^2,
        \label{eqn: diff_amp}
    \end{equation}
 where,  the average over the $3 \times 3$ polarization combinations of the initial $W$'s is considered in 
 $|\bar{\mathcal{A}}|^2$ and the factor $1/2$ due to the two identical final Higgs bosons is also included.  The predictions within the HEFT for this differential cross section with respect to $\cos\theta$ are shown in the plots of Fig. \ref{angular_plots} for the three CM energies,   $\sqrt{s}=500$ GeV (upper left),  $\sqrt{s}=1000$ GeV (upper right) and $\sqrt{s}=3000$ GeV (lower).  In these plots we explore the BSM effects from $a$ and $b$ and set one particular correlation given by $\Delta b= \Delta a /2$ for definiteness and just as an example.  We also include the SM prediction (black line) for comparison with respect to the BSM predictions.  
 The SM prediction gives a parabola with two maxima at the extreme values  $\cos\theta=\pm 1$ and a unique minimum at  
 $\cos\theta=0$.  In contrast,  the BSM prediction in the HEFT develops two minima which manifest in the plots with the larger energies (1 TeV and 3 TeV) symmetrically at both sides of the central point  $\cos\theta=0$.  The location of these minima gets closer to the extremes $\cos\theta=\pm 1$ for the largest energies.  Also for the largest energies, it is manifested the appearance of a flat and long plateau in the central region of these plots.  These commented features of the BSM HEFT predictions can be understood in terms of the dominant amplitude which,  as we have already reminded,  is ${\cal A}^L$.  The behaviour at high energies of ${\cal A}^L$ in Eq. \ref{eqn: total_amp_long} explains the appearance of the commented plateau,  in the shape of the differential cross section with respect to $\cos\theta$ and relates the height of this plateau with the value of $(a^2-b)$.  One can then conclude that the probability that the Higgs bosons get scattered in the central region for $\cos \theta$ increases in the HEFT as we stray from the condition $b = a^2$.  This is an important result, and will have phenomenological consequences for colliders, as we will see in the following sections.
 \begin{figure}[!t]
    \centering
        \includegraphics[scale=0.21]{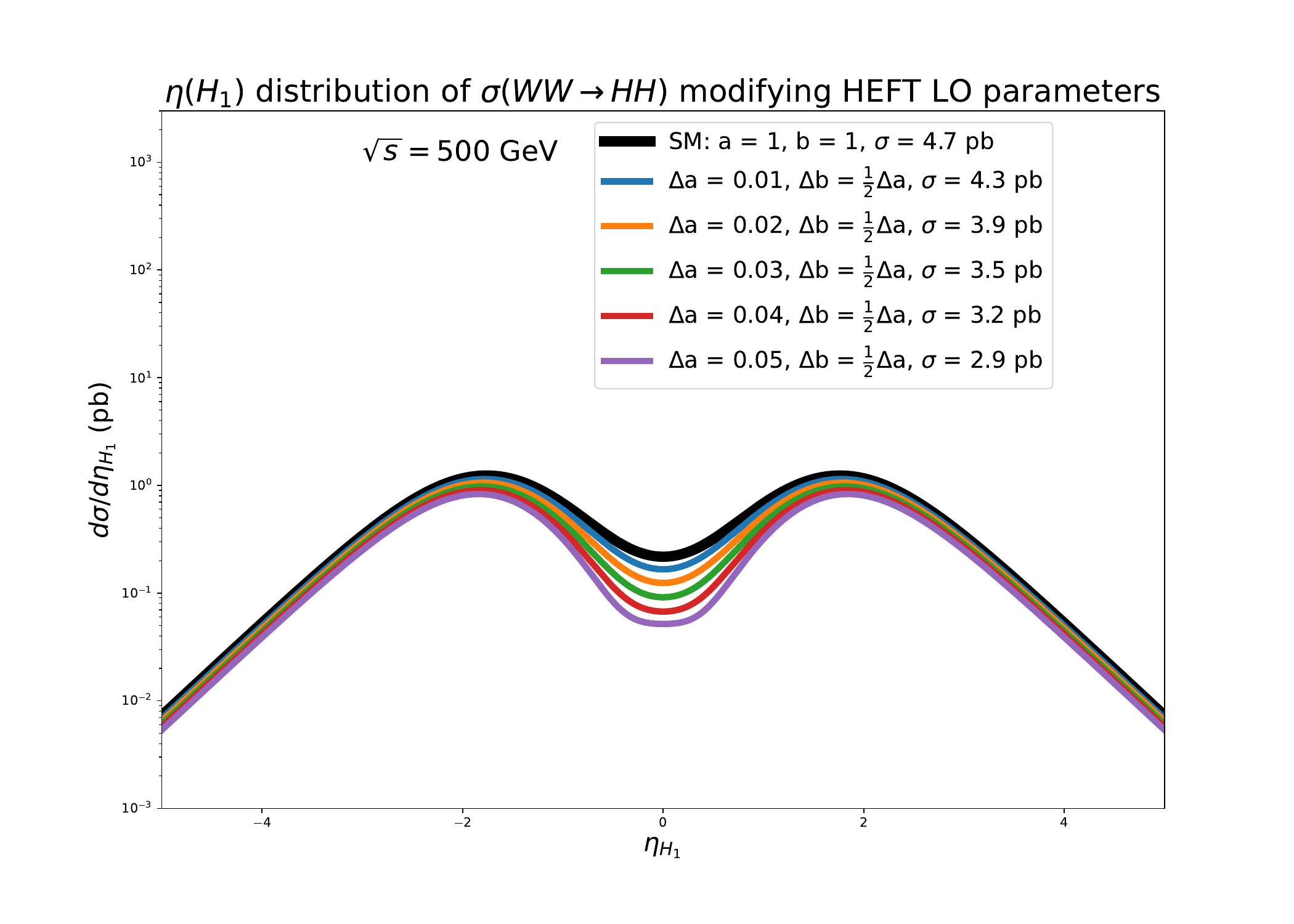}
        \includegraphics[scale=0.21]{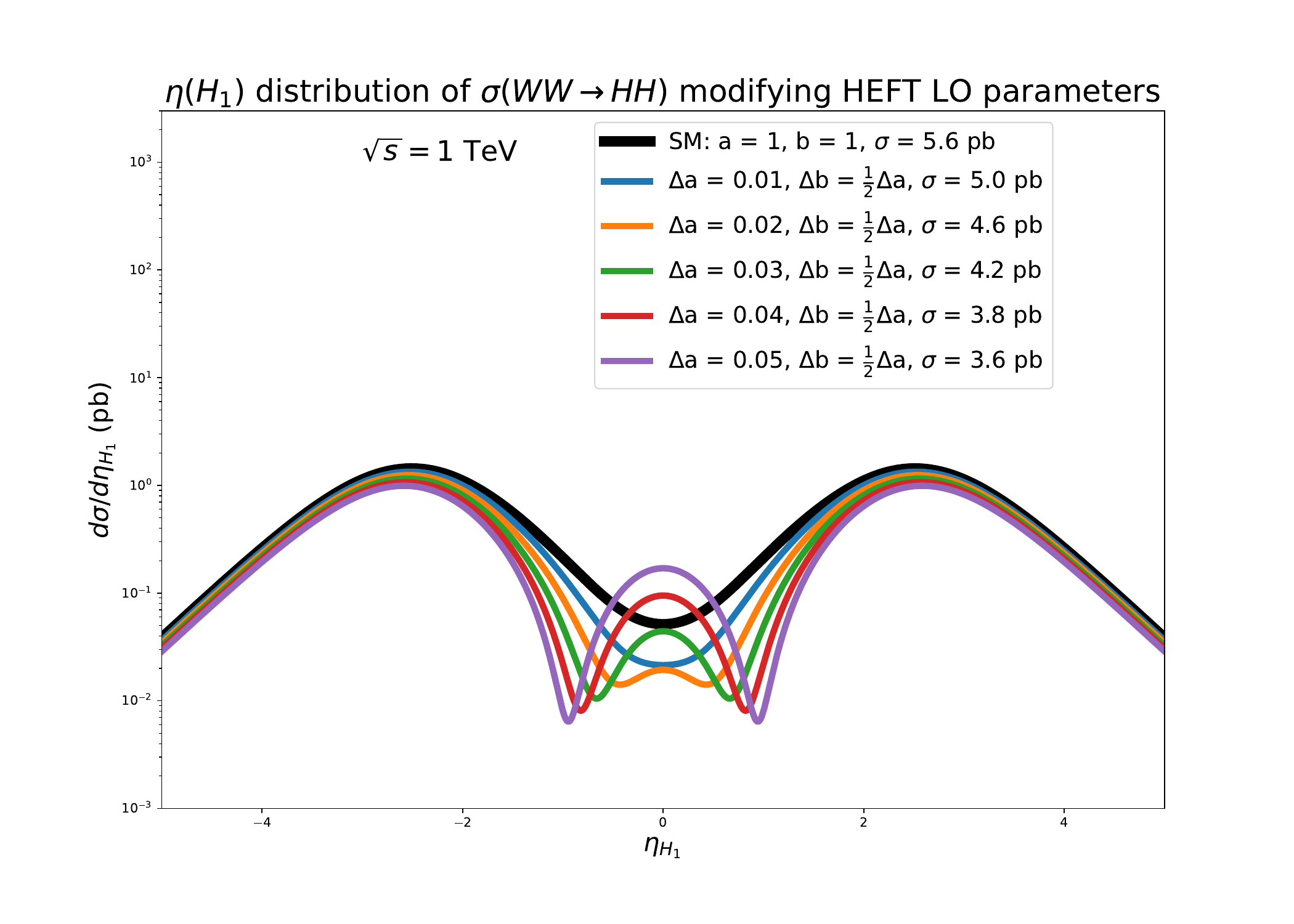}
        \\\hspace{0.1cm}
        \includegraphics[scale=0.4]{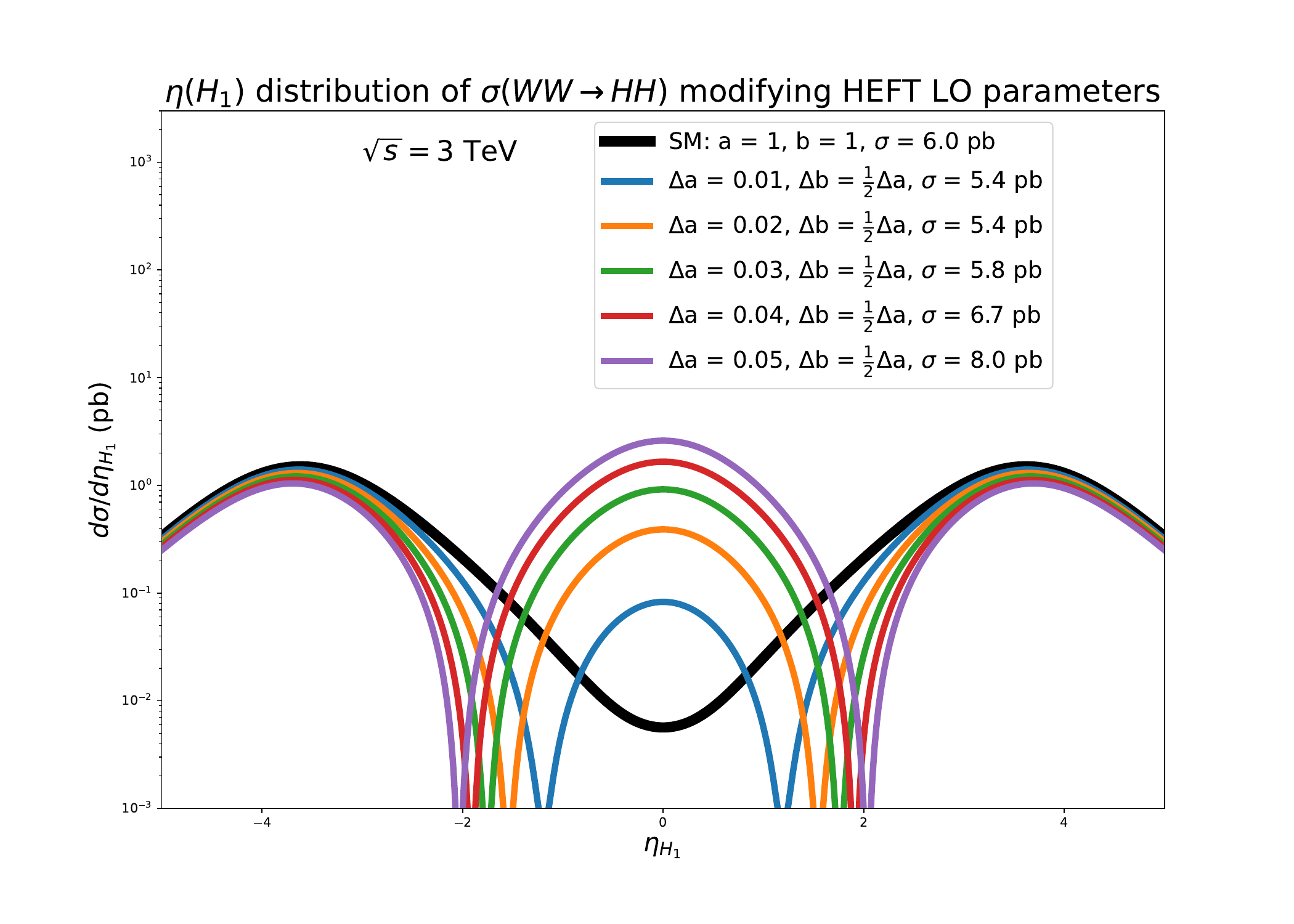}
    \caption{Predictions from the HEFT for the differential cross section of the process $W^+W^-\rightarrow HH$ with respect to $\eta_{H1}$, at center-of-mass energies of 500 GeV (upper left plot),  1 TeV (upper right plot) and 3 TeV (lower plot).  Several values of $a$ and $b$ are explored.  The corresponding $\Delta a$ and $\Delta b$ are assumed here to be correlated by  $\Delta b = \frac12 \Delta a$.}
    \label{angular_plots_eta}
    \end{figure} 
    
In relation to $\cos \theta$, there exist other interesting variables like the pseudorapidity and the transverse momentum of the final particles,  for the search of BSM effects in the process $W^+W^-\rightarrow HH$.  We explore here the pseudorapidity of  the final Higgs bosons $\eta_{H}$, which  is related to the angle $\theta$ by:
    \begin{equation}
        \eta_{H} \, = \, - \log \left( \tan (\theta/2) \right).
        \label{eqn: eta}
    \end{equation}
    The differential cross sections are therefore also related, 
    \begin{equation}
                \frac{\text{d} \sigma}{\text{d} \eta_{H}}=(\sin^2\theta)   \frac{\text{d} \sigma}{\text{d} \cos\theta} \,.
      \label{eqn: d_eta-versus-d_cos}
    \end{equation}
     Fig. \ref{angular_plots_eta} shows the plots for the differential cross section of the process $W^+W^-\rightarrow HH$ with respect to the pseudorapidity of one of the Higgs bosons,  $\eta_{H1}$,  again for the three energy cases,  500 GeV (upper left), 1000 GeV (upper right) and 3000 GeV (lower).  
   \begin{figure}[!t]
    \centering
        \includegraphics[scale=0.45]{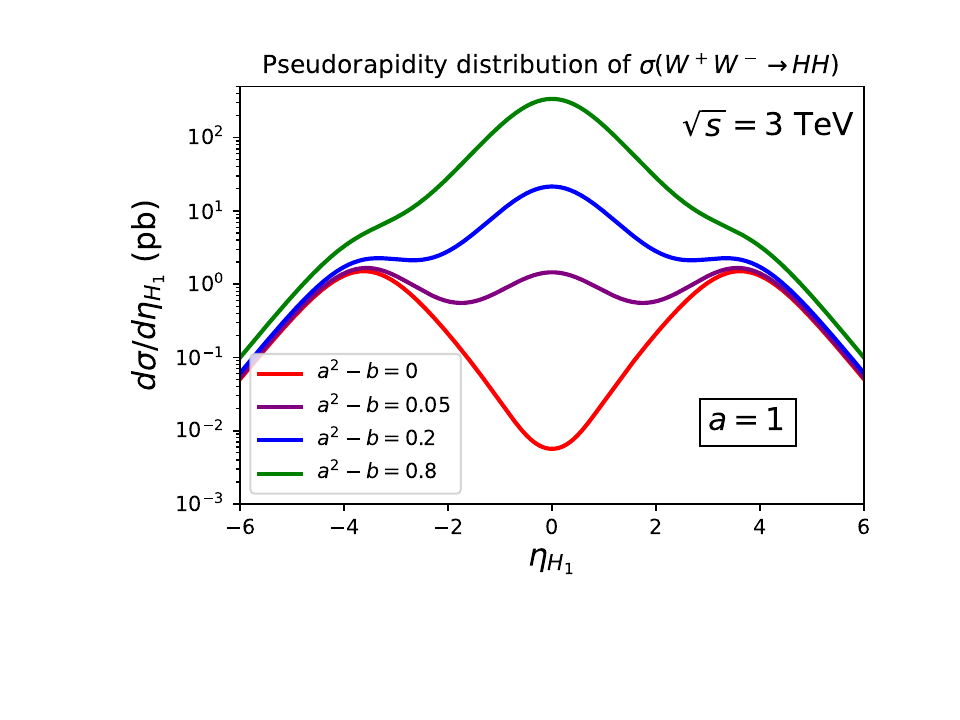}
        \includegraphics[scale=0.45]{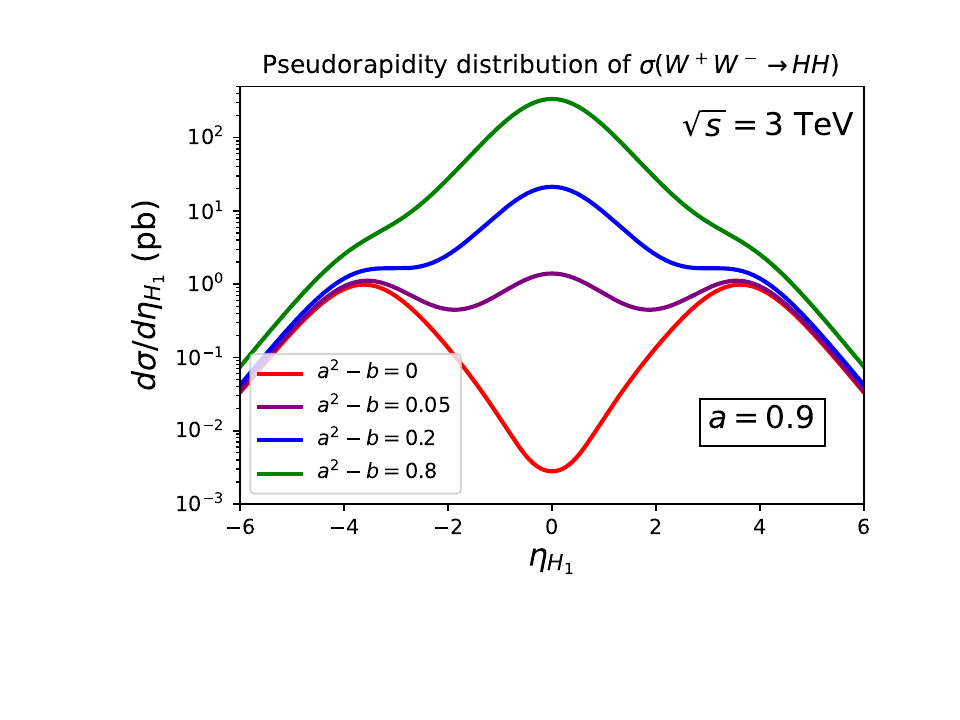}
        \\\hspace{0.1cm}
       \includegraphics[scale=0.45]{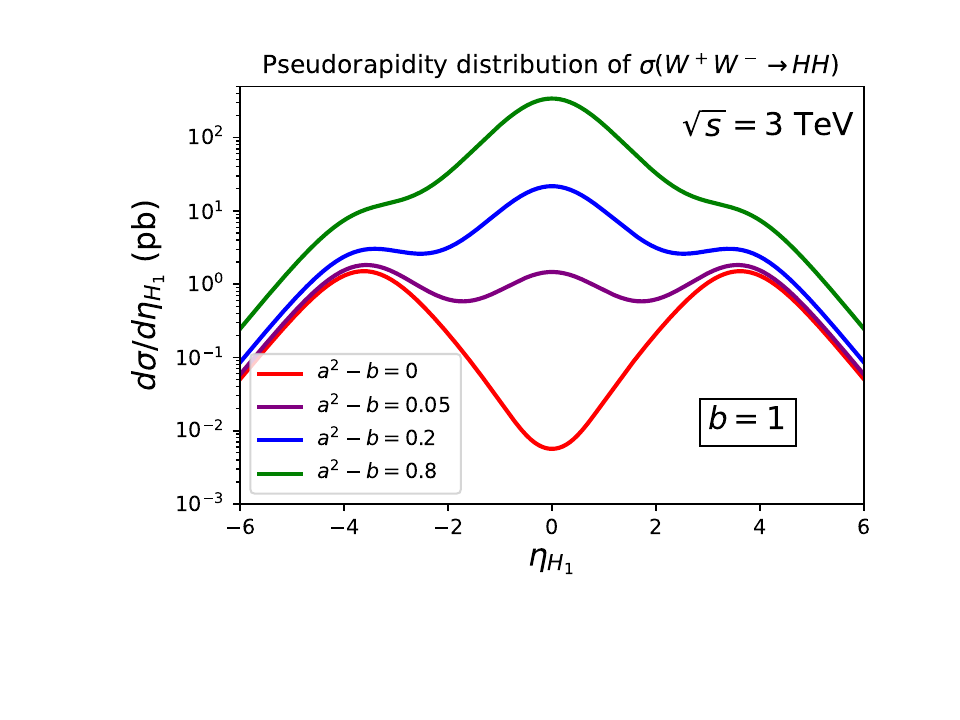}
           \includegraphics[scale=0.45]{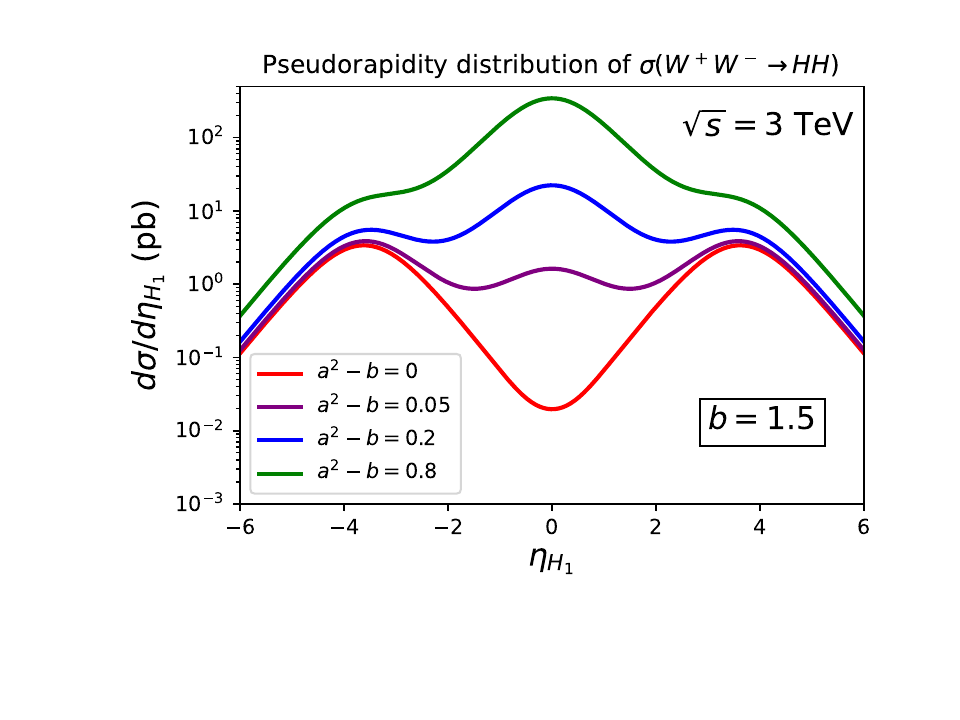}
    \caption{Predictions from the HEFT for the differential cross section of the process $W^+W^-\rightarrow HH$ with respect to $\eta_{H1}$,  for several values of the combination $a^2-b$.  The separate $a$ and $b$ values have been set to: $a=1$ (upper left plot),  $a=0.9$ (upper right plot),  $b=1$ (lower left plot) and $b=1.5$ (lower right plot). The center-of-mass energy is set in all plots to 3000 GeV.}
    \label{dsigma-detaH-a2-b}
    \end{figure}    
  We explore various values of $a$ and $b$ and set again the correlation $\Delta b = \frac12 \Delta a$ as an example.  
 First we see that small departures in $a$ and $b$ produce a noticeable change of behaviour with respect to the SM prediction (lines in black) at high energies.  Whereas the case with 500 GeV displays the same shape as in the SM,  with a minimum at the central value $\eta_{H1} = 0$ (which corresponds to $\theta=\frac{\pi}{2}$),  the two other cases with higher energies display the appearance of a new local maximum at this central value.   In particular,   at the 3 TeV plot this maximum is more clear and it can be converted into a global maximum for some $a$ and $b$ values.  Indeed,  this maximum gets higher as we stray from the SM with deviations in $a$ and $b$ whenever $a^2 \neq b$.  The interpretation of this feature is as follows:  the existence of this maximum at $\eta_{H1} = 0$ implies that, whenever $b \not= a^2$,  the Higgs bosons of the final state tend to travel in a perpendicular direction to the incoming $W$ bosons,  meaning that they have a high-transversality behaviour. 
 These are truly BSM features which are worth to further explore.  
 In particular, they will have interesting phenomenological BSM implications for colliders, which we will present in the next sections, where we will analyse the process $e^+e^- \to HH \nu \bar\nu$. 
  
Finally,  in order to illustrate clearly the role played by the particular combination of parameters $(a^2-b)$ in the transversality of the Higgs bosons produced via  this $W^+W^-\rightarrow HH$ subprocess,  we show in Fig. \ref{dsigma-detaH-a2-b} the differential cross-section with respect to $\eta_{H1}$,  for several values of the combination $(a^2-b)$ and for several settings of the separate values of $a$ and $b$.  Specifically, we explore here the values $a^2-b=0,0.05,0.2,0.8$ together with the separate settings for $a$ and $b$ given by:  $a=1$ (upper left plot),  $a=0.9$ (upper right plot),  $b=1$ (lower left plot),  and $b=1.5$ (lower right plot).  The appearance of a maximum in all plots at $\eta_{H1}=0$ and the height of this maximum is clearly correlated with the size of 
$(a^2-b)$,  as expected.  The larger  the value of $(a^2-b) \neq 0$ is,  the higher is the maximum in this differential cross section.  And this is true for all the separate values of $a$ and $b$  explored.  This behaviour with $(a^2-b)$ is a key point for the posterior analysis of the full process in the following sections.

\section{Exploring correlations between  $\kappa_V$ and $\kappa_{2V}$ in total and differential cross sections for $e^+e^- \to HH \nu \bar \nu$}
\label{xsectionee}
    \begin{figure}[!t]
    \centering
   \includegraphics[height=0.27\textheight]{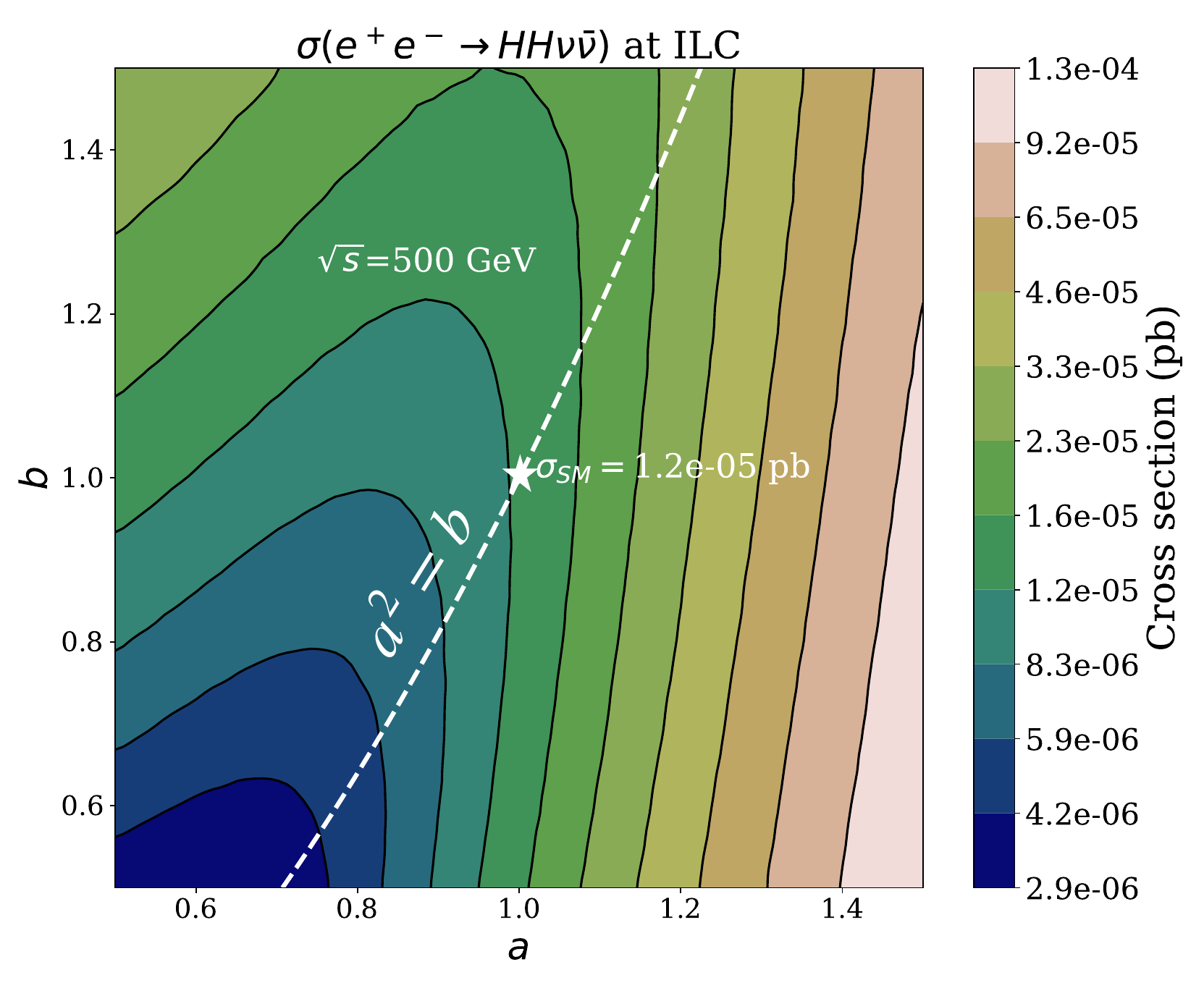}
   \includegraphics[height=0.27\textheight]{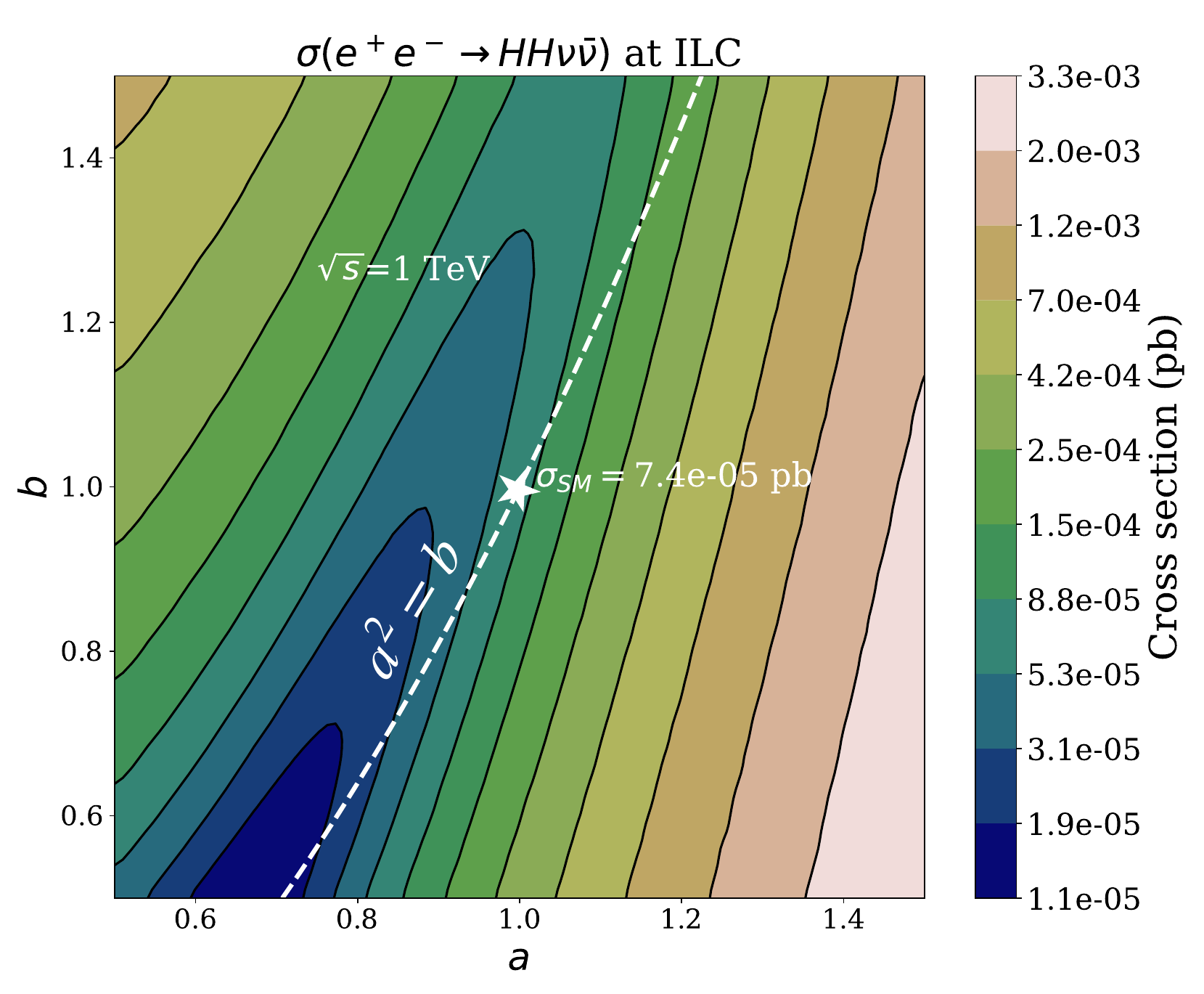}\\\hspace{0.1cm}
    \includegraphics[height=0.5\textheight]{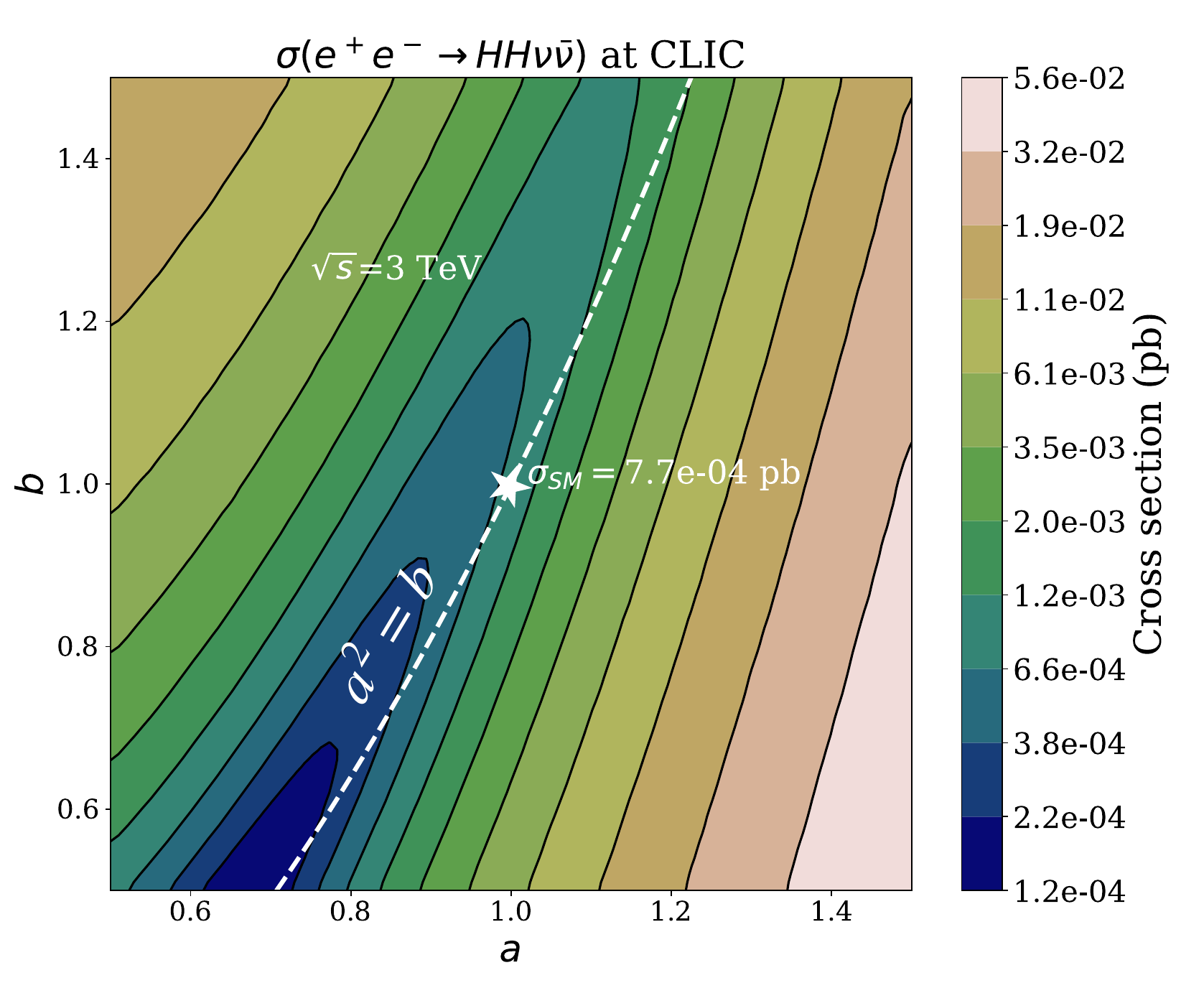}     
 \caption{Predictions from the HEFT for the contourlines of  cross section $\sigma (e^+ e^- \rightarrow HH\nu \bar{\nu}$) (in pb) in the $(a,b)=(\kappa_V, \kappa_{2V})$ plane,  for the three planned energies at future $e^+e^-$ colliders  of 500 GeV (upper left panel), 1 TeV (upper right panel) and 3 TeV (lower panel).  The SM cross section predictions are also displayed at the $(a,b)=(1,1)$ point marked here with a star.  The dashed white line in each plot represents the correlation between the HEFT parameters given by $a^2=b$.   All predictions shown here include a cut on the missing transverse energy from the final $\nu \bar{\nu}$ of $ \slashed{E}_T > 20 \, \text{GeV}$.}
    \label{contour_plots}
    \end{figure}
In this section 
we explore the phenomenological consequences of assuming different correlations between $a$ and $b$ both in the total and some selected differential cross sections for the $e^+e^-$ colliders.  
Specifically we study the most relevant process, where a pair of  Higgs bosons and a pair of neutrino-antineutrino are produced: $e^+e^- \to HH \nu \bar \nu$.  For this computation we use  \textsc{MadGraph5} (MG5) \cite{Alwall:2014hca} which generates and accounts for all the Feynman diagrams contributing to the full scattering process,  $e^+e^- \to HH \nu \bar \nu$.   All the participating diagrams are taken into account and, for completeness,  are displayed in the Appendix \ref{Diagrams}. These include diagrams with $WW$ fusion configuration (diags. 5, 6, 7, and 8) and  diagrams with an intermediate  $Z$ boson that decays to $\nu \bar \nu$ (diags. 1, 2, 3 and 4). 
Just the first configuration allows to explore both $\kappa_V$ and $\kappa_{2V}$ simultaneously at tree level in the colliders. In addition, the latter configuration is known to be highly subdominant as compared to the $WW$ fusion ones,  in the case of  $e^+e^-$ colliders with TeV energies.  For a comparison of these two contributions in the SM case,  see for instance Ref.~\cite{Gonzalez-Lopez:2020lpd}.

Our predictions of the contourlines for the total cross section
 $\sigma(e^+e^- \to HH \nu \bar \nu)$ in the $(a,b)$ plane  and  for the three selected energies
$\sqrt{s}=500$ GeV (upper left), $1000$ GeV (upper right) and $3000$ GeV (lower) are shown in Fig. \ref{contour_plots}.  In these plots we have also included the line that defines the correlation $a^2-b=0$ (dashed white line) and the explicit prediction for the SM cross section at the $(a,b)=(1,1)$ point. 
Regarding these three contourplots we first notice the good agreement found in the numerical results with Ref.~\cite{Gonzalez-Lopez:2020lpd}.  As expected, the size of the cross section for this $e^+e^- \to HH \nu \bar \nu$ process within the region of the $(a,b)$ parameter space explored in this figure is in general larger for the larger energy colliders, reaching sizeable values at 
CLIC(3000 GeV) of up to around 0.056 pb in the lower right corner of this plot near $(a,b)=(1.5,0.5)$.  The corresponding cross section values in this region for ILC(500 GeV) and ILC(1000 GeV) are $1.3 \times 10^{-4}$ pb and $3.3 \times 10^{-3}$ pb , respectively.  These BSM cross sections should be compared with the corresponding SM predictions which are $1.2 \times 10^{-5}$ pb,
$7.4 \times 10^{-5}$ pb and $7.7\times 10^{-4}$ pb for ILC(500 GeV),  ILC(1000 GeV) and CLIC(3000 GeV) respectively.  It is clear that CLIC with the highest energy of 3000 GeV will offer the best sensitivity to the 
$\kappa_V$ and $\kappa_{2V}$ parameters, improving it respect to the present sensitivity at the LHC. 
  
   Second,  the appearance of a minimum located in the dark blue area is clearly manifested in the lower left quadrant of these three plots,  showing small cross section values of around
 $2.9 \times 10^{-6}$ pb,  $1.1 \times 10^{-5}$ pb,  and  $1.2 \times 10^{-4}$ for 500 GeV,  1 TeV and 3 TeV collider energies,  respectively.   Third,  the line with $a^2=b$ tends to be more parallel to the contourlines when the energy is increased from 500 to 3000 GeV,  and this line enters in the region of minimum cross section at the lower left quadrant.  This confirms our expectations from the analysis of the subprocess in Sec. \ref{WBF-HH}.  At the higher energies,  the WBF diagrams dominate,  and the cross section behaviour of the full process with $(a^2-b)$ reproduces approximately the corresponding  behaviour of the subprocess.  In particular,  the subprocess is dominated by the $LL$ amplitude where we have seen that the largest contribution of ${\cal O}(s)$ vanishes for $(a^2-b)=0$. Consequently,  for the highest energy of 3000 GeV,  we find the dashed line almost parallel and close to the contourlines of lower cross section.  This does not happen in the 500 GeV case because the WBF diagrams do not dominate and the $Z$ mediated diagrams are instead the most relevant ones.  The case of  1000 GeV is in between the two other cases,  but yet the WBF dominate at these energies and we can see the dashed line better aligned with the contourlines than in the 500 GeV case. 

\begin{figure}[!t]
    \centering
        \includegraphics[height=0.28\textheight]{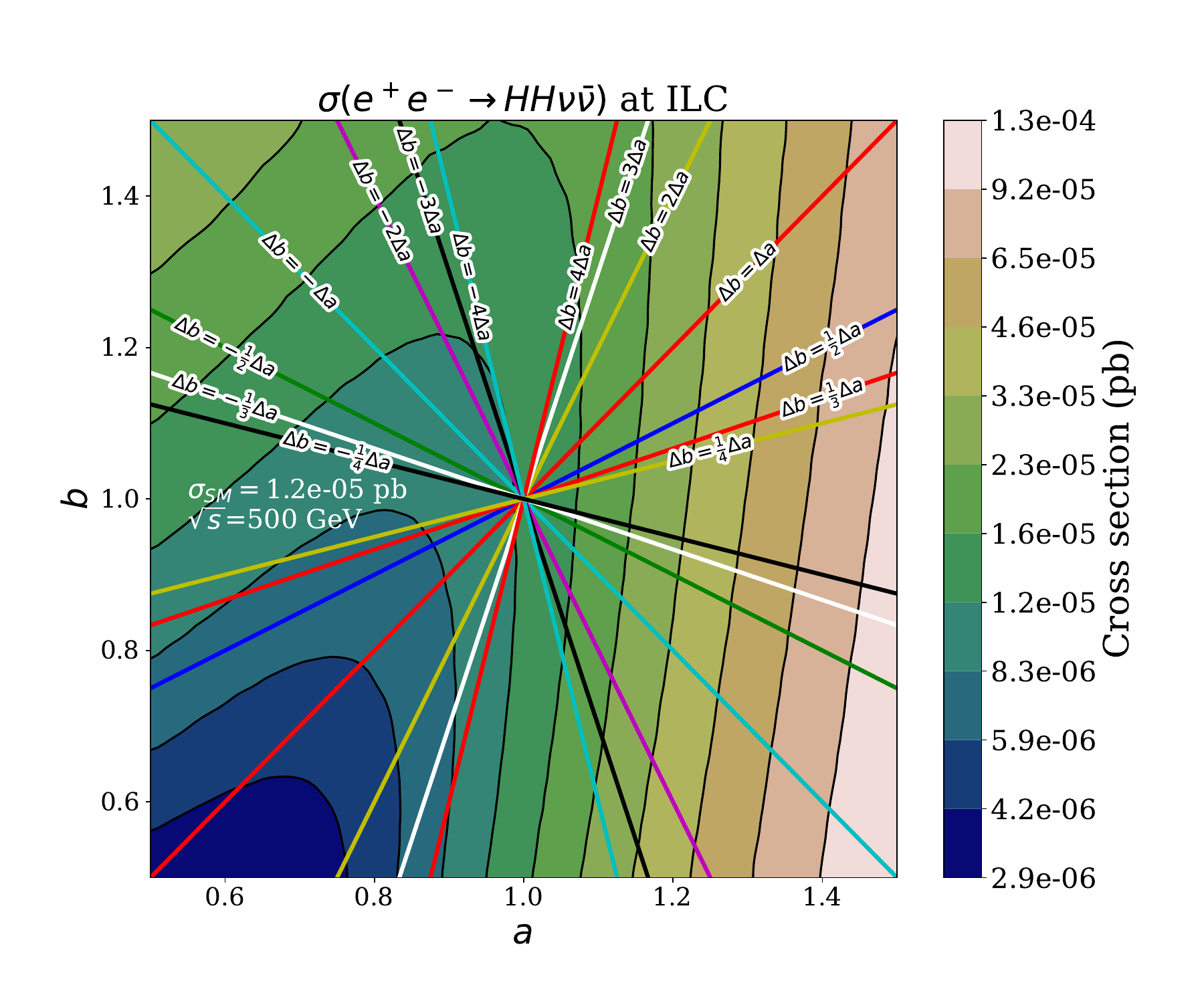}
        \includegraphics[height=0.28\textheight]{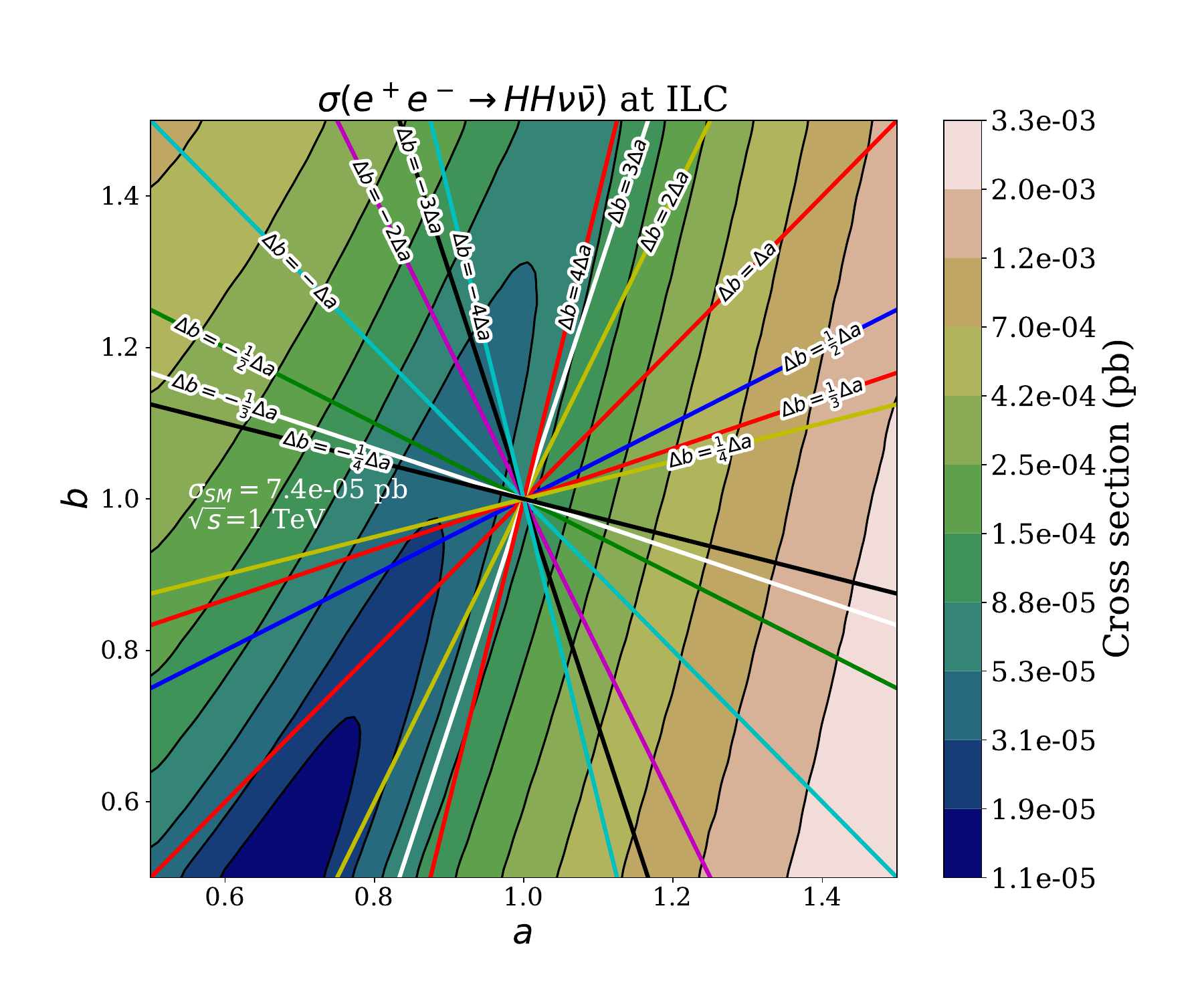}\\\hspace{0.1cm}
        \includegraphics[height=0.5\textheight]{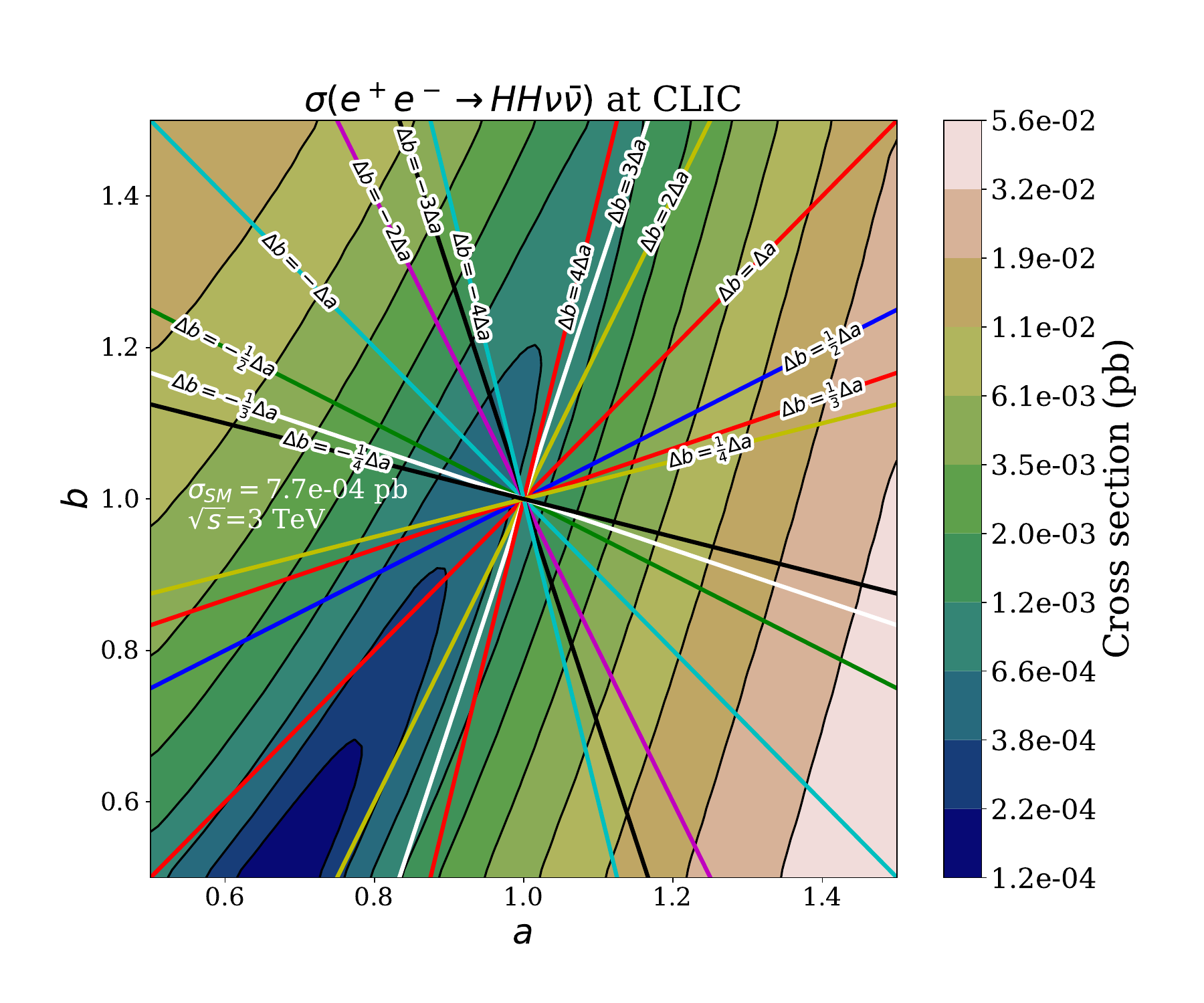}
    \caption{Comparison of the predictions from the HEFT for contourlines of cross section $\sigma (e^+ e^- \rightarrow HH\nu \bar{\nu}$) (in pb), in the $(a,b)=(\kappa_V, \kappa_{2V})$ plane,  with the different correlation hypotheses between the HEFT parameters given by $\Delta b = C \Delta a$  with $C= \pm \frac{1}{4},  \pm \frac{1}{3},  \pm \frac{1}{2},  \pm 1,  \pm 2,  \pm 3,  \pm 4$ (coloured straight lines crossing at $(a,b)=(1,1)$). The three $e^+e^-$ collider cases are displayed:  500 GeV (upper left panel), 1 TeV (upper right panel) and 3 TeV (lower panel). The SM cross section predictions are also displayed.  All predictions shown here include a cut on the missing transverse energy from the final $\nu \bar{\nu}$ of $ \slashed{E}_T > 20 \, \text{GeV}$.}
    \label{contour_plots_star}
    \end{figure}
We next wish to contrast the previous results of the cross section in the ($a,b$) plane with other specific correlations different than $(a^2-b)=0$.  For this comparison we define the correlations in terms of the coefficient deviations,  i.e.  in terms of $\Delta a$ and $\Delta b$ which are assumed here to be linearly related by
$\Delta b = C \Delta a$ where $C$ is a given correlator factor. 
The results of this comparison are shown in Fig. \ref{contour_plots_star} where the various coloured straight lines crossing at the SM point $(a,b)=(1,1)$ define our various choices for the correlator factor $C$. Concretely,  we explore here 14  different correlations given by 
$C= \pm \frac{1}{4},  \pm \frac{1}{3},  \pm \frac{1}{2},  \pm 1,  \pm 2,  \pm 3,  \pm 4$ that correspond to the 14 coloured lines in Fig. \ref{contour_plots_star}.  We recall that $\Delta b= 2 \Delta a$ is the linearized version of $a^2=b$.  Therefore this particular line in  Fig. \ref{contour_plots_star} approaches the dashed line of the previous Fig. \ref{contour_plots}.  We also recall that this $\Delta b= 2 \Delta a$  case has been found in some models like dilaton models,  and in models with iso-singlet mixing. The case $\Delta b=-2 \Delta a$ has been found in the 2HDM.  And the case $\Delta b= 4 \Delta a$ has been found in SMEFT,  in MCHM,  and in SILH. 
The most important conclusion from this figure is the following:  in BSM scenarios where the departures from the SM prediction at $(a,b)=(1,1)$ are obtained by moving through the $(a, b)$ plane in these different directions given by the straight coloured lines,  we find that the reach to a given contourline with fixed cross section is done with different sensitivities for each correlation.  For instance,   in the upper left quadrant of the 3 TeV plot,   the contourline of $\sigma=1.2 \times 10^{-3}$ pb,   which is about two times the SM value,   is reached first (i.e. with the shortest distance) by the $\Delta b= -\frac{1}{2} \Delta a$ line,  and then by others,  in the order $\Delta b= -\frac{1}{3} \Delta a$,   $\Delta b= -\frac{1}{4} \Delta a$,   $\Delta b= - \Delta a$,  $\Delta b= - 2 \Delta a$,  $\Delta b= - 3 \Delta a$ and $\Delta b= - 4 \Delta a$.   In general,  the more perpendicular lines to the main axis of these parabolic curves  the easier is to test such correlation hypotheses.  Then,  in this upper left quadrant of the CLIC plot the correlation  $\Delta b= -\frac{1}{2} \Delta a$ is expected to be the easiest to be tested.  In the lower left quadrant,  the  correlation that is the easiest to test is $\Delta b=\frac{1}{4} \Delta a$,  then in order $\Delta b=\frac{1}{3} \Delta a$,  $\Delta b=\frac{1}{2} \Delta a$ and  $\Delta b= \Delta a$.   Other correlations like   $\Delta b= 2 \Delta a$,  $\Delta b=3 \Delta a$ and $\Delta b= 4 \Delta $ seem very difficult to be tested because the corresponding lines lay close to the contourlines with the lowest cross section.  We will discuss in more detail the accessibility to those correlations in the next section,  once we determine the realistic final state particles,  i.e.  after the decays of the final $HH$ pairs. 

\begin{figure}[!t]
    \centering
        \includegraphics[height=0.182\textheight]{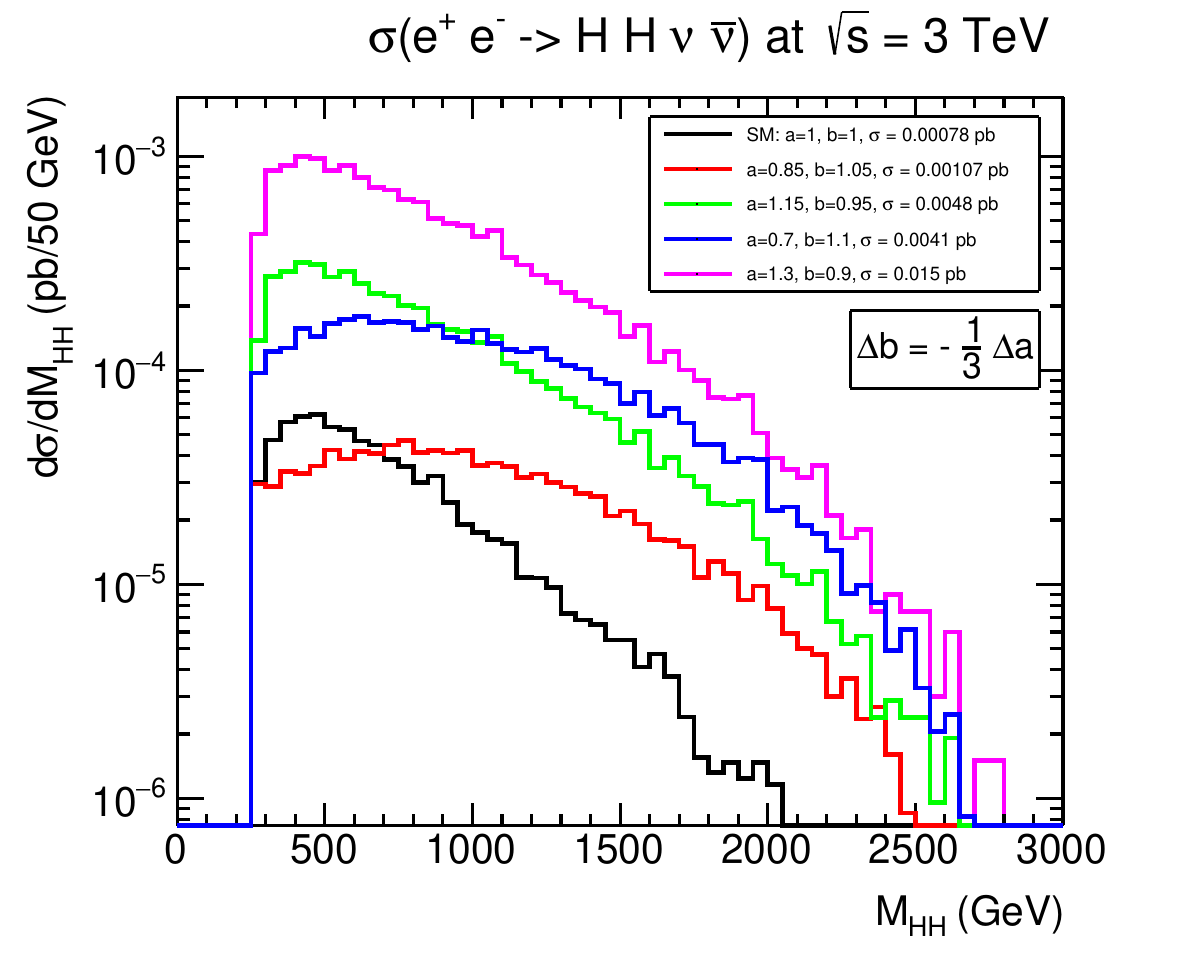}
        \includegraphics[height=0.182\textheight]{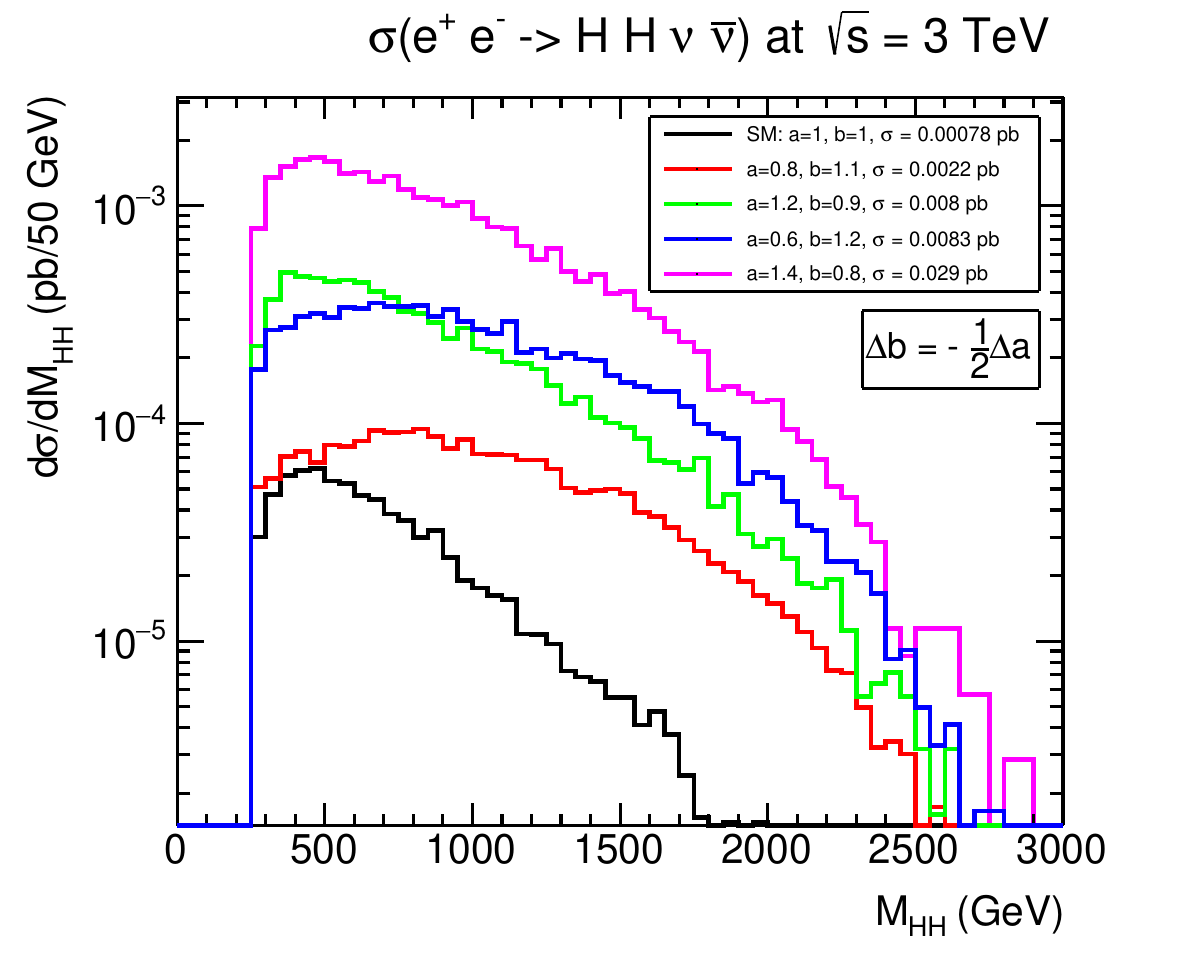}
        \includegraphics[height=0.182\textheight]{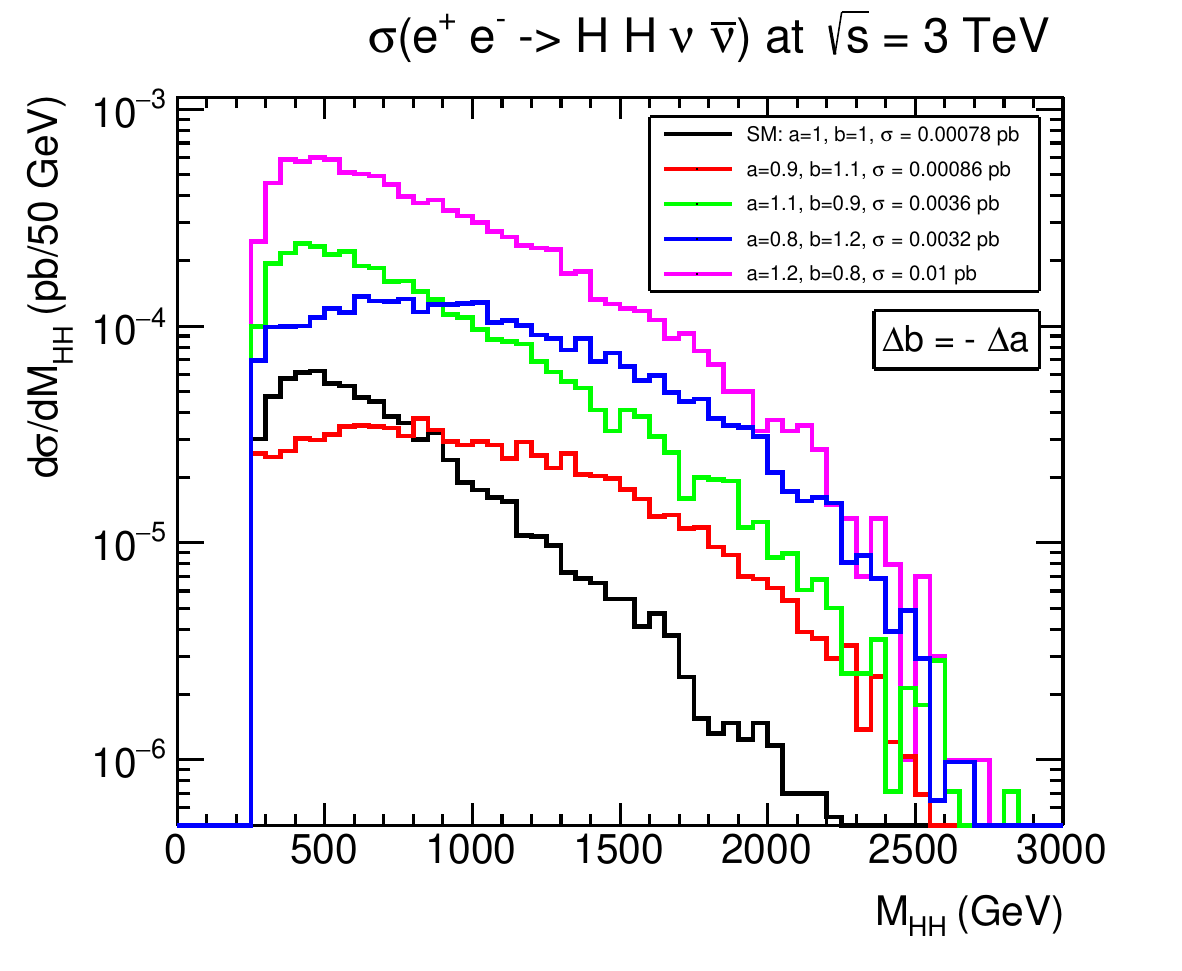}\\\hspace{0.1cm}
        \includegraphics[height=0.182\textheight]{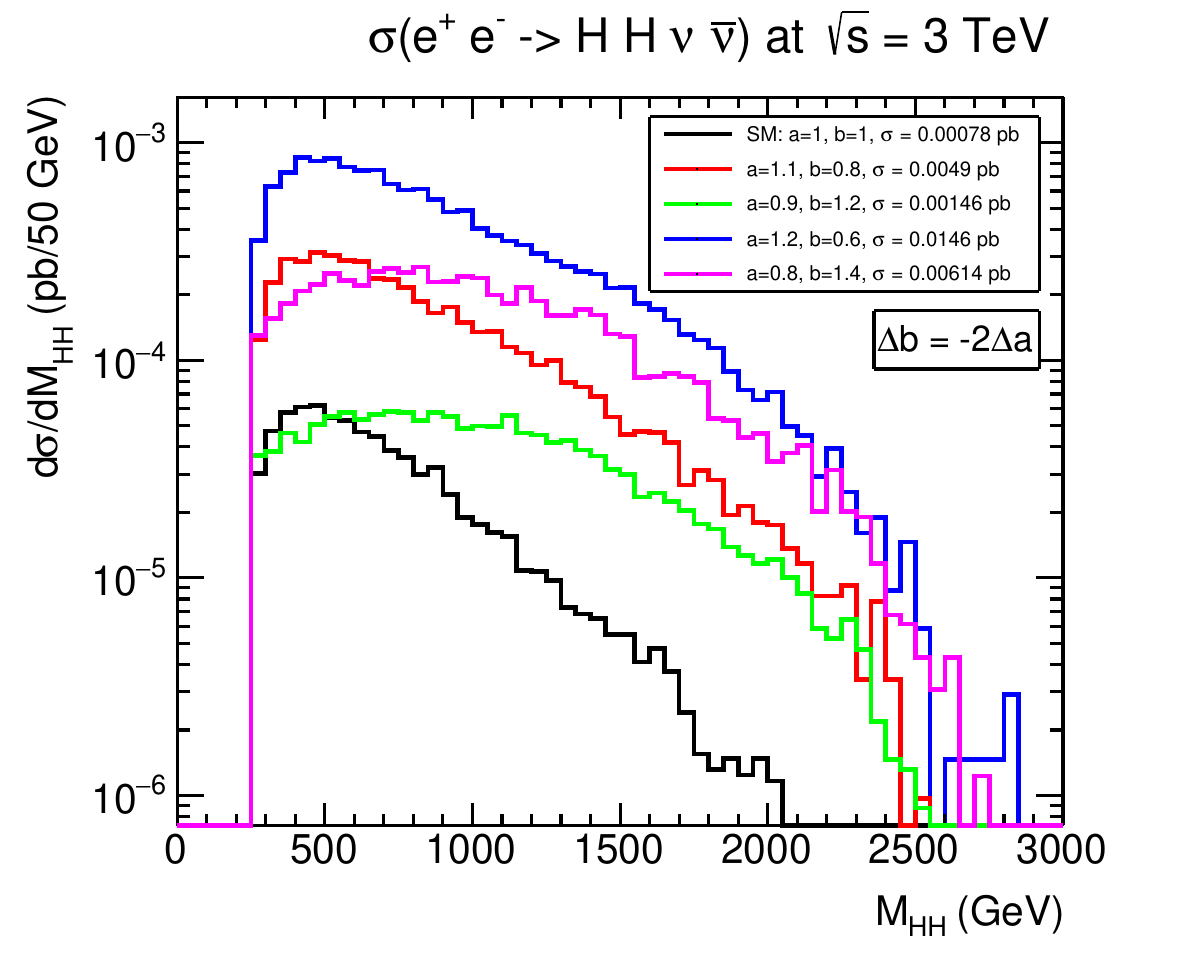}
        \includegraphics[height=0.182\textheight]{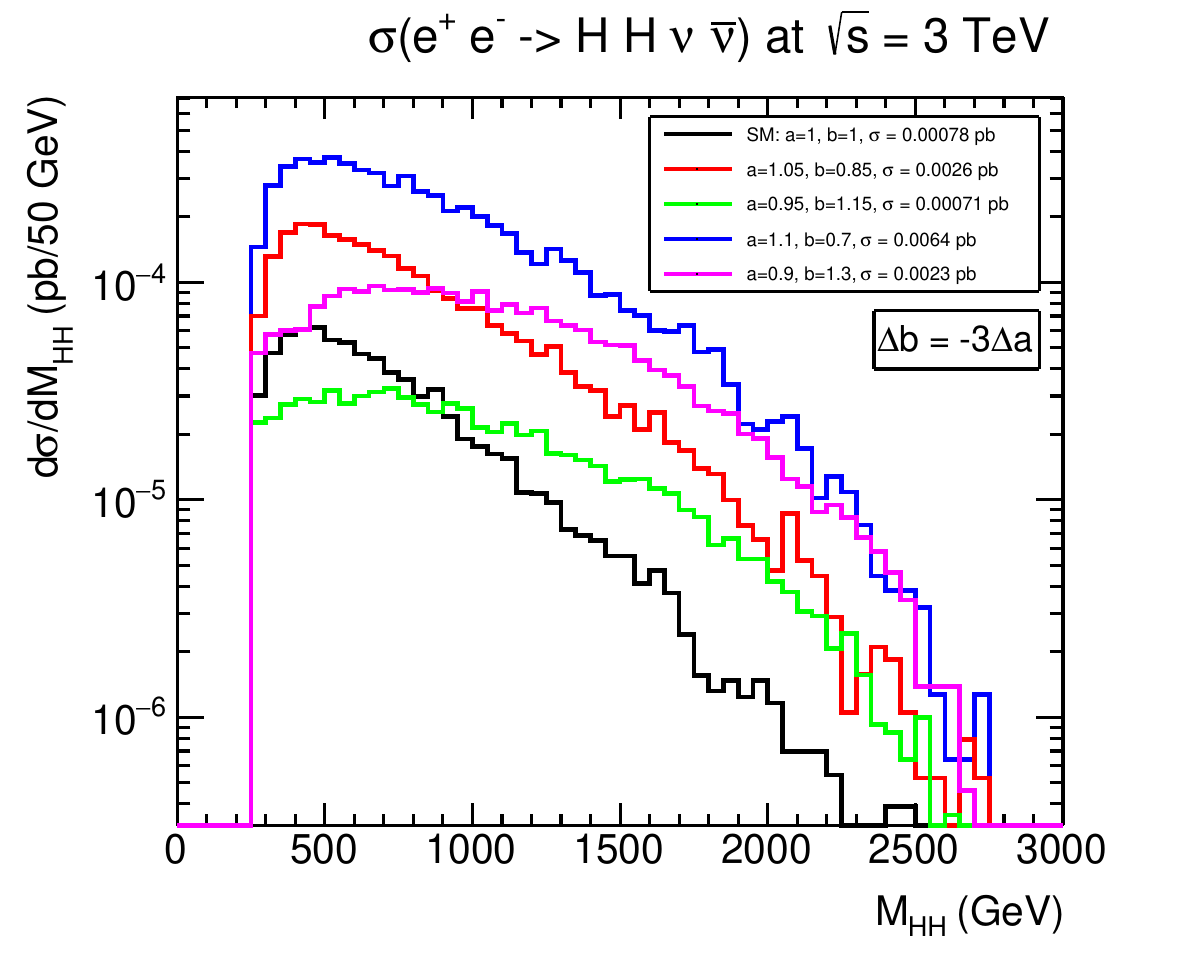}
        \includegraphics[height=0.182\textheight]{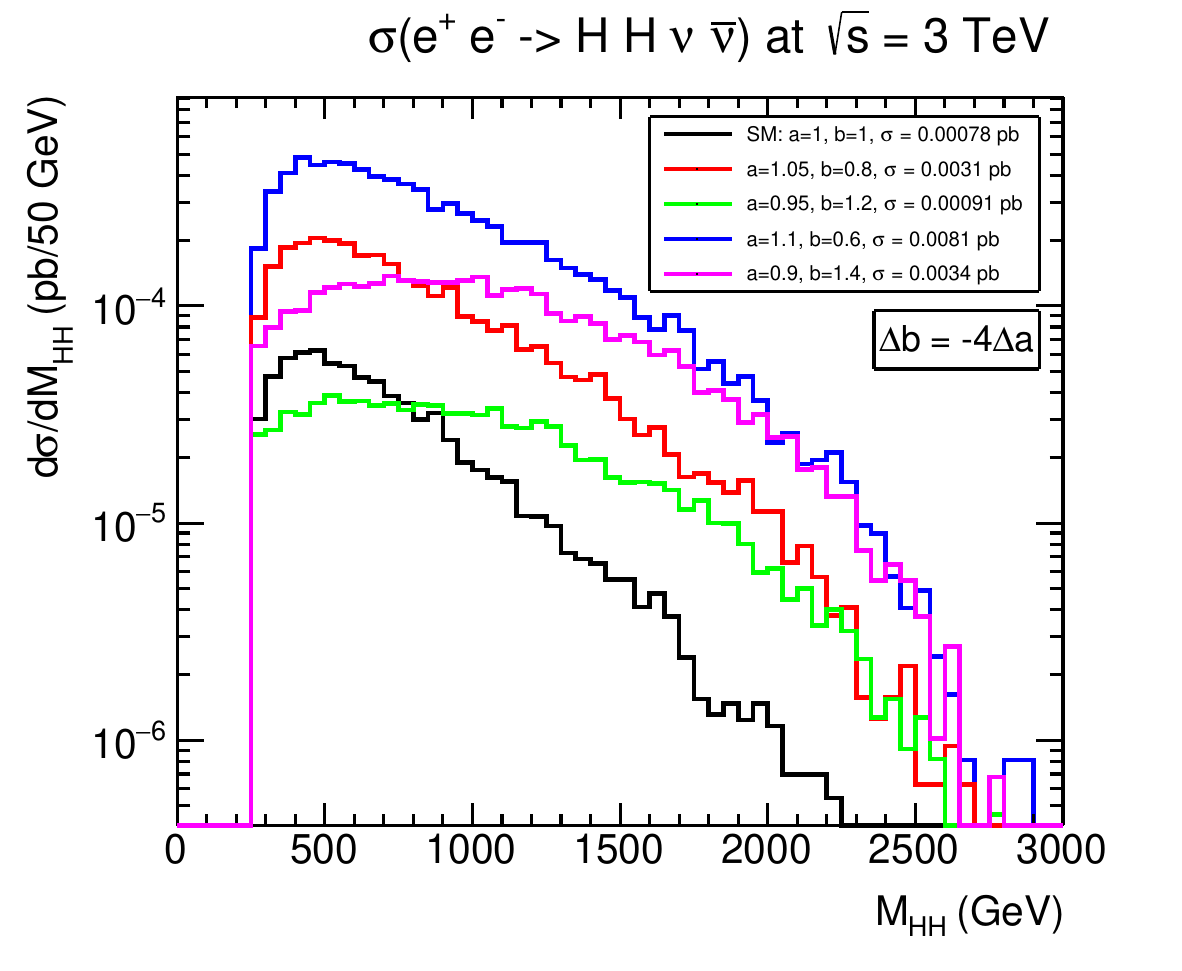}\\\hspace{0.1cm}
        \includegraphics[height=0.182\textheight]{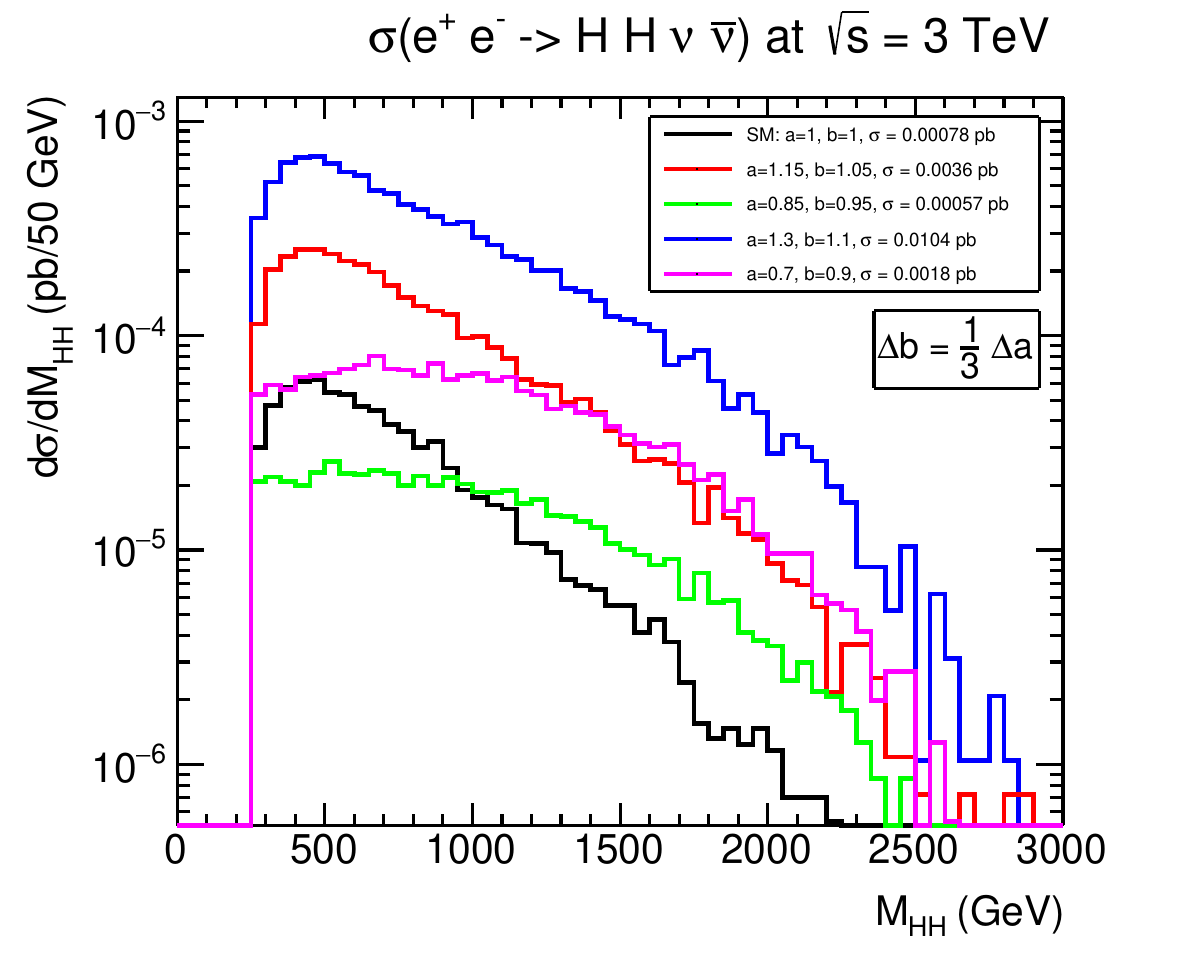}
        \includegraphics[height=0.182\textheight]{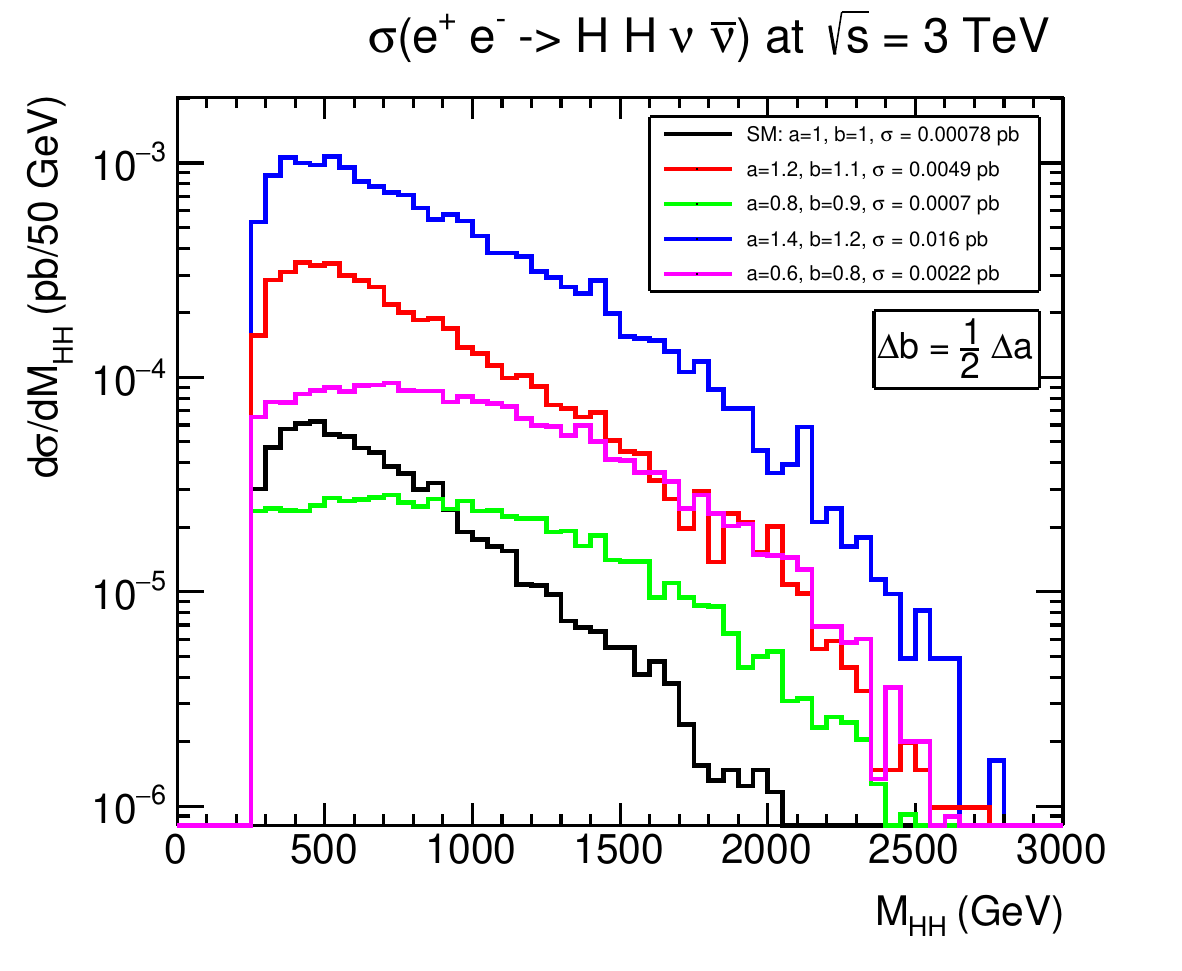}
        \includegraphics[height=0.182\textheight]{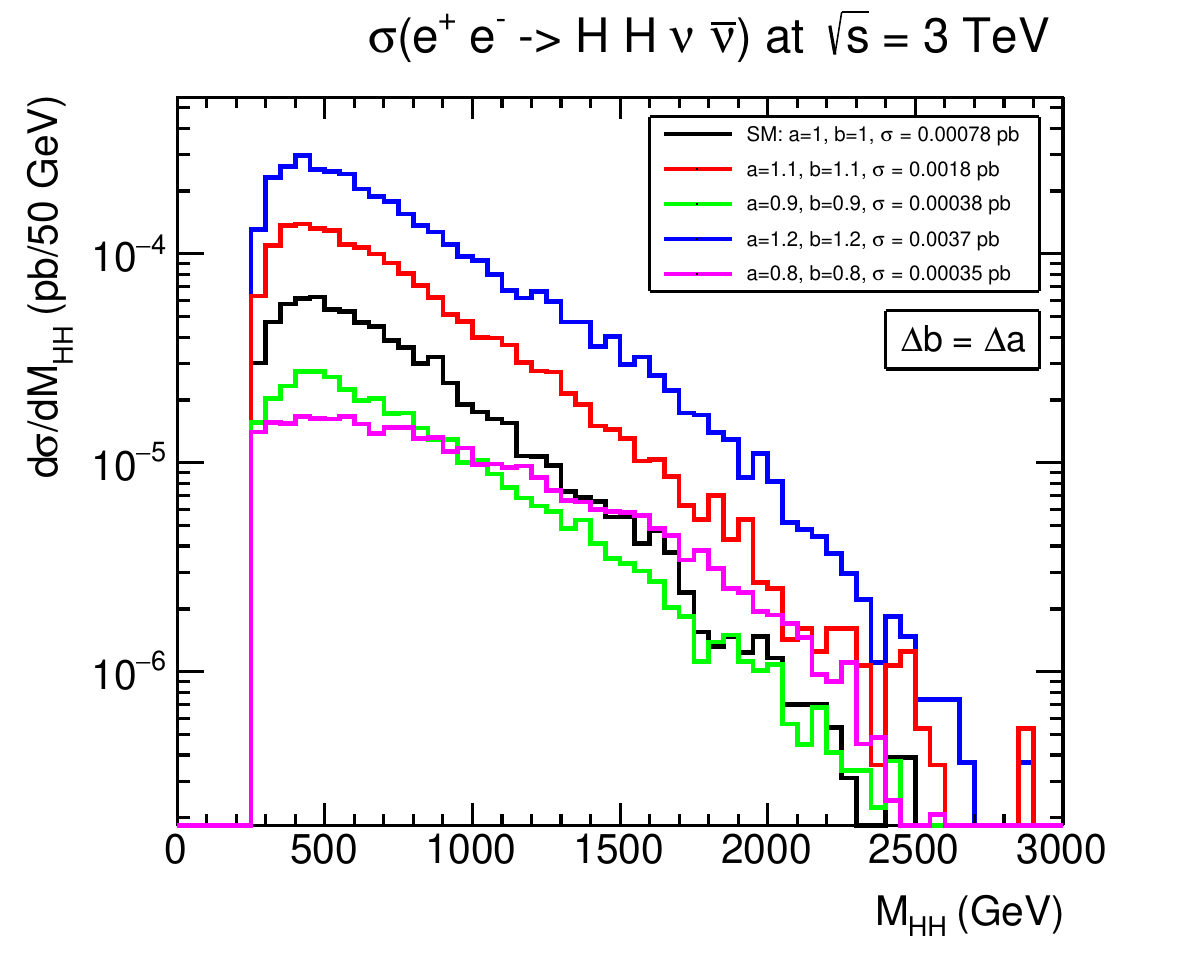}\\\hspace{0.1cm}
        \includegraphics[height=0.182\textheight]{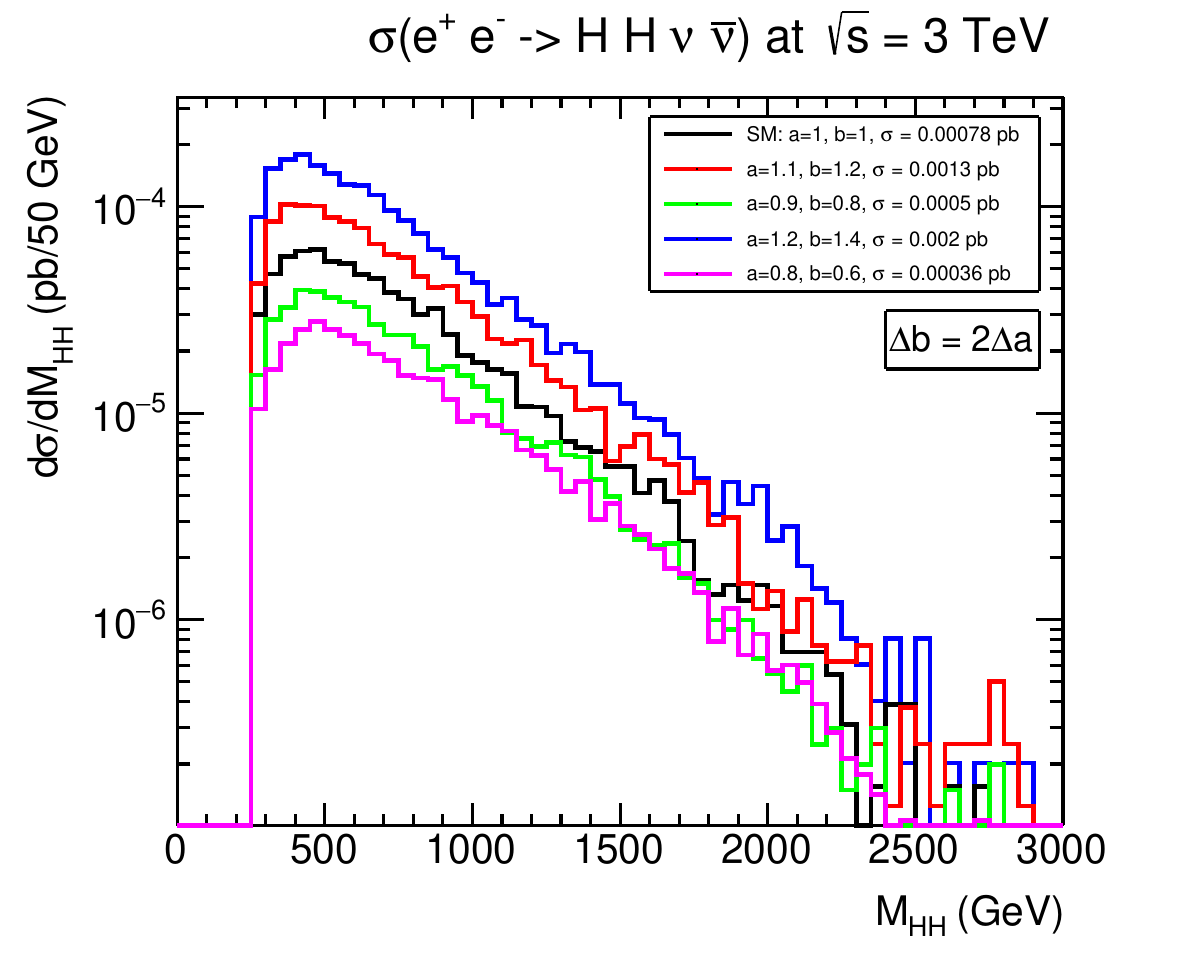}
        \includegraphics[height=0.182\textheight]{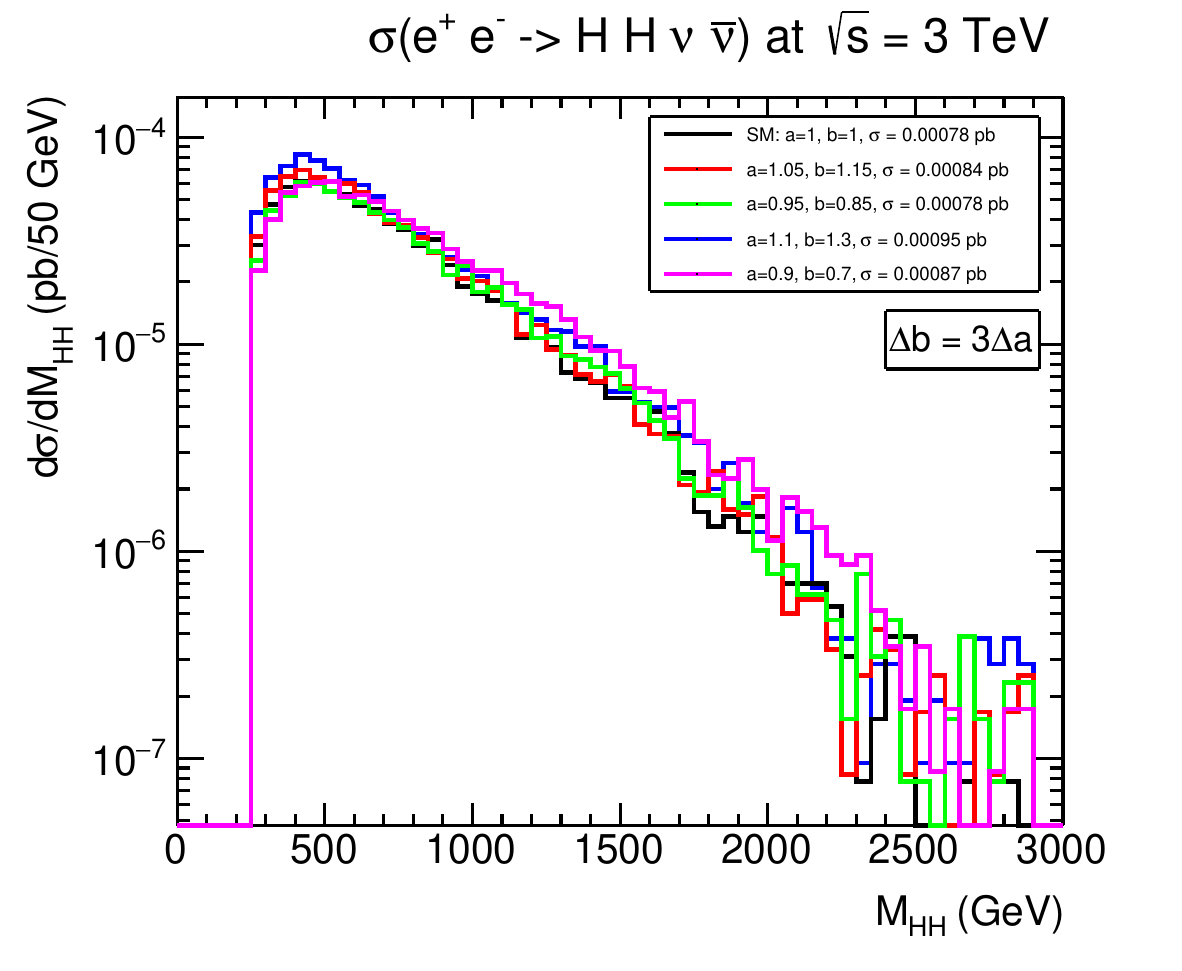}
        \includegraphics[height=0.182\textheight]{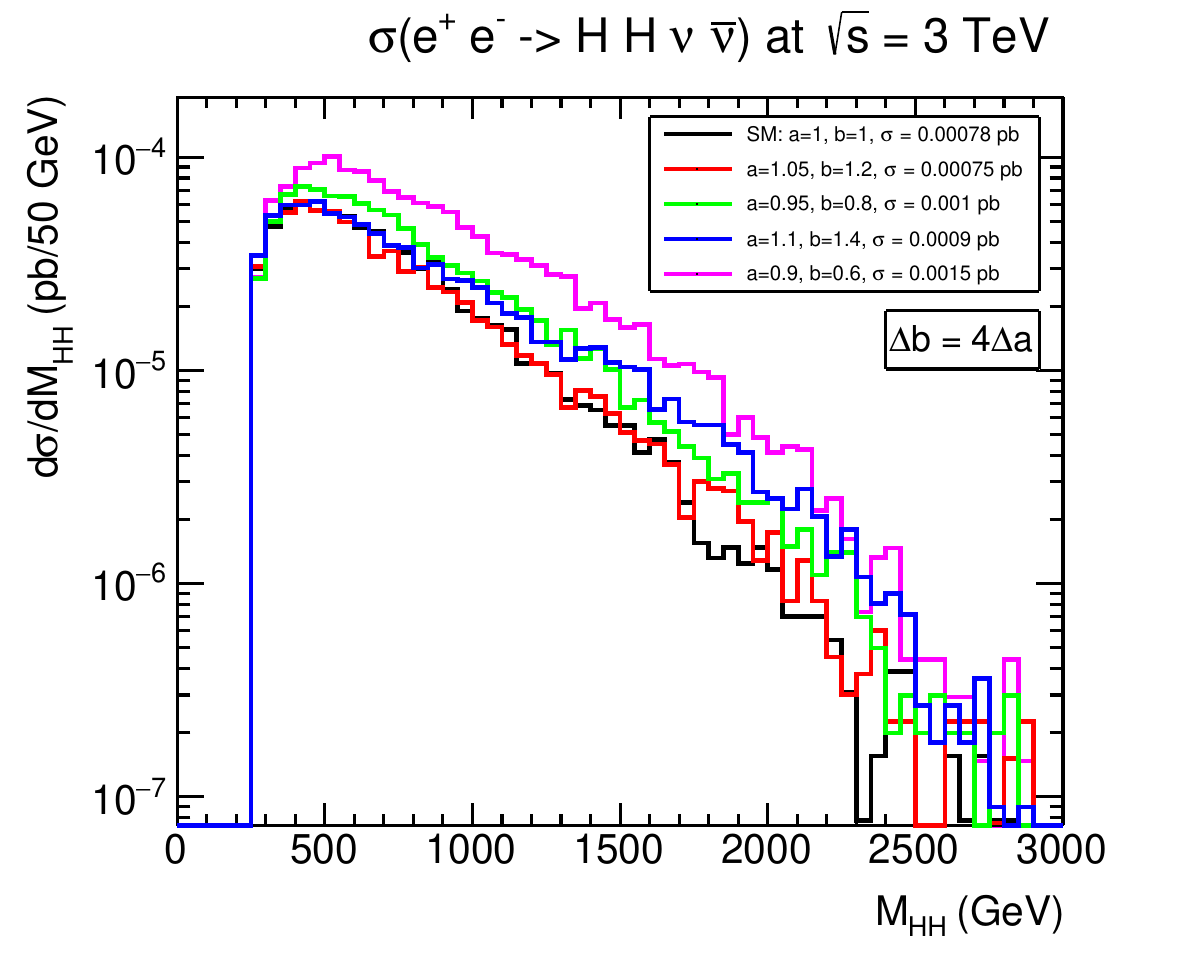}
    \caption{Predictions for the behaviour of the $e^+ e^- \rightarrow HH\nu \bar{\nu}$ cross section distribution with respect to the Higgs pair invariant mass, $M_{HH}$, in the HEFT and assuming a correlation between the HEFT-LO parameters parameterized by the equation $\Delta b = C \Delta a$, for the cases $C = \pm \frac13$, $\pm \frac12$, $\pm 1$, $\pm 2$, $\pm 3$ and $\pm 4$.  The BSM predictions are given by the coloured lines. The SM predictions
    are also included for comparison (black lines). The center-of-mass energy is set at 3 TeV.}
    \label{hist: MHH_corr}
    \end{figure}

\begin{figure}[!t]
    \centering
        \includegraphics[height=0.182\textheight]{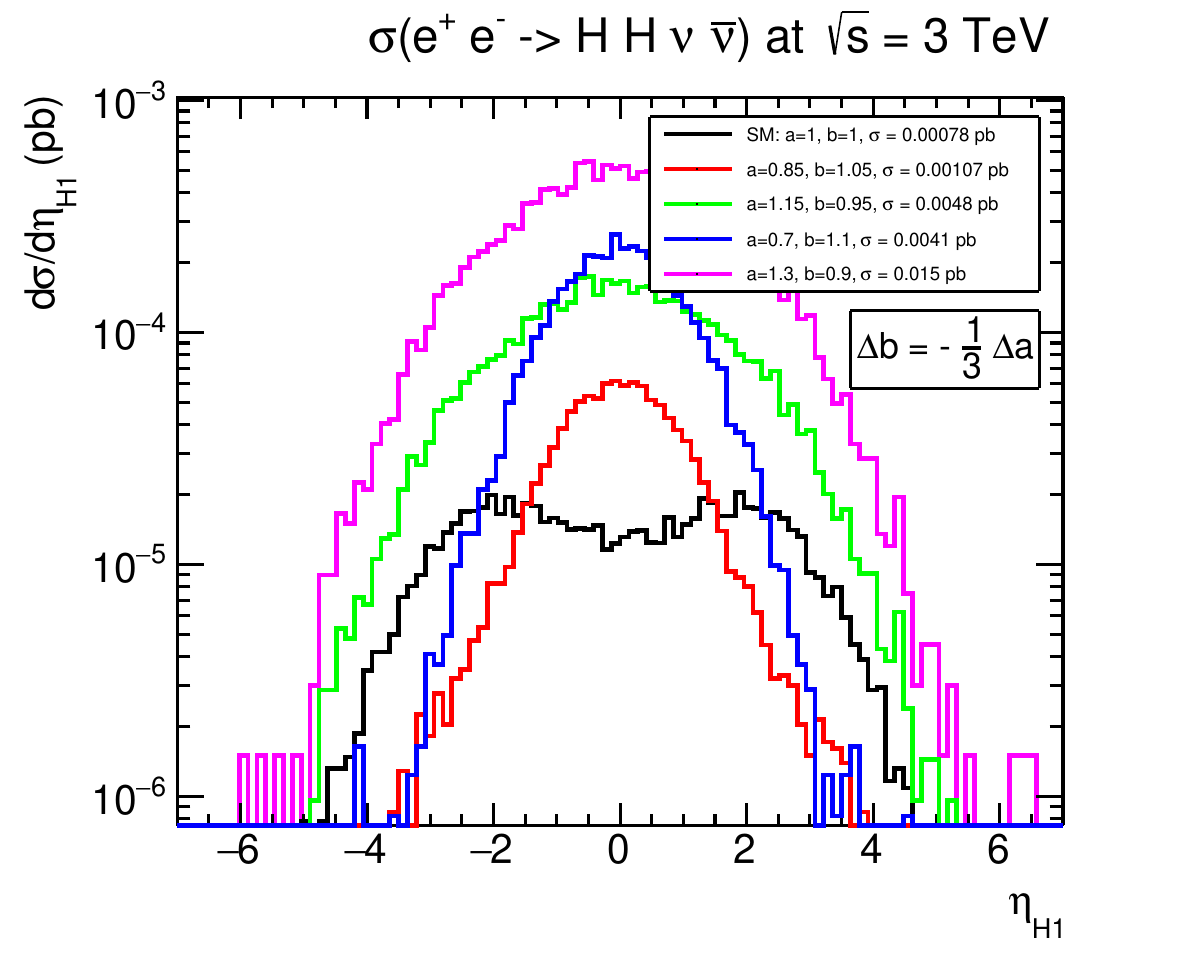}
        \includegraphics[height=0.182\textheight]{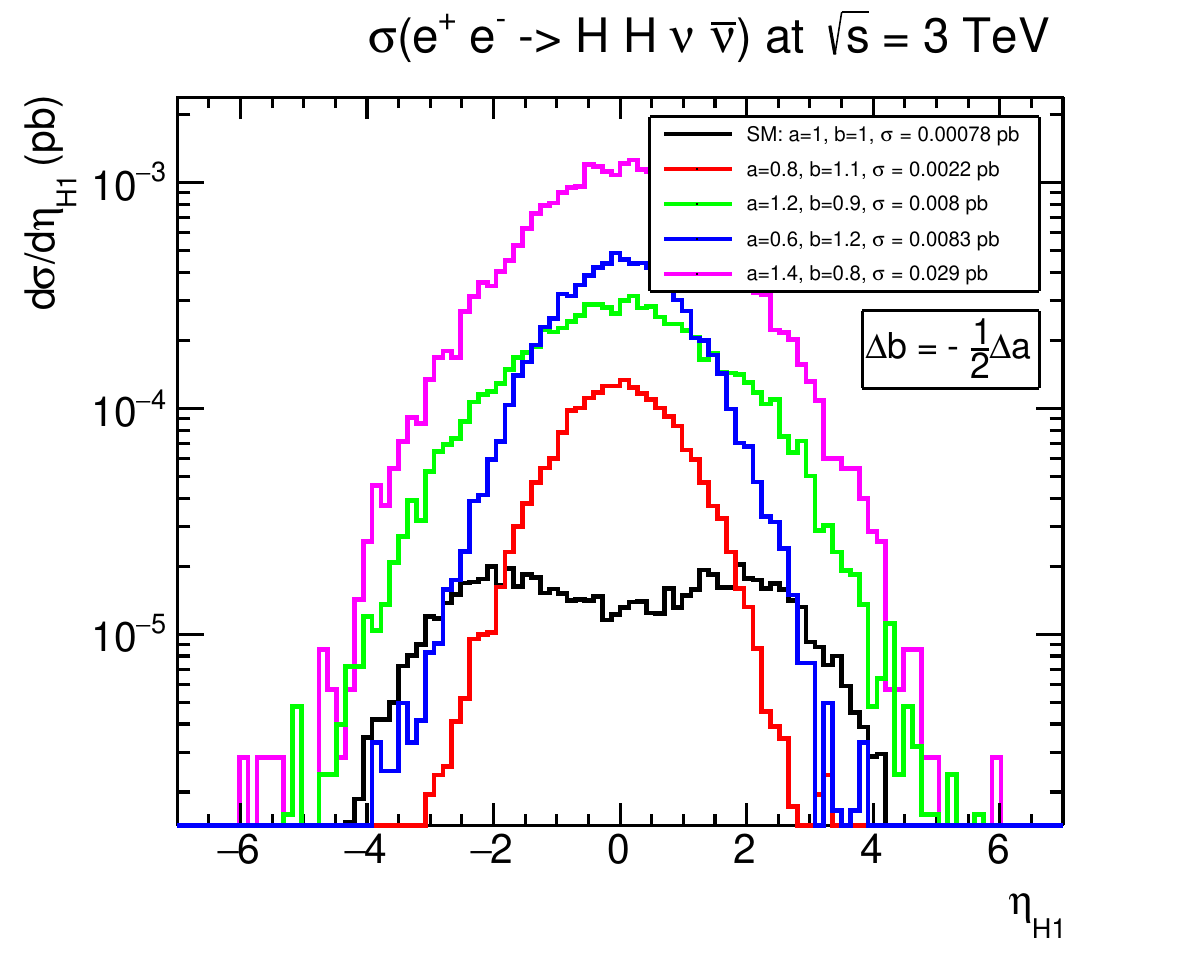}
        \includegraphics[height=0.182\textheight]{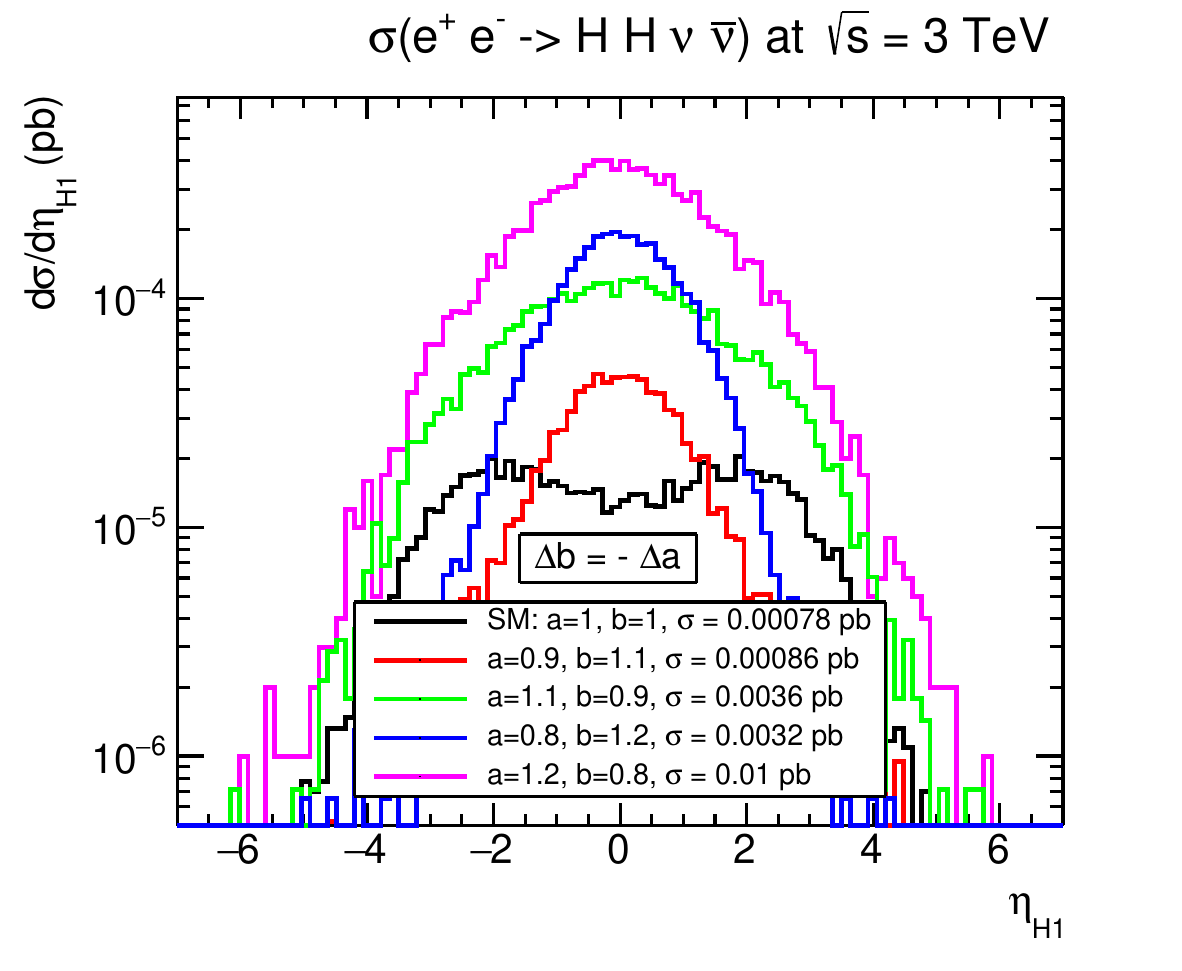}\\\hspace{0.1cm}
        \includegraphics[height=0.182\textheight]{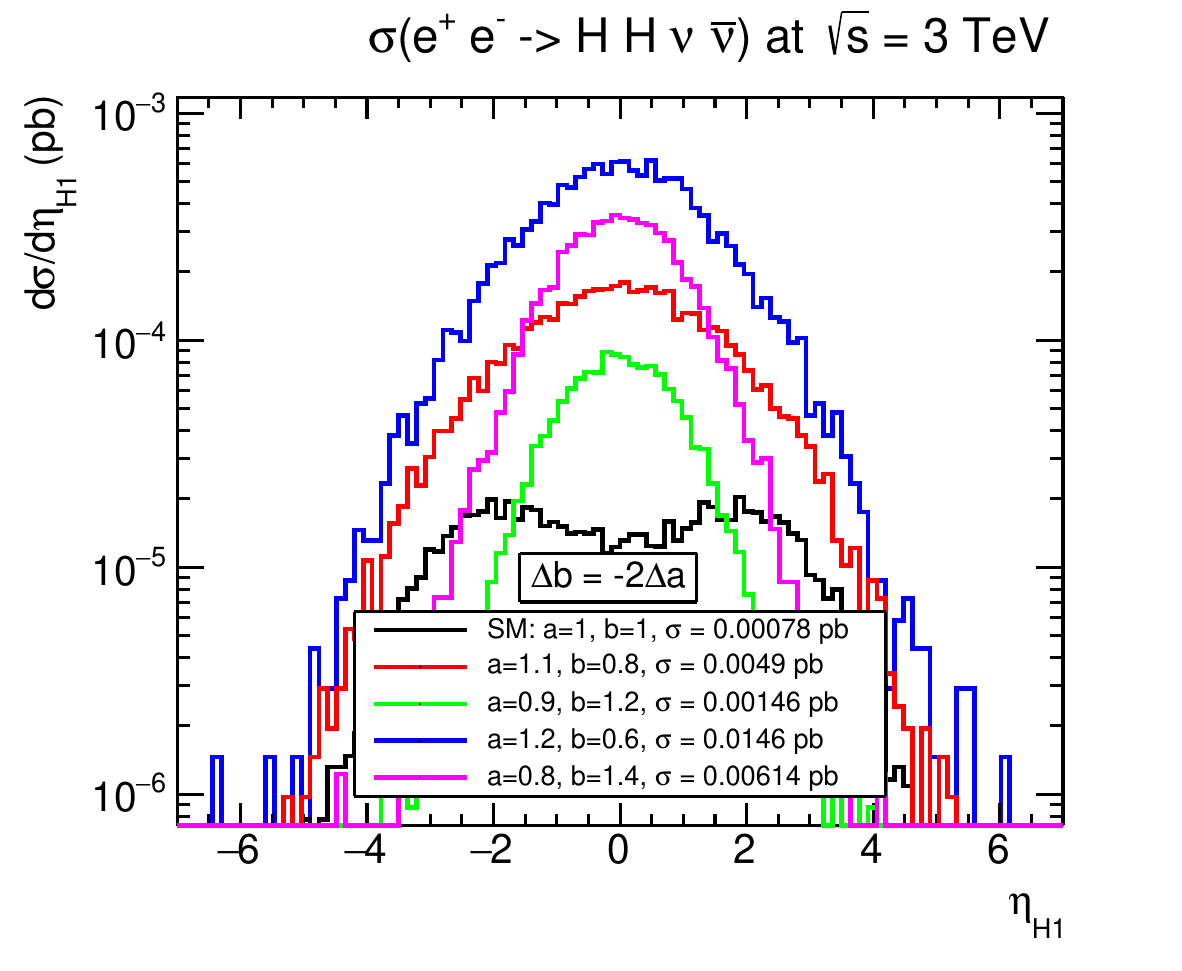}
        \includegraphics[height=0.182\textheight]{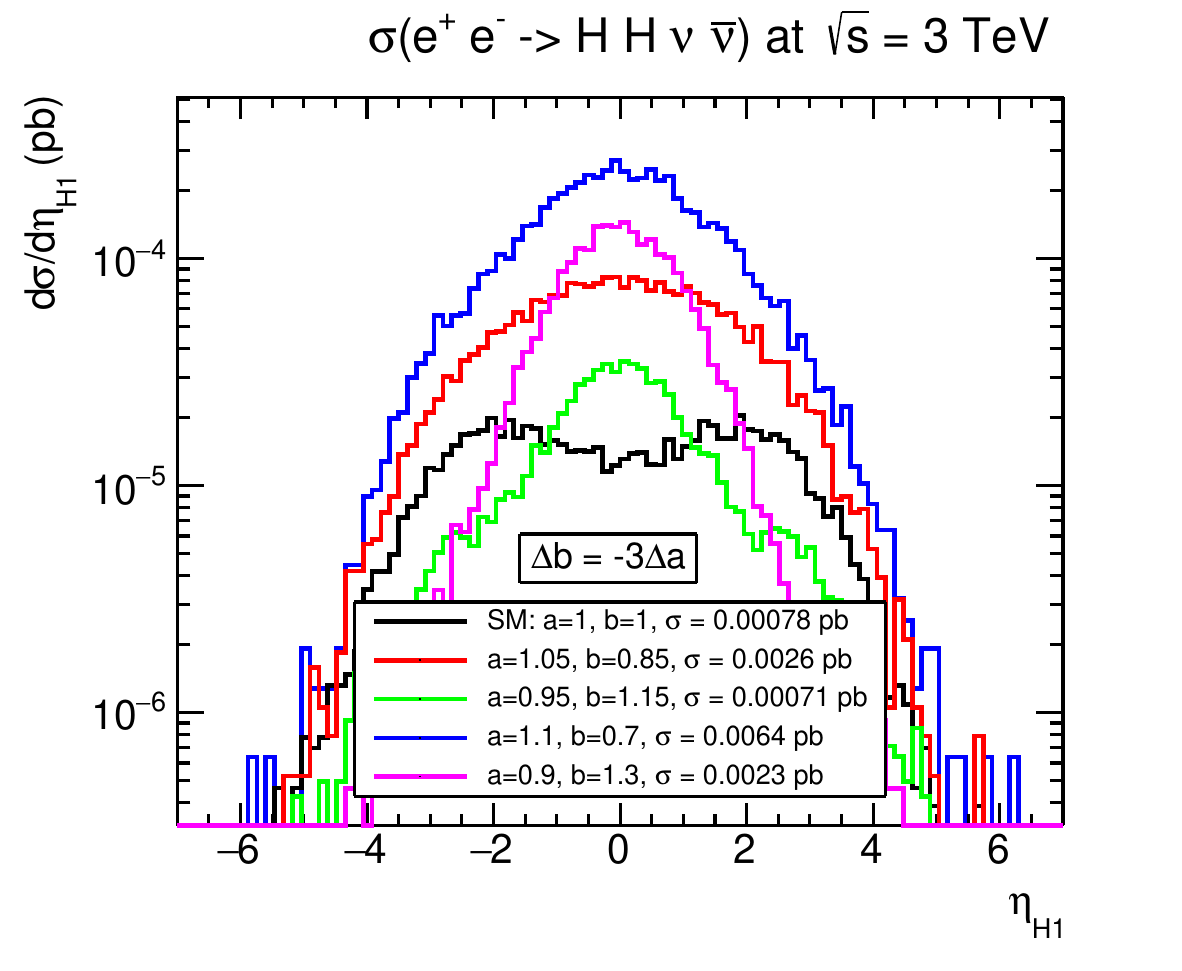}
        \includegraphics[height=0.182\textheight]{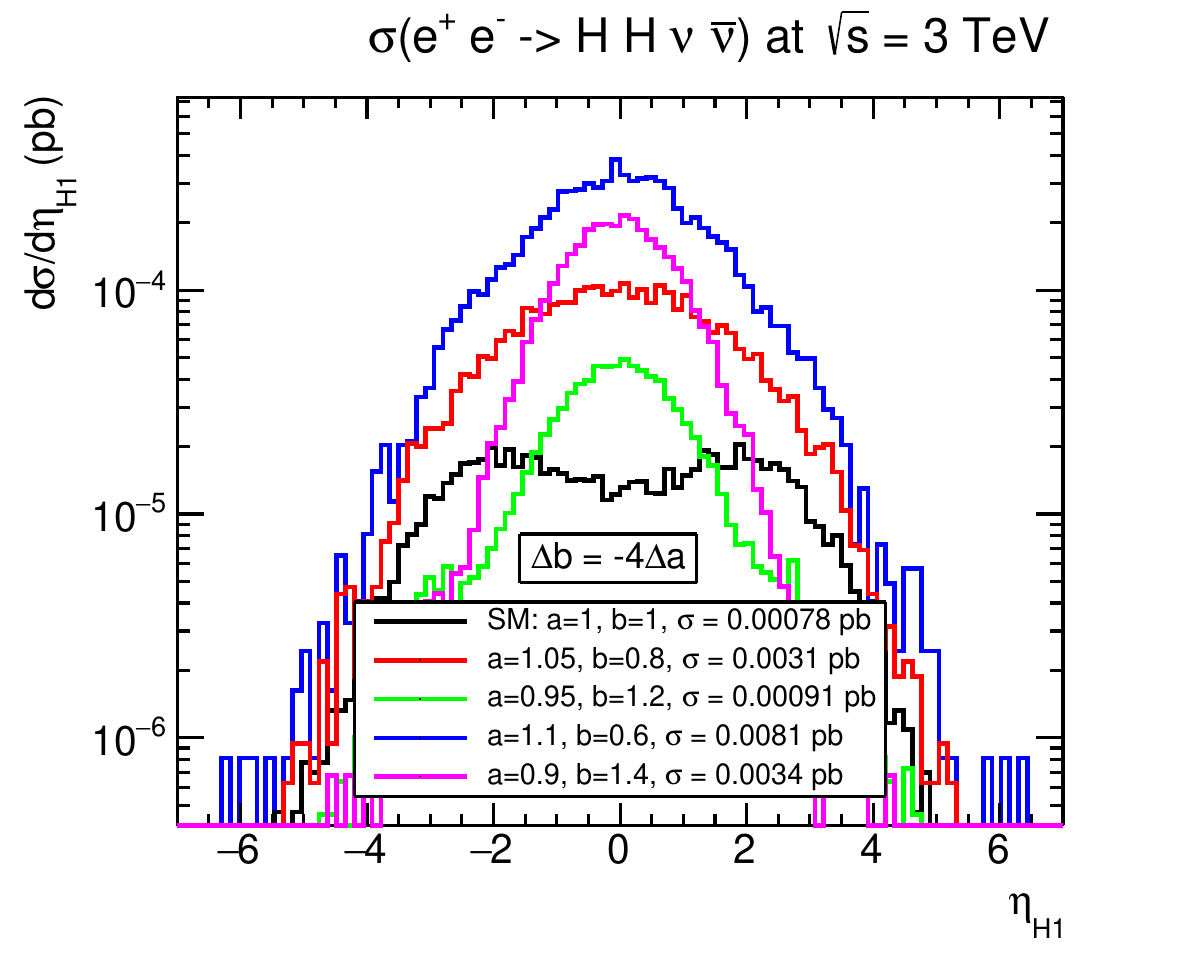}\\\hspace{0.1cm}
        \includegraphics[height=0.182\textheight]{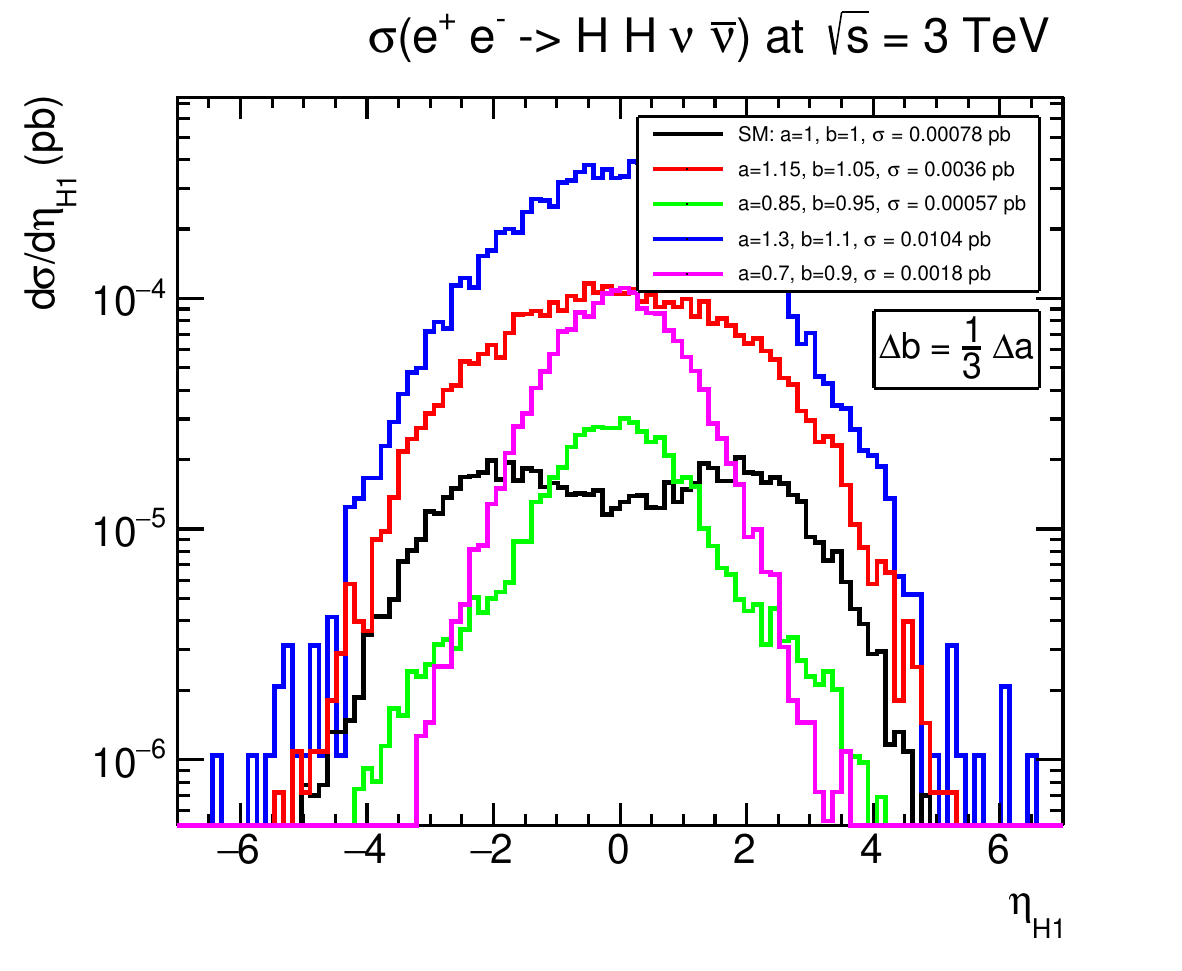}
        \includegraphics[height=0.182\textheight]{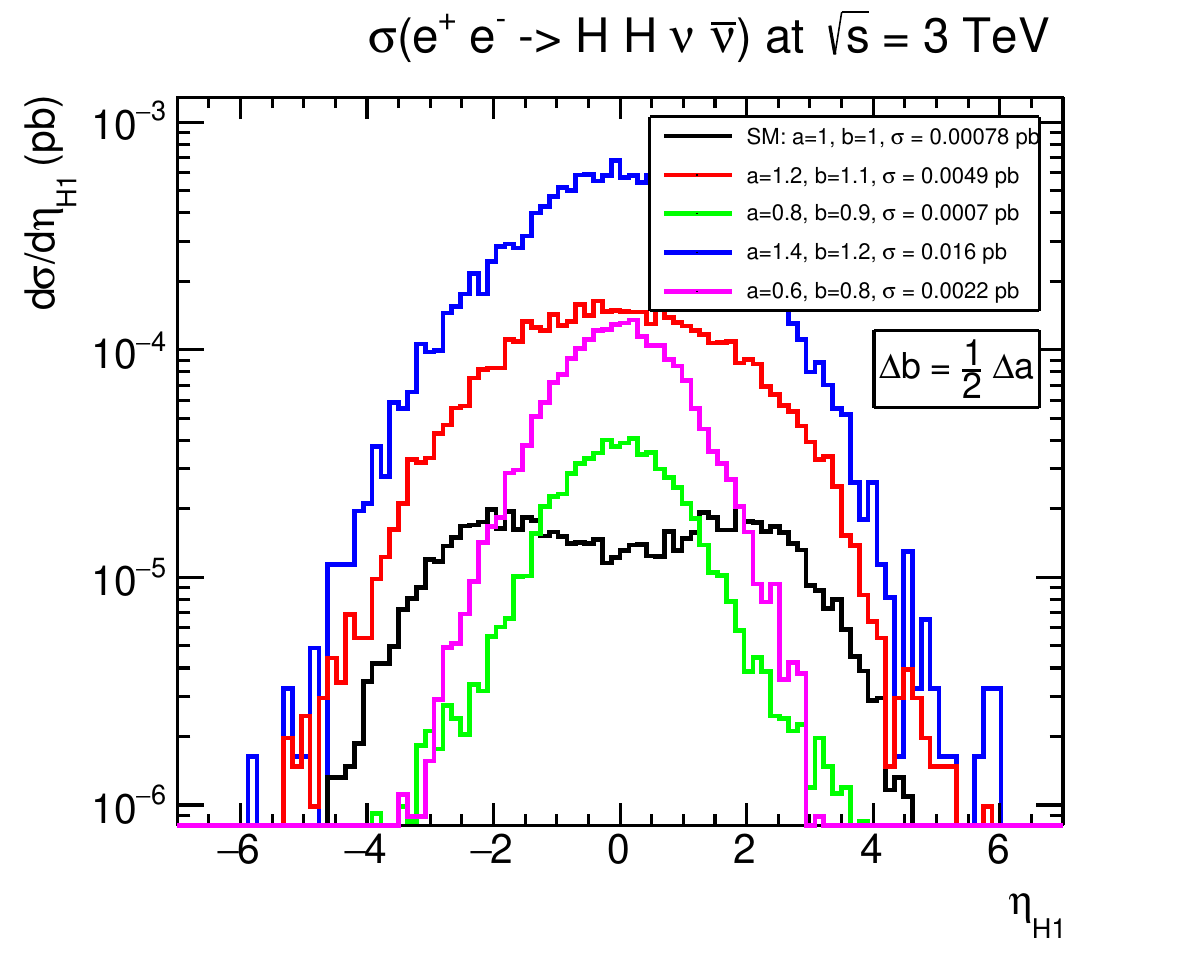}
        \includegraphics[height=0.182\textheight]{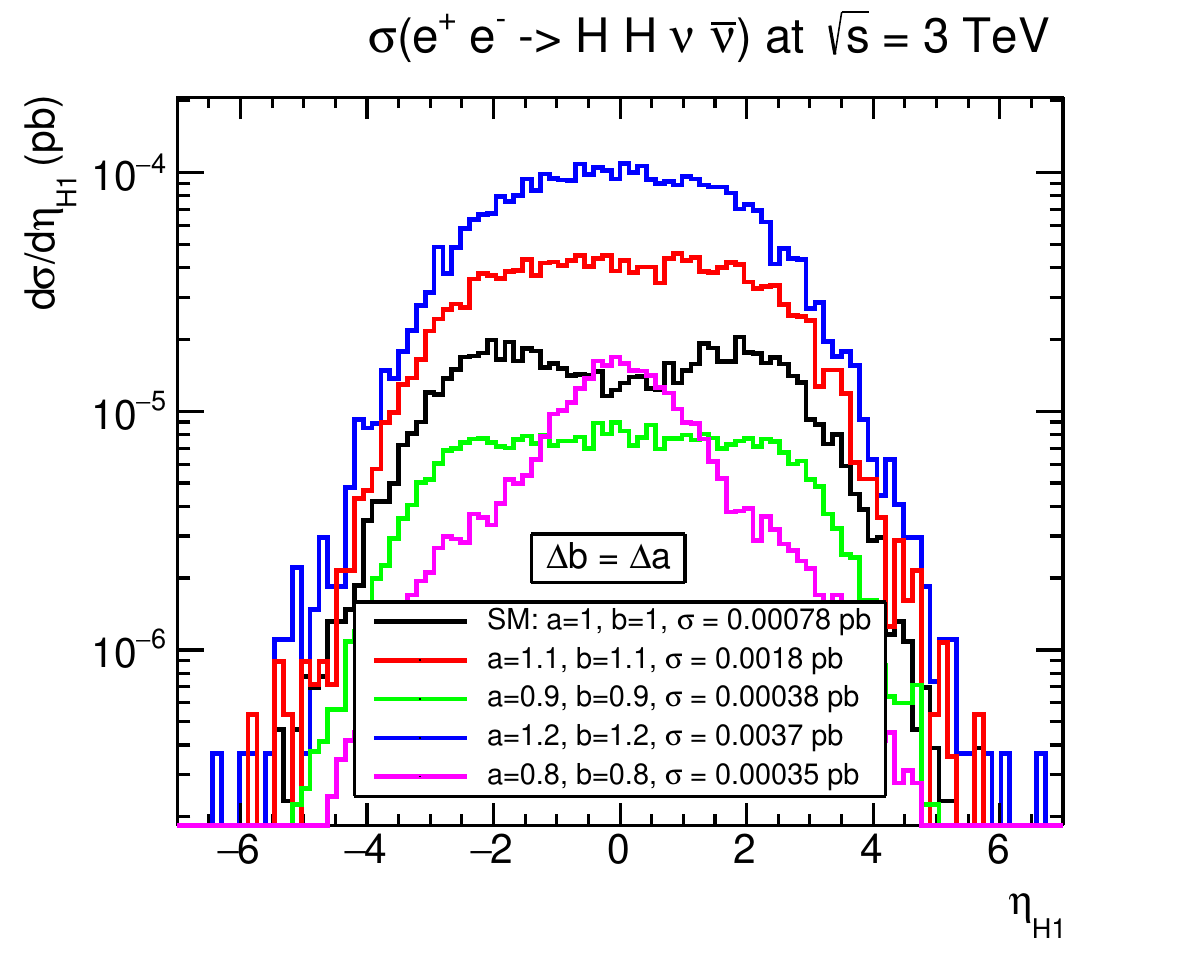}\\\hspace{0.1cm}
        \includegraphics[height=0.182\textheight]{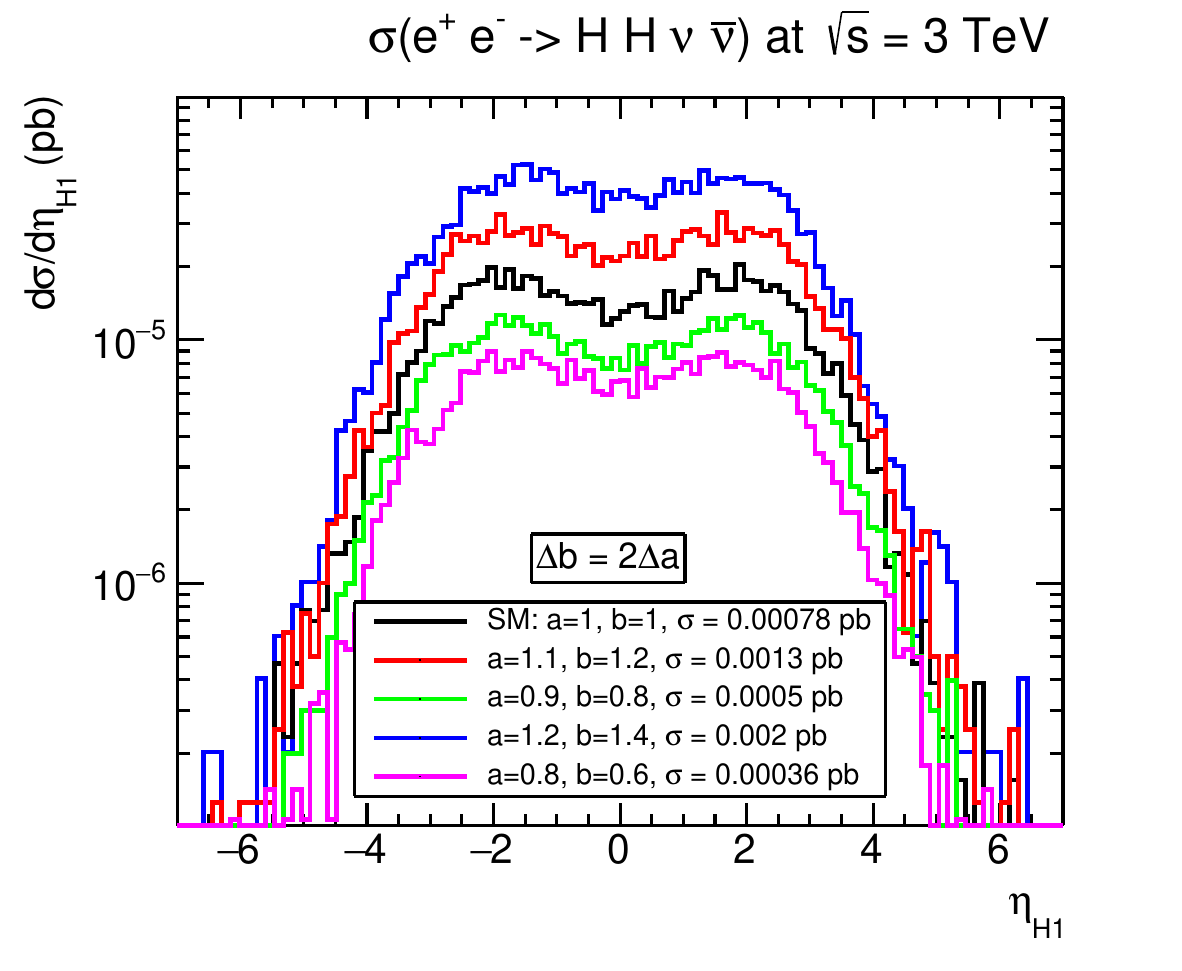}
        \includegraphics[height=0.182\textheight]{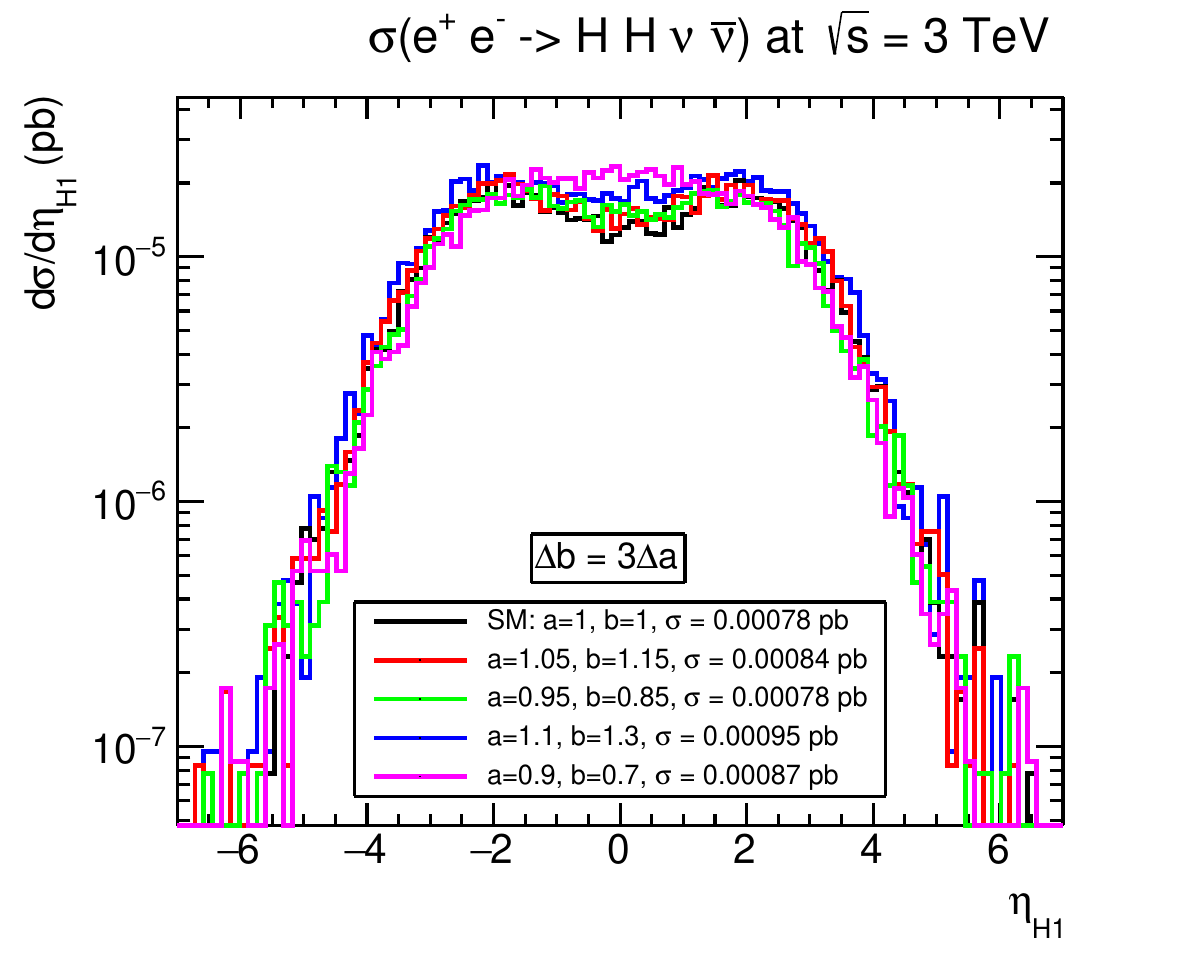}
        \includegraphics[height=0.182\textheight]{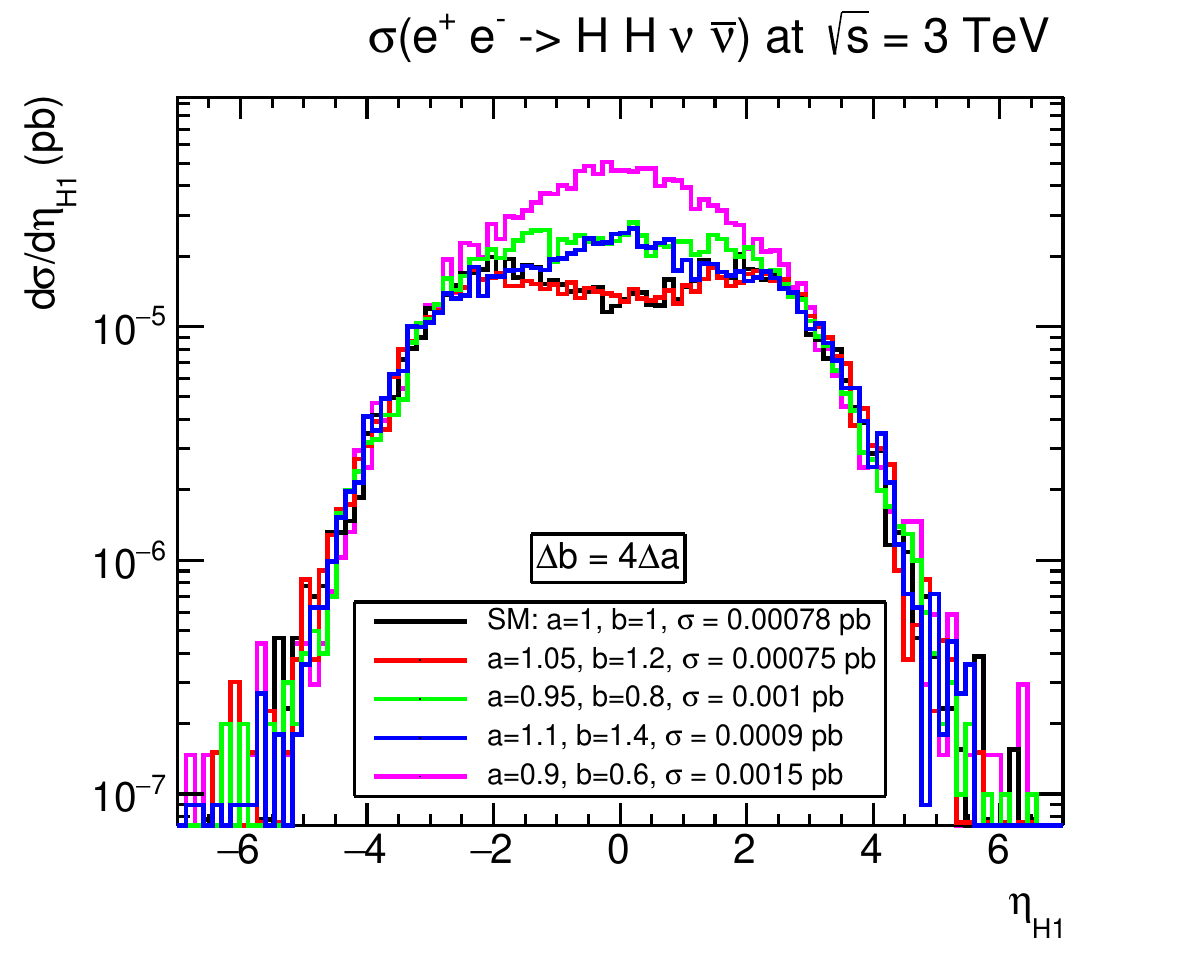}
    \caption{Predictions for the behaviour of the $e^+ e^- \rightarrow HH\nu \bar{\nu}$ cross section distribution with respect to the pseudorapidity of one of the Higgs bosons, $\eta_{H1}$, in the HEFT and assuming a correlation between the HEFT-LO parameters parameterized by the equation $\Delta b = C \Delta a$, for the cases $C = \pm \frac13$, $\pm \frac12$, $\pm 1$, $\pm 2$, $\pm 3$ and $\pm 4$. The SM prediction is shown in black for comparison. The center-of-mass energy is set at 3 TeV.}
    \label{hist: Eta_corr}
    \end{figure}

\begin{figure}[!t]
    \centering
        \includegraphics[height=0.182\textheight]{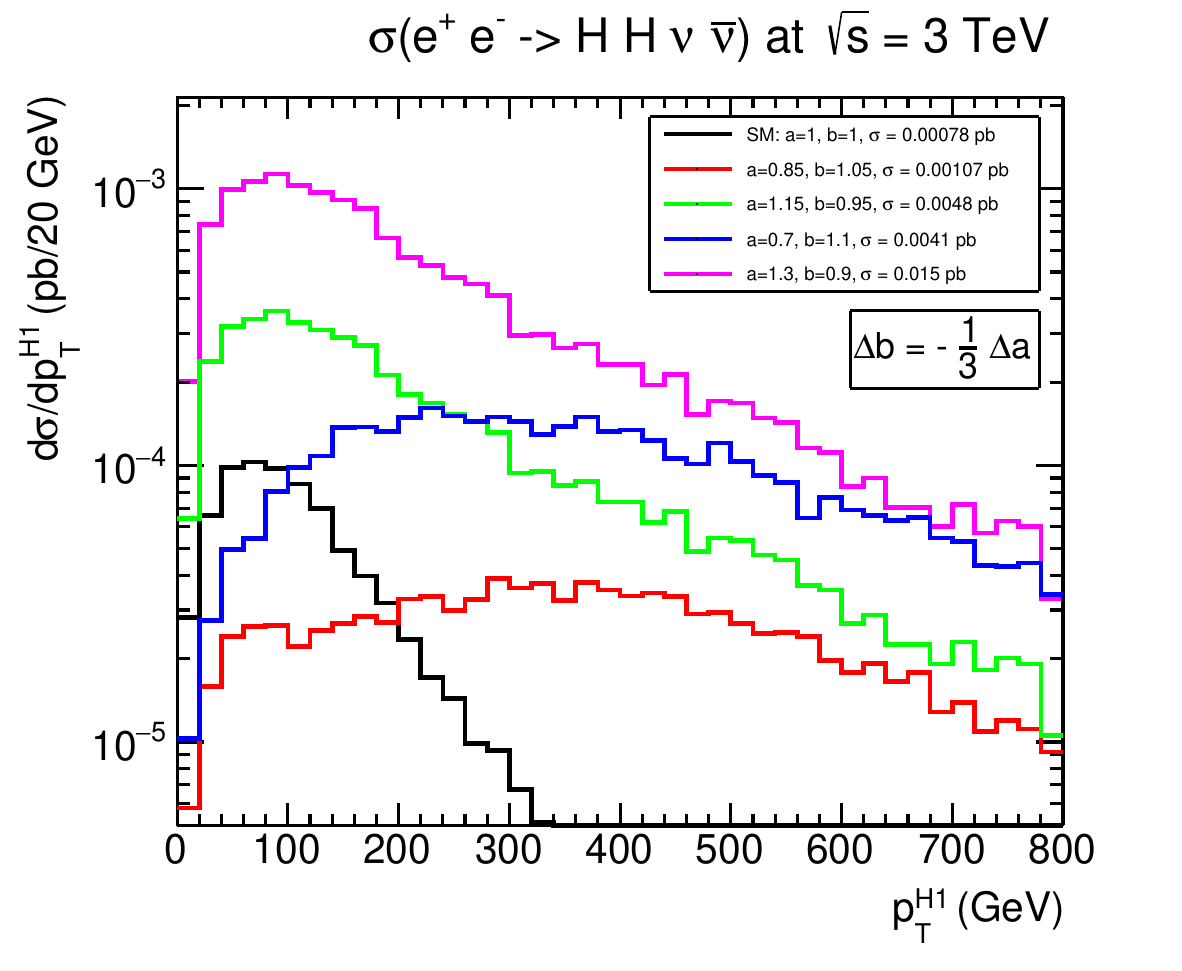}
        \includegraphics[height=0.182\textheight]{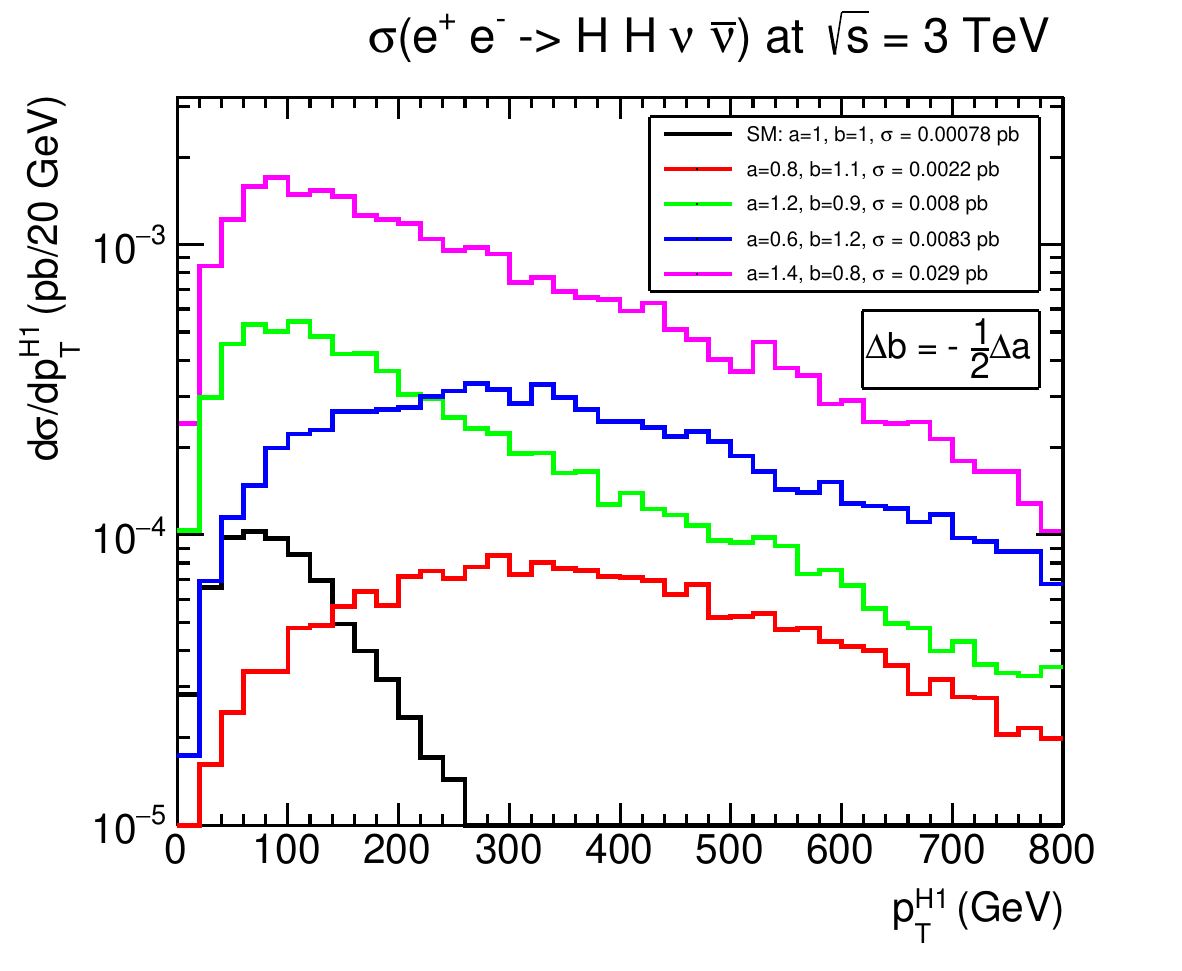}
        \includegraphics[height=0.182\textheight]{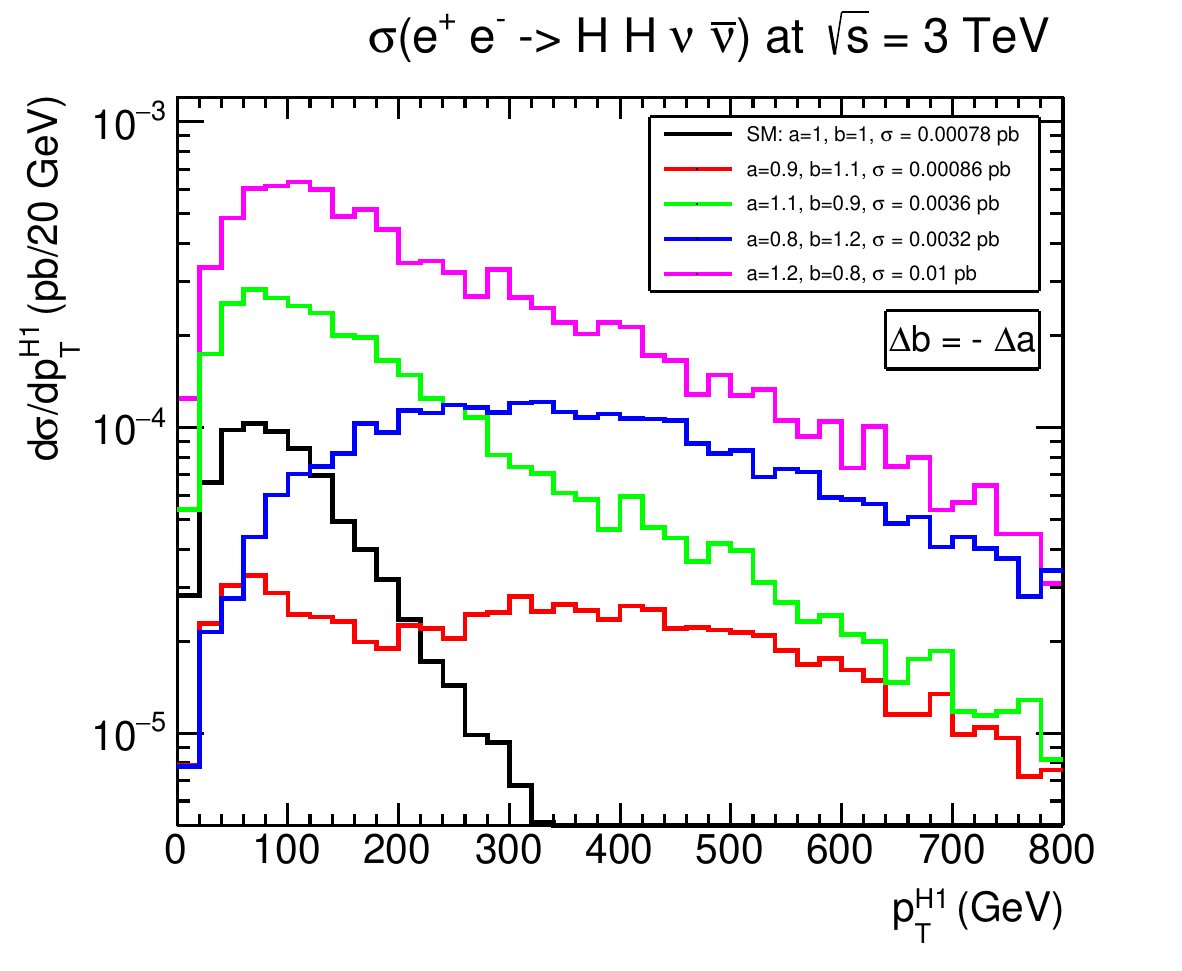}\\\hspace{0.1cm}
        \includegraphics[height=0.182\textheight]{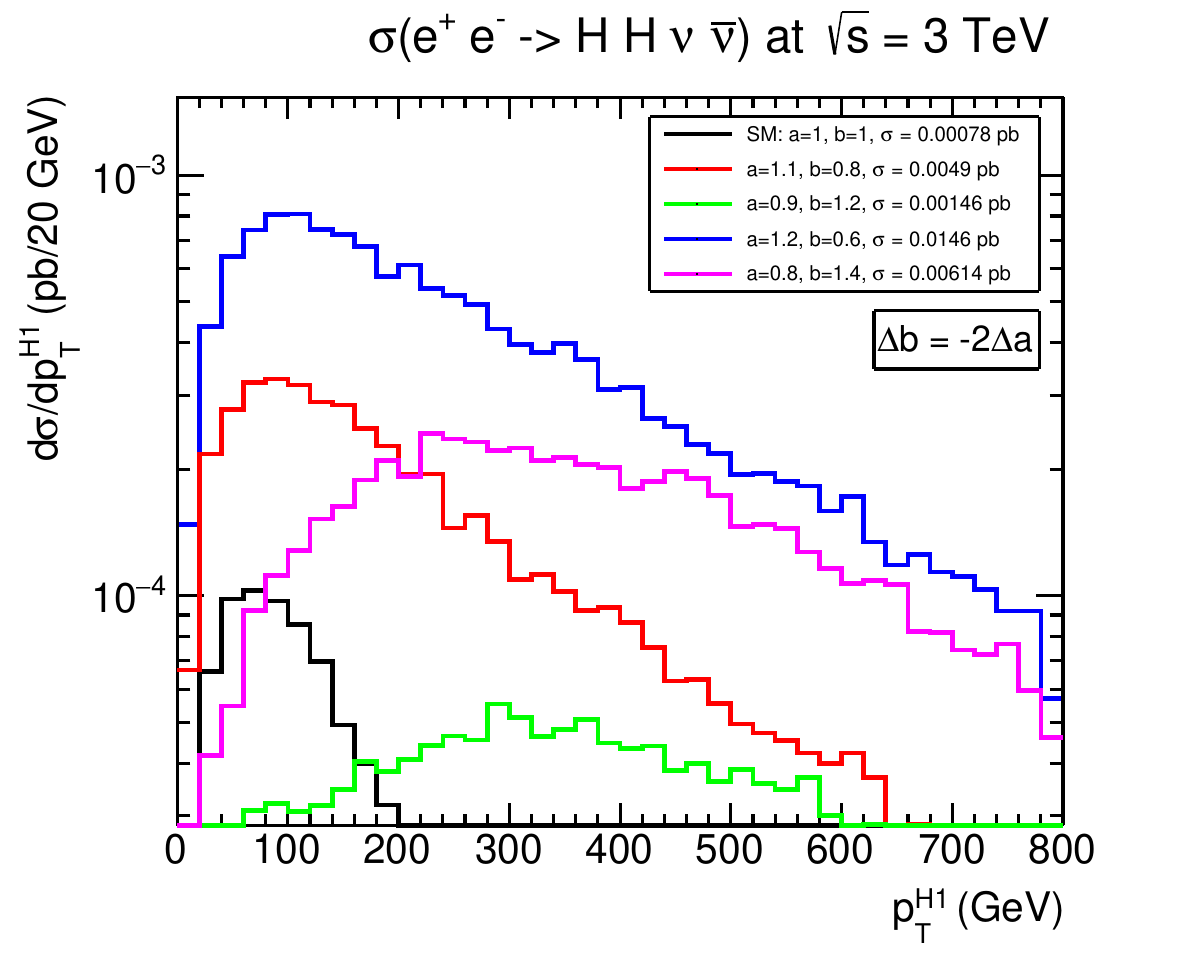}
        \includegraphics[height=0.182\textheight]{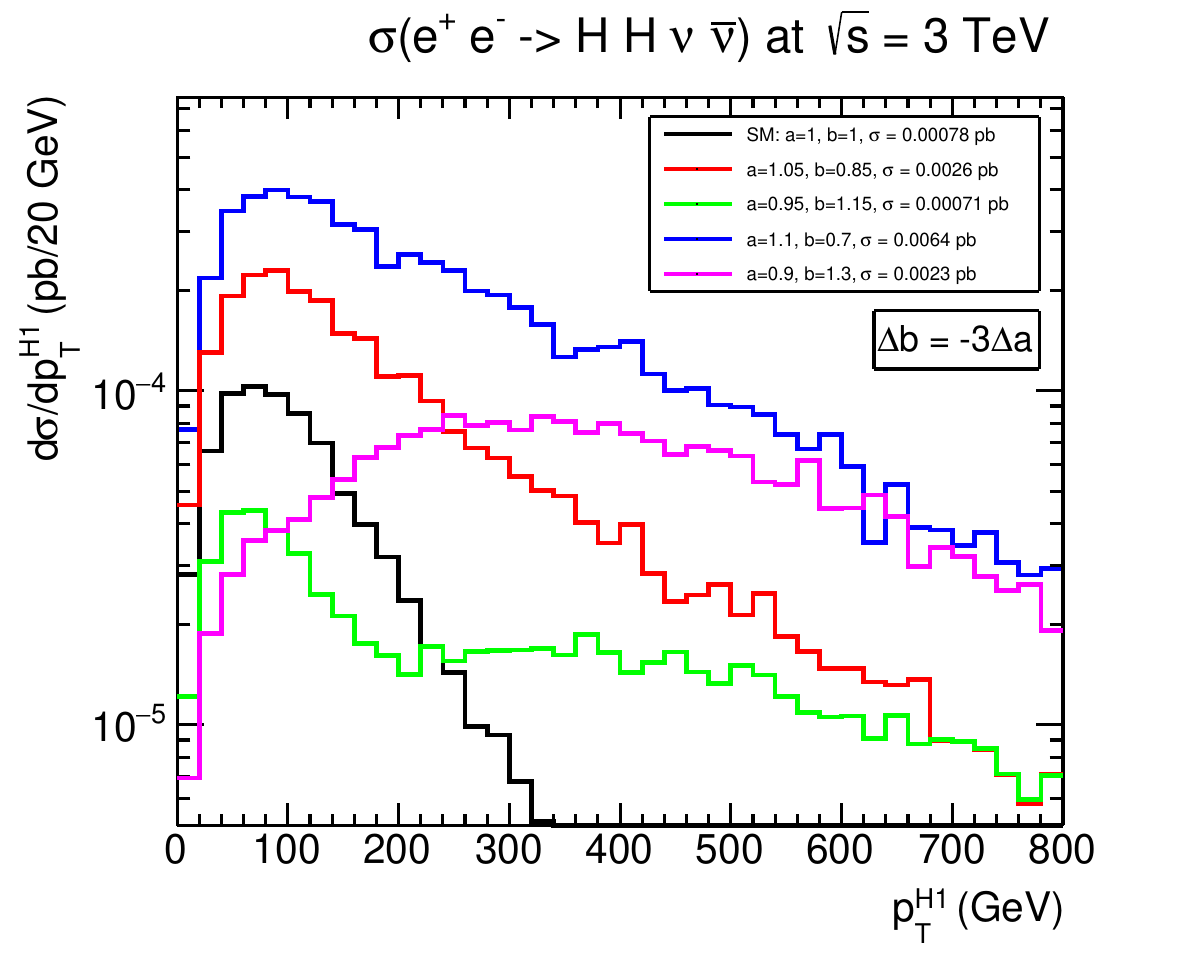}
        \includegraphics[height=0.182\textheight]{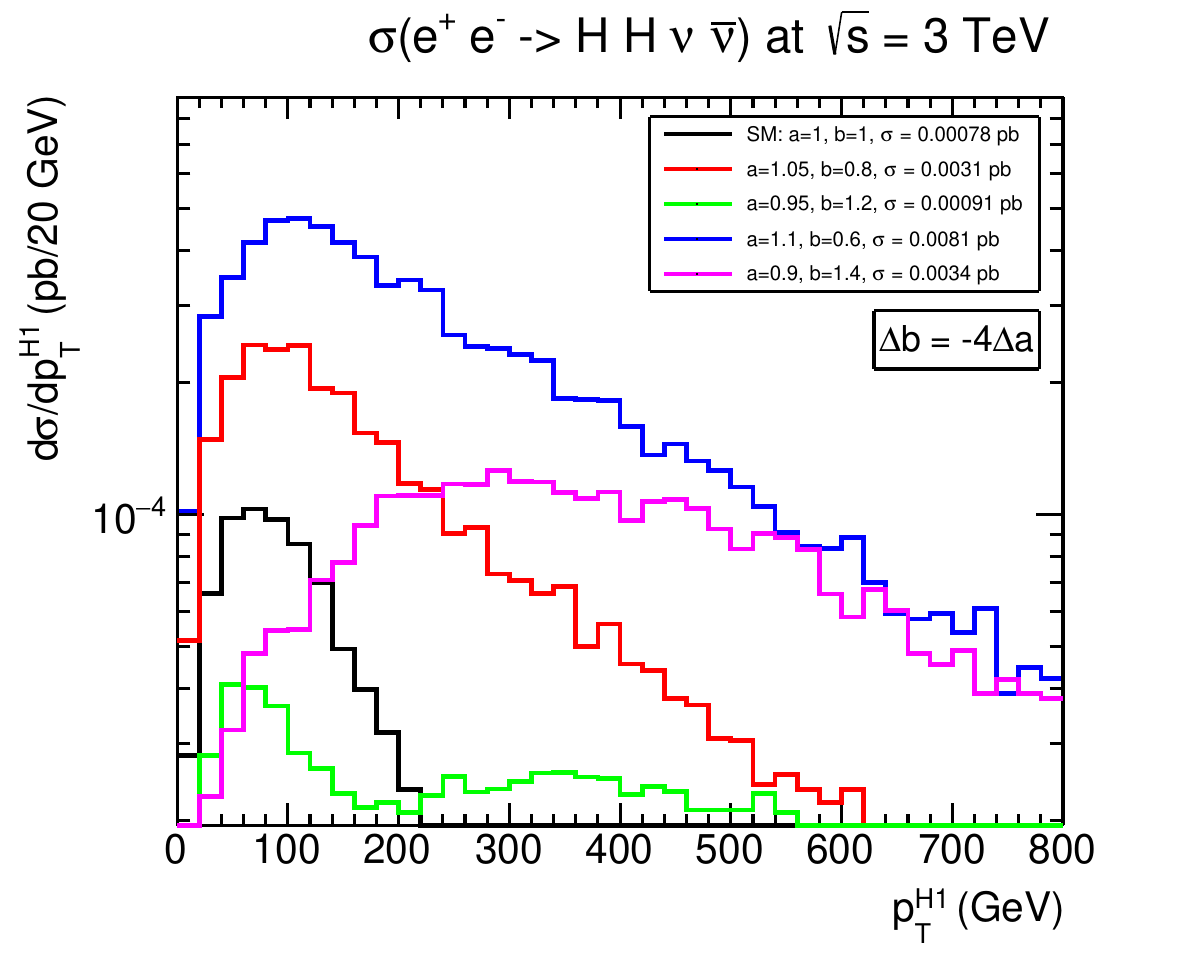}\\\hspace{0.1cm}
        \includegraphics[height=0.182\textheight]{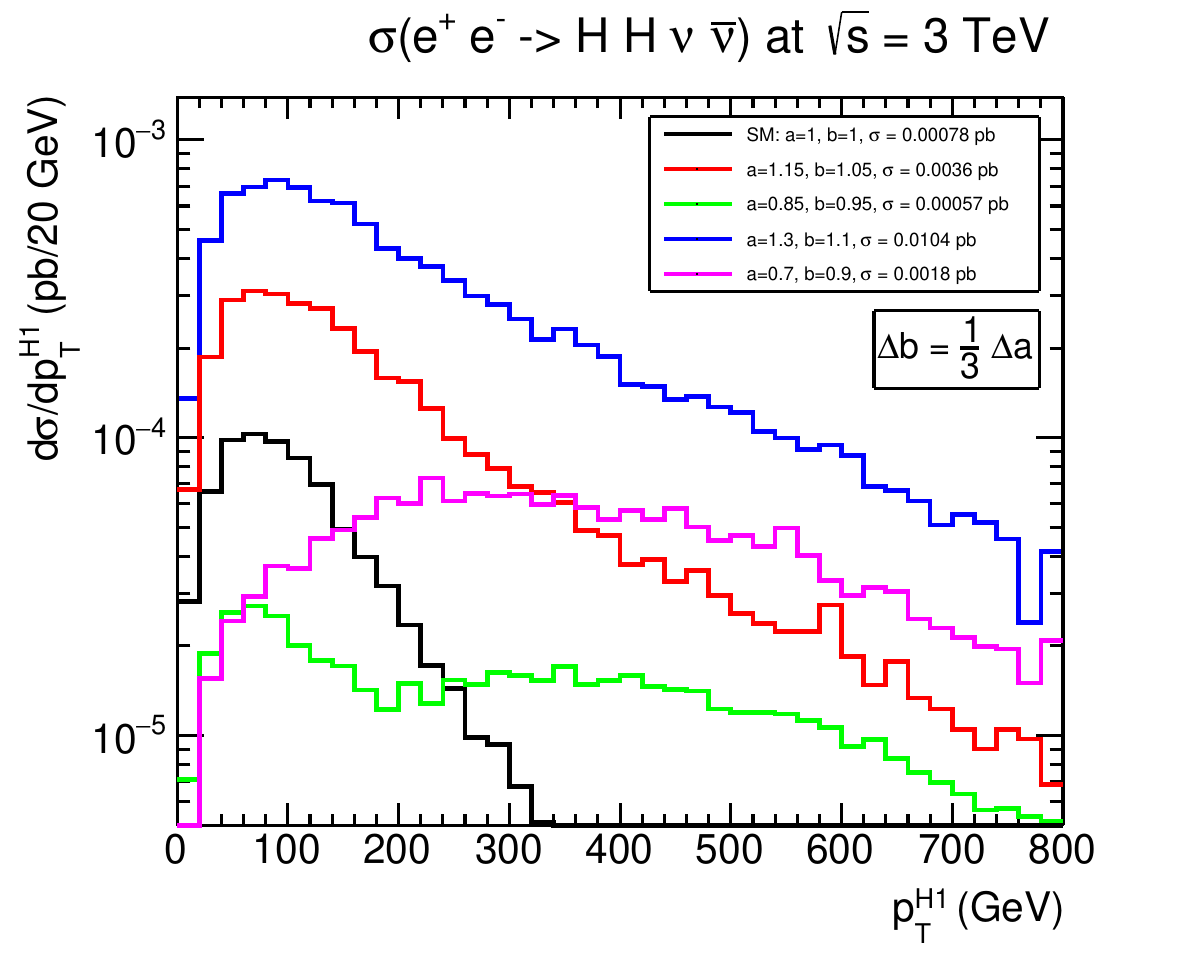}
        \includegraphics[height=0.182\textheight]{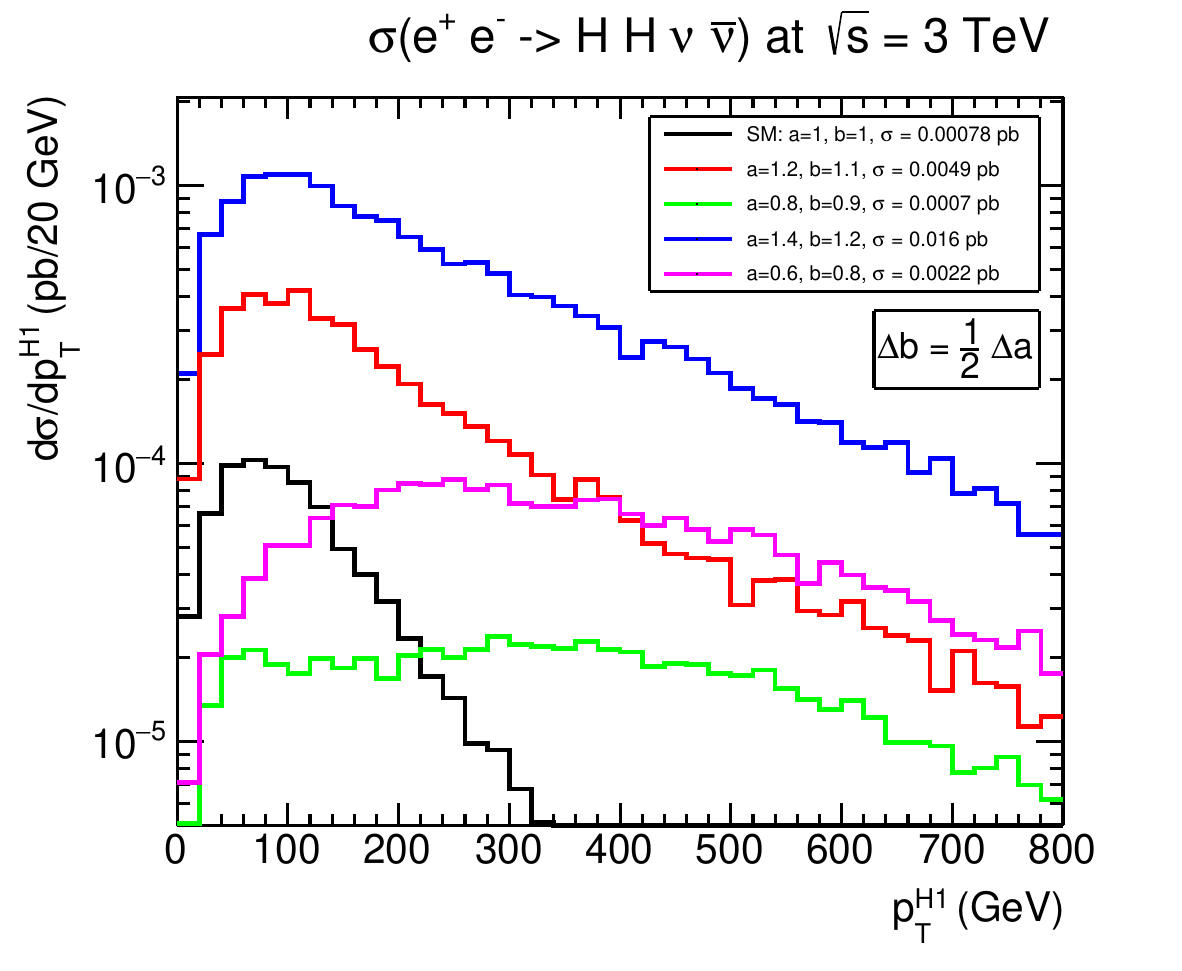}
        \includegraphics[height=0.182\textheight]{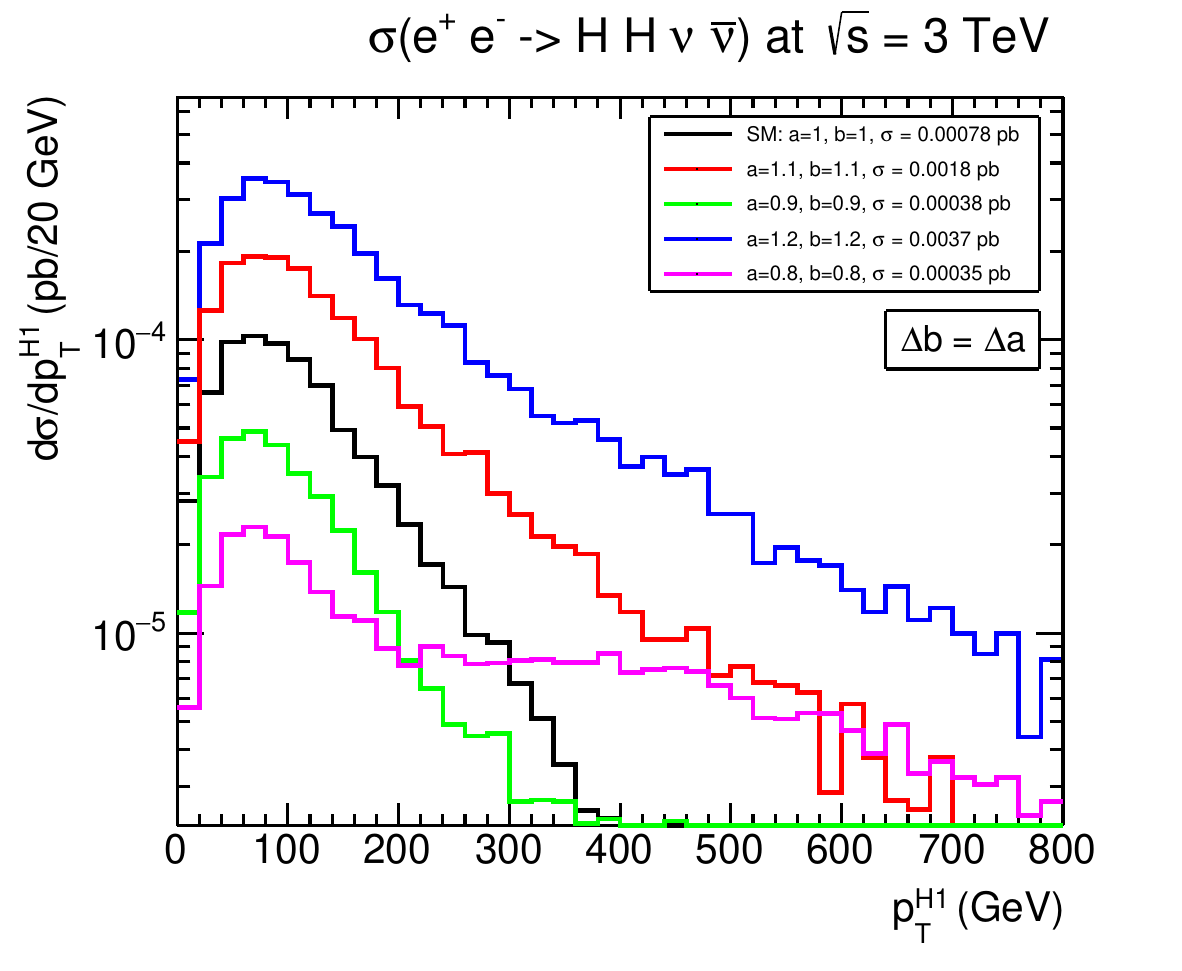}\\\hspace{0.1cm}
        \includegraphics[height=0.182\textheight]{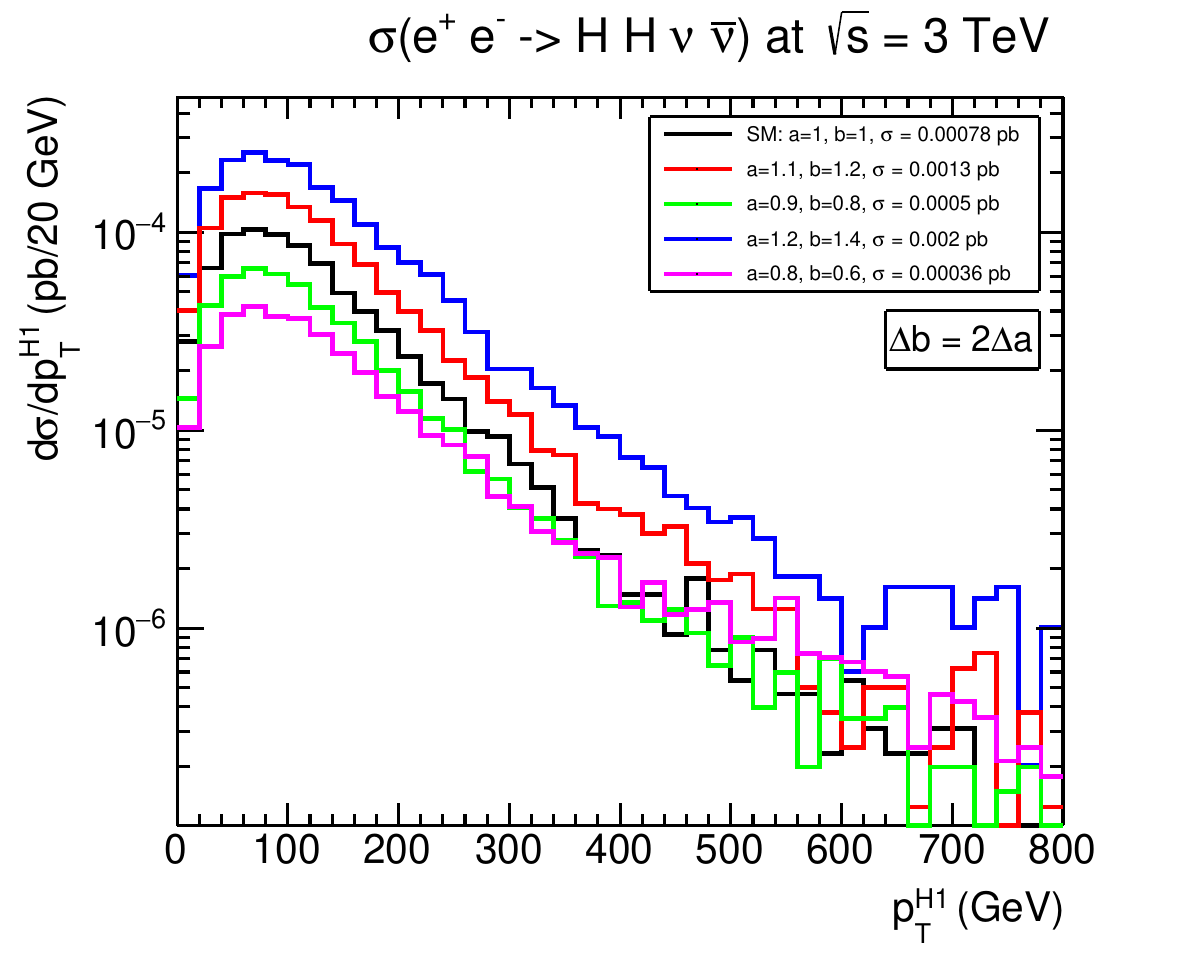}
        \includegraphics[height=0.182\textheight]{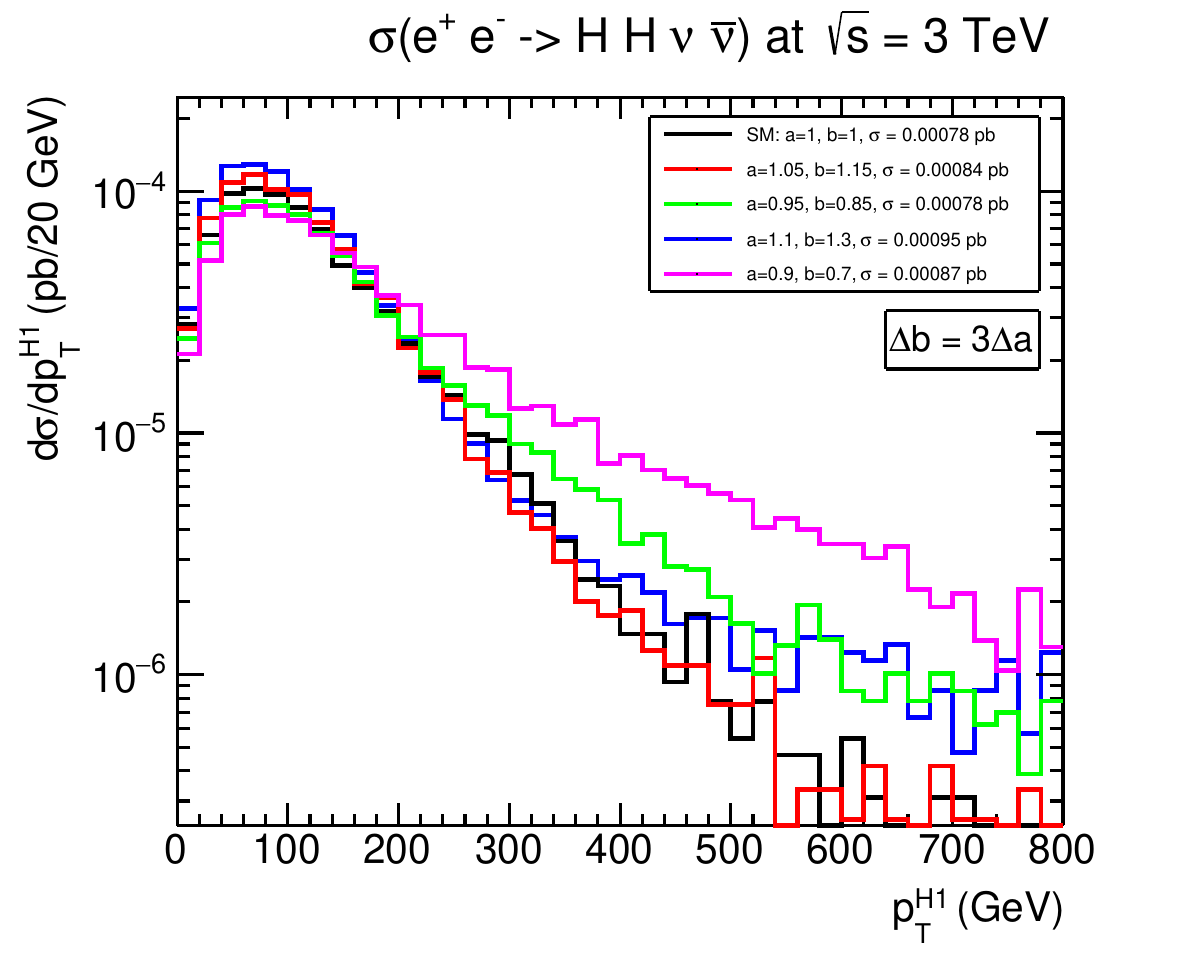}
        \includegraphics[height=0.182\textheight]{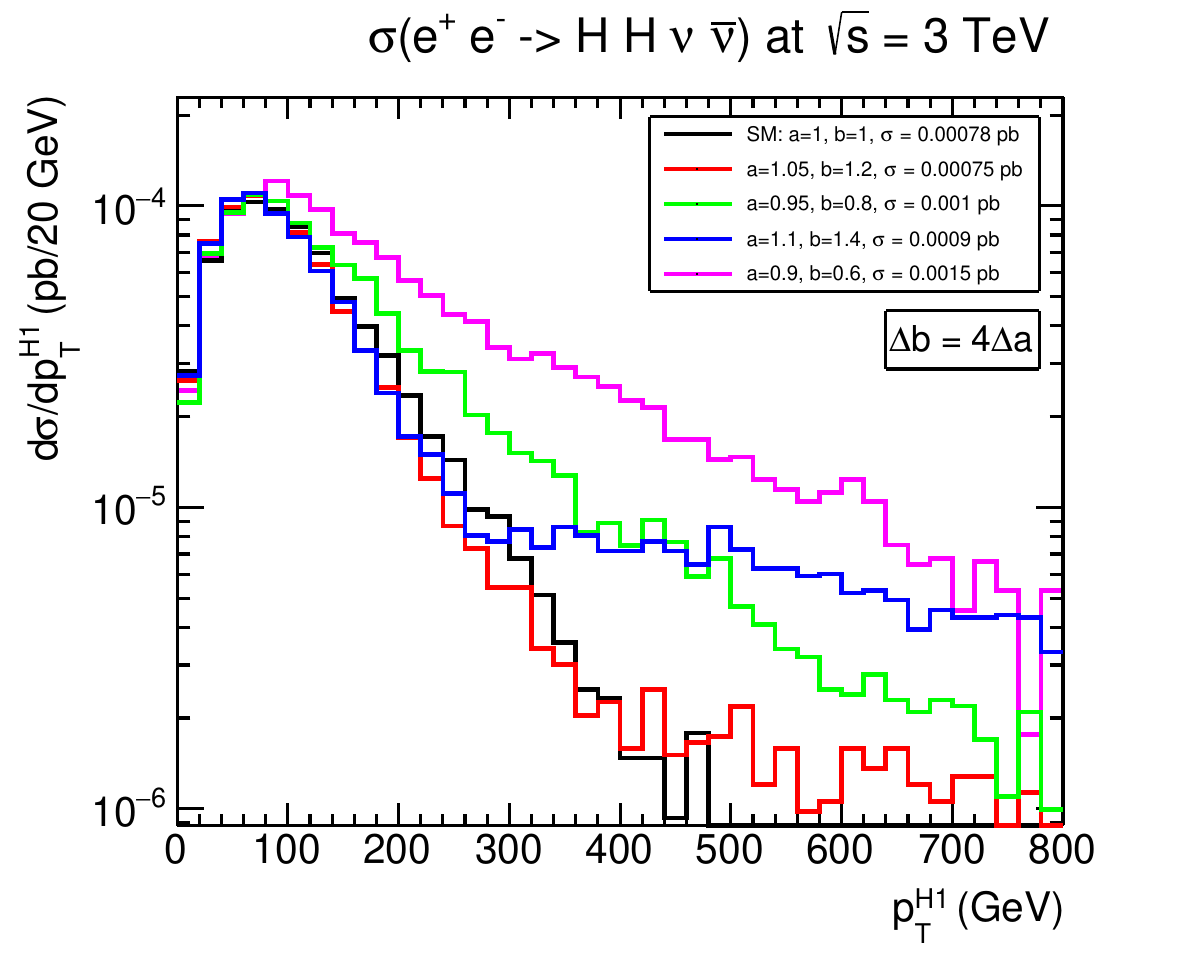}
    \caption{Predictions for the behaviour of the $e^+ e^- \rightarrow HH\nu \bar{\nu}$ cross section distribution with respect to the transverse momentum of one of the Higgs bosons, $p_T^{H1}$, in the HEFT and assuming a correlation between the HEFT-LO parameters parameterized by the equation $\Delta b = C \Delta a$, for the cases $C = \pm \frac13$, $\pm \frac12$, $\pm 1$, $\pm 2$, $\pm 3$ and $\pm 4$. The SM prediction is shown in black for comparison. The center-of-mass energy is set at 3 TeV.}
    \label{hist: Pt_corr}
    \end{figure}

In the final part of this section we explore the implications of these correlations in some selected differential cross sections of  the process $e^+e^- \to HH \nu \bar \nu$.  This selection is motivated by our findings at the subprocess level,  $WW \to HH$ in section \ref{WBF-HH} where the subprocess energy and the angular variables of the final state play an important role in enhancing the sensitivity to $\kappa_V$,  $\kappa_{2V}$ and their possible correlations.  
We choose here to explore the differential cross sections with respect to the following variables: 1) the invariant mass of the $HH$ pair,  2) the pseudorapidity of one of the final $H$,  and 3) the transverse momentum of one of the final $H$.  
We set for this study the case with the highest energy,  i.e.   we take $\sqrt{s} =3$ TeV.   Our predictions for $d\sigma/dM_{HH}$,  $d\sigma/d\eta_H$,  and $d\sigma/d p_H^T$ are displayed in Figs. \ref{hist: MHH_corr},  \ref{hist: Eta_corr} and \ref{hist: Pt_corr} respectively.  In each plot we set  one of the 12 correlations assumed here  for $\Delta b = C \Delta a$,  with  $C = \pm \frac13$, $\pm \frac12$, 
$\pm 1$, $\pm 2$, $\pm 3$,  and $\pm 4$,  and include the predictions for four selected ($a$,  $b$) points  fulfilling that particular correlation. These predictions are represented by the four colored lines in each plot.
The SM predictions are also included in all plots for comparison (black lines).   

We start commenting first our results for $d\sigma/dM_{HH}$ in Fig. \ref{hist: MHH_corr}.  As a common feature in these plots,  we see that the BSM predictions (coloured lines) depart from the SM predictions (black lines) in many cases quite significantly.  The largest departures are produced for the largest $\Delta a$ and $\Delta b$ values whenever $a^2 \neq b$.   If we focus on the large $M_{HH}>800$ GeV region of these plots we see that there are big enhancements in the BSM predictions  in all the plots of the first three rows of this figure,  whose lines are always above the SM line.  The departures are less evident in the cases $\Delta b =2 \Delta a $,  $\Delta b=3 \Delta a$ and $\Delta b =4 \Delta a$ which are drawn in the last row of this figure.  In fact,  for $\Delta b =2 \Delta a $ we see that the profiles of these lines are quite similar to each other and to the SM one.  Indeed it is just in this case where some of the coloured lines are below the SM line,  providing a very difficult to disentangle BSM signal. This is clearly related to the previous observation that the cross section of the subprocess has a minimum close to the $a^2-b=0$ line,  which as we have said is in correspondence with 
$\Delta b =2 \Delta a $. 
    
 Next, we comment on our results  for $d\sigma/d\eta_H$ in Fig. \ref{hist: Eta_corr}.  Again, all plots in the first three rows show large deviations of the BSM predictions (coloured lines) with respect to the SM ones (black lines).   Whereas the SM lines have the typical shape of the WBF configuration with two maxima at about $\eta_{H1}= \pm 2$ and a minimum at $\eta_{H1}= \pm 0$,  the BSM lines show in contrast a unique maximum at $\eta_{H1}= \pm 0$, which is very prominent in some cases.  This peculiar behaviour of the BSM signal  was already anticipated in our study of the subprocess in Sec. \ref{WBF-HH} that indicated the high transversality of the final Higgs bosons.  The highest and narrowest picks of these distributions are found in the cases with $C<0$ and in particular in the plots with $C=-1/2$ and $C=-1/3$ where the distance in pb between the height of the pick in the BSM lines and the minimum in the SM can be as large as two orders of magnitude.   It is clear that these particular correlations will give rise to clear BSM signals with high transversality of the final Higgs bosons (and their decays products) with respect to the beam.  Therefore, they will be easier to test at future colliders.  In contrast,  the lines in the plots of the last row in this figure do not differentiate much from the SM line.  The shape of the coloured lines in the case $\Delta b =2 \Delta a$ is like in the SM, with two maxima at about $\eta_{H1}= \pm 2$ and one minimum at  
 $\eta_{H1}=  0$,  and the case $\Delta b =3 \Delta a$  is nearly indistinguishable from the SM.  Then these two correlations do not provide much transversality in the produced $H$'s.    
  
Finally, we comment on  our results for  $d\sigma/d p_H^T$ in Fig. \ref{hist: Pt_corr}.  This figure confirms the main features found in the previous figure regarding the high transversality of the final Higgs bosons in the BSM  and provides the most clear outcome.  All the BSM lines in all the plots of this figure present a departure at high $p_T$ with respect to the SM line,  and they all have a shape where the falling of the tails at large transverse momentum is smoother than in the SM case.  This common feature indicates again the high transversality of the final Higgs bosons in the BSM signals.  The only exception to this behaviour occurs in  the lower left plot that corresponds to the correlation  $\Delta b =2 \Delta a$, where a similar shape respect to the SM in this distribution is found.  We also learn from this figure that this differential cross section 
$d\sigma/d p_H^T$ will be a good discriminator for the sensitivity to $(\kappa_V, \kappa_{2V})$ and their possible 
    correlations.  In particular,  we find the largest departures with respect to the SM at 
$p_H^T >200$ GeV for the case $\Delta b = -1/2 \Delta a$, in concordance with our previous findings.

In summary,  from our study of the correlations in the three differential cross sections in this section we have learnt that quite distinctive BSM signals appear in the three cases $d\sigma/dM_{HH}$,  $d\sigma/d\eta_H$ and  $d\sigma/d p_H^T$,  characterizing the departures from the SM prediction by significant enhancements mainly: 1) in the large $M_{HH}>500$ GeV region,  2) in  the central rapidity region around $\eta_{H1}=  0$ and 3) at large transverse momentum $p_H^T >200$ GeV. These enhancements appear in all cases of correlations studied except for those cases close to the particular  assumption $a^2=b$ (like $\Delta b = 2 \Delta a$) where the distributions are closer to the SM ones and therefore these will be the most difficult to test experimentally.   In contrast,  the particular correlation $\Delta b = -1/2 \Delta a$  appears like the easiest one to be tested since it provides the largest deviations with respect to the SM predictions.

\section{Accessibility to correlations between $\kappa_V$ and $\kappa_{2V}$ at future $e^+e^-$ colliders}
\label{correlations}
In this section we finally explore the accessibility to the effective couplings $\kappa_V$ and $\kappa_{2V}$ and their correlations at future $e^+e^-$ colliders in a more realistic scenario, where the final Higgs bosons decay to $b \bar b$ pairs which is their dominant decay channel.  Therefore, we study here the full process 
\begin{equation}
e^+e^- \to HH \nu \bar \nu \to b \bar b  b \bar b \nu \bar \nu \,,
\end{equation}
 and perform a more realistic analysis of the BSM departures with respect to the SM by looking at events with 4 $b$-jets and missing transverse energy $\slashed{E}_T$ from the final neutrinos which are not detected at colliders.  For this study of the (4 $b$-jets + $\slashed{E}_T$) events we use again \textsc{MadGraph5} (MG5) \cite{Alwall:2014hca}. 

To simplify our study,  the $b$-jets from the final state will be defined at  the parton level, i.e.  we will not consider here any showering effects.   Instead,  and in order to take into account in an approximate way the $b$-tagging  effects in our estimates of the event rates later we have introduced a reducing factor $\epsilon_b^4$, where $\epsilon_b = 0.8$ is our assumed $b$-tagging efficiency factor for each $b$-jet which is the usually assumed value in the literature (see for instance,  \cite{Contino:2013gna}).  Regarding the background estimates, we have checked here that the dominant background is given by the same process $e^+e^- \to HH \nu \bar \nu \to b \bar b  b \bar b \nu \bar \nu$ as predicted within the SM.  We have also checked with MG5 that these SM background rates correspond to the HEFT predictions for 
$e^+e^- \to HH \nu \bar \nu \to b \bar b b \bar b \nu \bar \nu$  with $a=b=1$.   Other potential SM backgrounds to  ($b \bar b b \bar b \nu \bar \nu$) events are expected to be either small or easily treatable.  In particular,  events from mixed EW-QCD with virtual gluons producing the $b \bar b$ final pairs and $Z$ bosons producing the $\nu \bar \nu$ pairs can be easily reduced by applying cuts on the final $M_{b \bar b}$ invariant mass  and final missing transverse energy $\slashed{E}_T$.  Events from tops production do not produce the same final state,  since neutrinos from the top decays will come together with charged leptons.  Events from $W's$ production can also provide the neutrinos of the final state, but again they will come together with charged leptons.  Of course,  some additional backgrounds will appear when some particles  in the final state escape detection.  However,  these more realistic detector effects are not considered here, and they are left for a future more refined work once the detector characteristics of these projects have been  fixed.  

For a more quantitative check of the main backgrounds to ($b \bar b b \bar b \nu \bar \nu$) events within the EW-SM, we have computed with MG5\footnote{When running MG5  we use the input parameter values: $m_Z= 91.19$ GeV,  $\Gamma_Z=2.44$ GeV,  $m_H=125$ GeV,  $\Gamma_H=4.07 \times 10^{-3}$ GeV,  
$\alpha(m_Z)=1/128$,  $m_b=4.7$ GeV,  ymb $=3.1$} the three most important ones which are $e^+ e^- \to XY \nu \bar \nu  \to b \bar b b \bar b \nu \bar \nu$ with the intermediate states $XY$ being $HH$,  $HZ$ and $ZZ$.  The results of the corresponding cross sections  within the SM are summarized in table \ref{table:backg}, where we have also included the reduction in the results after some selected cuts are applied.  First,  we have checked that,  for the case without cuts ,  the total rates found for the final ($b \bar b b \bar b \nu \bar \nu$) events,  are approximately equal to the $HH \nu \bar \nu$ rates multiplied by the corresponding branching ratios of the Higgs boson and $Z$ boson to $b\bar b$ pairs,  given by $58\%$  and $15\%$ respectively.   Second,  we have also checked that  the last cut on the invariant mass of the $b \bar b$ pair is very efficient in eliminating practically the SM backgrounds from $XY$ being $HZ$ and $ZZ$.  In contrast, the event rates from $XY=HH$ are not practically affected by this cut on $M_{b \bar b}$, as expected.   Therefore, in our following study of the accessibility to $\kappa_{V}$ and $\kappa_{2V}$ and to their potential correlations in ($b \bar b b \bar b \nu \bar \nu$) events, we will assume that $XY=HH$ within the SM provides the main background to our BSM signal from the HEFT with $(a,b) \neq (1,1)$. 

\begin{table}[b!]
\begin{center}
\vspace{.2cm}
\begin{tabular}{ c @{\extracolsep{0.8cm}} c @{\extracolsep{0.8cm}} c @{\extracolsep{0.8cm}} c }
\toprule
\toprule
Cuts & $\sigma(XY=HH)({\rm pb})$ & $\sigma(XY=HZ)({\rm pb})$  & $\sigma(XY=ZZ)({\rm pb})$\\
\midrule
No cuts & $2.6 \times 10^{-4}$  & $7.7 \times 10^{-4}$  & $1.1 \times 10^{-3}$\\
Cuts in Eq. \ref{eqn: cuts} &$6.0\times 10^{-5}$   &   $ 2.4 \times 10^{-4}$  & $3.7 \times 10^{-4}$\\
$120< |M_{b \bar b}({\rm GeV})|<130$ & $6.0  \times 10^{-5}$    & $ 1.6 \times 10^{-6}$ & negligible \\
\bottomrule
\bottomrule
\end{tabular}
\caption{\small Predictions for
$\sigma(e^+e^- \to XY \nu \bar \nu \to b \bar b b \bar b \nu \bar \nu)$(pb) within the SM for various intermediate states $XY=HH$,  $HZ$ and $ZZ$ and for various choices of cuts.  In the first row there are no cuts applied.  In the second row the set of cuts in Eq. \ref{eqn: cuts} is applied.   In the last row,  the cuts in Eq. \ref{eqn: cuts} and the specified cut on the invariant mass of each bottom antibottom pair of the final state are applied. The total energy is set here to $\sqrt{s}=3 \,  {\rm TeV}$. }
\label{table:backg}
\end{center}
\end{table}


In order to simplify the analysis of the ($b \bar b b \bar b \nu \bar \nu$) events, we will then consider just some basic requirements.  In order to guarantee the detection of the final particles, we will implement some minimal detection cuts to the $b$-jets and require a minimum in the missing energy.  This is motivated by the well known configuration of the radiated fermions in the processes that are mediated dominantly  by WBF.  These are characterized by final fermions (neutrinos in the present case) that are produced in the backward and forward directions respect to the beam with transverse momentum being related to the mass of the vector boson participating in the WBF subprocess, i.e. the $W$ mass here.   The specific values for the cuts assumed here are summarized in the following: 
    \begin{equation}
        p_T^b > 20 \, \text{GeV}, \hspace{5mm} |\eta^b| < 2, \hspace{5mm} \Delta R_{bb} > 0.4, \hspace{5mm} \slashed{E}_T > 20 \, \text{GeV}, 
        \label{eqn: cuts}
    \end{equation}
    where $p_T^b$ and $\eta^b$ are the transverse momentum and the pseudorapidity of the $b$-jets, $\slashed{E}_T$ is the missing transverse energy from the $\nu \bar \nu$ pairs and $\Delta R_{bb} = \sqrt{(\Delta \eta_{bb})^2 + (\Delta \phi_{bb})^2}$, where $\Delta \eta_{bb}$ and $\Delta \phi_{bb}$ are the separations in pseudorapidity and azimuthal angle between the $b$-jets. The cuts that are used in this work are based on the ones used in \cite{Gonzalez-Lopez:2020lpd} and \cite{Abramowicz:2016zbo}.  In addition,  as we have said above,  in order to take into account in an approximate way the $b$-tagging effects in our estimates of the rate events we have introduced a reducing factor $\epsilon_b^4$, where $\epsilon_b = 0.8$ is our assumed $b$-tagging efficiency factor for each $b$-jet. 
    
\begin{figure}[!t]
    \centering
\includegraphics[height=0.28\textheight]{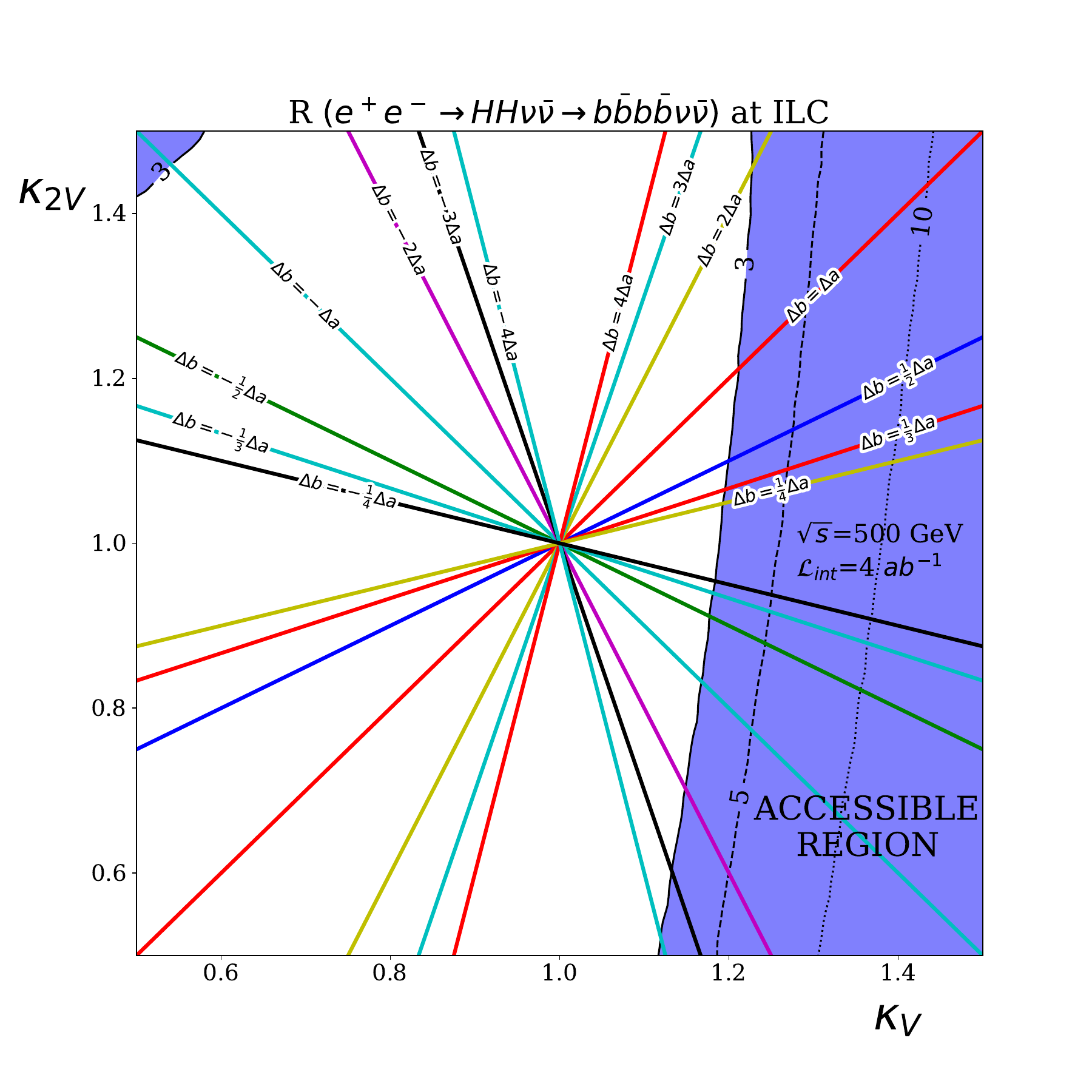} \includegraphics[height=0.28\textheight]{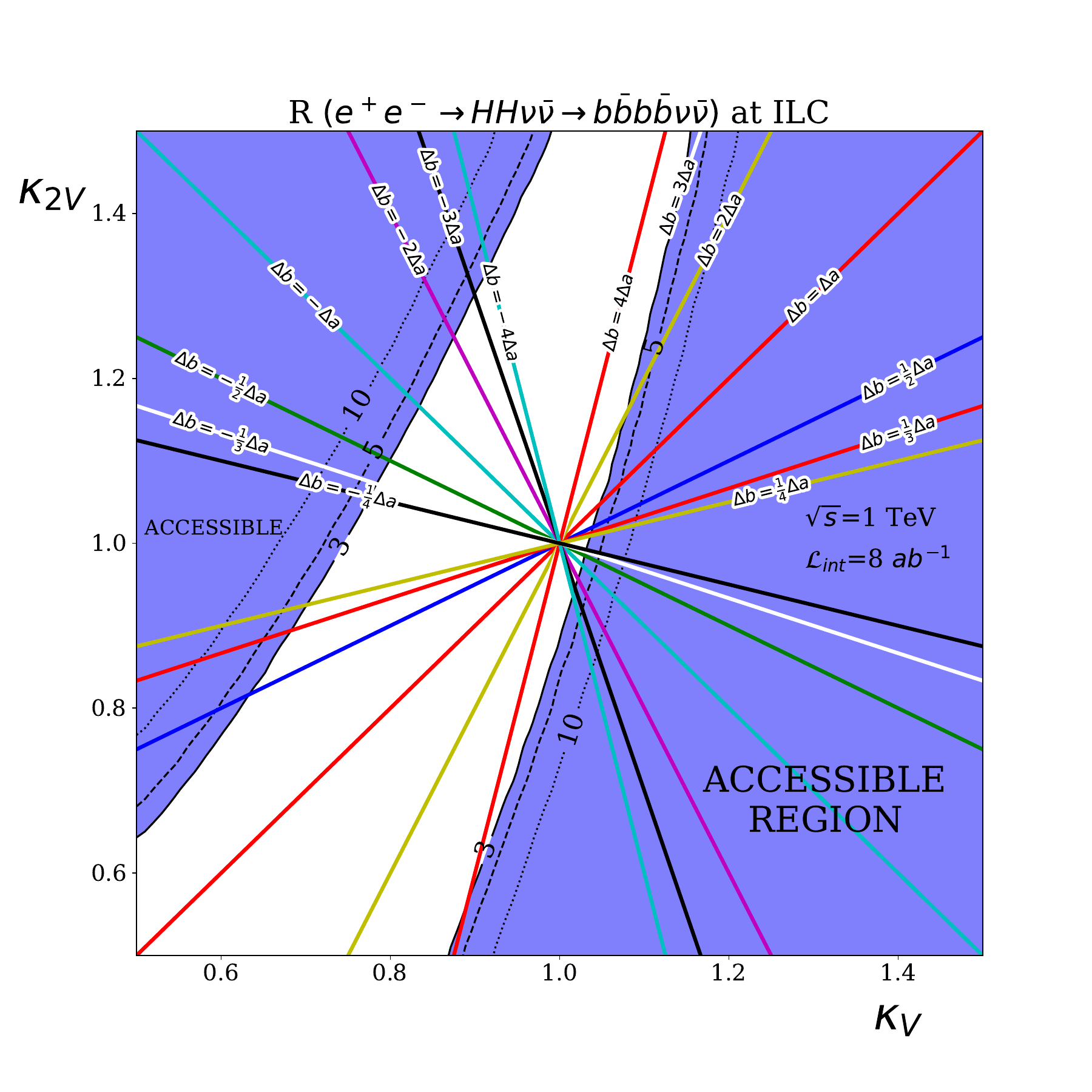}\\\hspace{0.1cm} \includegraphics[height=0.5\textheight]{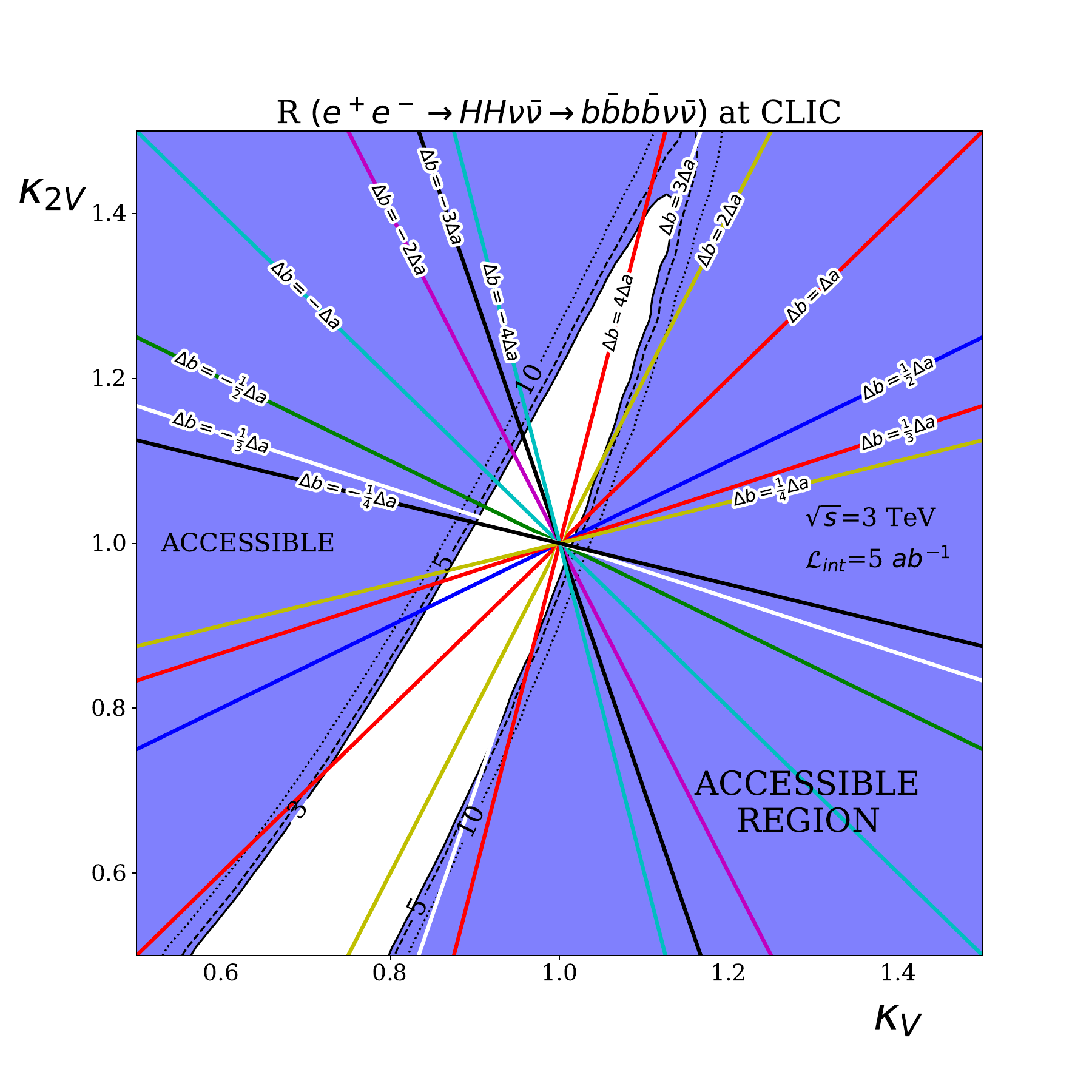}        
 \caption{Predictions from the HEFT for contour lines of the parameter $R$ defined in Eq. \ref{eqn: R}, which serves as an estimate of the potentially accessible regions in the $(a, b)$ plane,  at the future 
    $e^+e^-$ colliders:  ILC $(500\, {\rm GeV},  4\, {\rm ab}^{-1})$ (upper left plot),  ILC $(1000\, {\rm GeV},  8\, {\rm ab}^{-1})$ (upper right plot), and CLIC  $(3000\, {\rm GeV},  5\, {\rm ab}^{-1})$ (lower plot).  The shaded areas in purple represent the accessible region defined by $R > 3$.  The coloured straight lines represent the different correlation hypotheses for $\Delta b= C \Delta a$ with $C=\pm 1/4,  \pm 1/3, \pm 1/2, \pm 1, \pm 2, \pm 3, \pm 4$.   }
    \label{contourRstar}
    \end{figure}

  Finally, in order to quantify the sensitivity to the BSM coefficients $\kappa_V$,  $ \kappa_{2V}$ and their possible correlations we have defined the ratio $R$ in terms of the number of (4 $b$-jets + $\slashed{E}_T$) events as: \par
    \begin{equation}
        R \, = \, \frac{N_{BSM} - N_{SM}}{\sqrt{N_{SM}}}.
        \label{eqn: R}
    \end{equation}
Here,  $N_{BSM}$ is the events rate from the BSM signal as computed from the HEFT for the given $\kappa_V=a$ and $ \kappa_{2V}=b$ parameters, and $N_{SM}$ is the events rate from the SM (which, as said above, corresponds to the HEFT prediction for $a=b=1$).  
    By requiring $R$ to be above some threshold in order to differentiate the BSM signal from the SM  prediction,  we can set an approximate criterion for the accessibility to these $\kappa_V$, and $ \kappa_{2V}$ coefficients . 
    In Fig. \ref{contourRstar} we show the contour lines for $R=3$ (solid line), 5 (dashed line) and 10 (dotted line),  for the three cases,  
    ILC$(500 \,{\rm GeV}, 4 \,{\rm ab}^{-1} )$ (upper left plot),  ILC  $(1000\, {\rm GeV}, 8\, {\rm ab}^{-1} )$ (upper right plot),  and CLIC  $(3000 \, { \rm GeV},  5\,{\rm ab}^{-1})$ (lower plot). 
     We have also coloured in purple the region for $R>3$, which corresponds to the largest accessible region of the parameter space.  
From these plots, we learn several features.  First, we see again that CLIC with the highest planned energy will provide the highest sensitivity to these $a=\kappa_V$, and $b=\kappa_{2V}$ parameters and it will definitely improve the present sensitivity at the LHC.   
For instance,  at CLIC(3000 GeV),  for the simple setting of $a = 1$, the accessible values (i.e.  those with $R>3$)  will be $b > 1.2$ and $b < 0.95$, and for the simple setting of $b = 1$, the accessible values  will be $a < 0.89$ and $a > 1.01$.  The corresponding values for the other collider options are worse.  At ILC(500 GeV),  for $a=1$ the reach in $b$ will be outside the explored interval in this plot, and for $b=1$ the reach in $a$ will be $a>1.19$.  At ILC(1000 GeV),  for $a=1$  the accessibility region will be $b > 1.5$ and $b < 0.91$, and for $b=1$ we find the accessibility  at $a<0.75$ and $a>1.03$.
 
Regarding our study of the correlations,  in Fig. \ref{contourRstar} we have included several lines for multiple correlation hypotheses that can be compared with the previously commented accessible regions in the $(a, b)$ plane.  One can see that by starting from the SM point and moving through the $(a, b)$ plane in order to explore the different BSM hypotheses by following the marked correlation lines one reaches the given sensitivity lines (i.e. the parabolas for a fixed value of $R$) sooner in some cases than in others of those considered in this figure. The shorter the distance to reach a given sensitivity line is, the easier will be to probe the given correlation hypotheses.  For instance, if we focus on the $a < 1$, $b > 1$ region, we see that the shortest path to reach the $R = 3$ parabola is provided by following the green line which corresponds to the correlation $\Delta b = - \frac12 \Delta a$, and the longest path occurs for the light-blue line corresponding to the correlation $\Delta b = - 4 \Delta a$.  If instead we focus on the $a < 1$, $b < 1$ region the shortest path is for $\Delta b = 4 \Delta a$ and the longest one is for $\Delta b = 2 \Delta a$ (which in fact does not even enter the accessible region for the studied range of the parameter space). Looking at the region with $a > 1$ and $b > 1$, the shortest path is for $\Delta b = \frac14 \Delta a$, and the longest one is for $\Delta b = 4 \Delta a$. Finally, at the region with $a > 1$ and $b < 1$, there are no remarkable differences in the corresponding distances. All these findings for the comparative sensitivities of the various correlation hypotheses agree qualitatively with our previous  results for the differential cross section distributions with $M_{HH}$, $\eta_{H1}$ and $p_T^{H1}$ presented in the previous section.  A further more refined analysis of those distributions could in any case improve the viability to test in a more quantitative and precise way the various different correlation hypotheses for the HEFT parameters.  But this is left for a future work. 
\newpage

\section{Conclusions}
\label{conclu}
   In this work, our goal has been focused on the study of the phenomenological consequences in double Higgs production at $e^+e^-$ future colliders coming from the most relevant leading order HEFT  $a$ and $b$ coefficients. These parametrize the BSM interactions/couplings of the Higgs particle with the $V$ ($W$ and $Z$) EW gauge bosons,  $HVV$ and $HHVV$ respectively, and they are identified with the usual $\kappa$-modifiers as $a=\kappa_V$ and $b=\kappa_{2V}$.  For our study of  the expected sensitivity to  $\kappa_V$ and $\kappa_{2V}$ via Di-Higgs production at $e^+e^-$ colliders we have considered both the total cross section and some selected differential cross sections.   Our proposal  of considering these selected differential cross-sections is motivated by our final goal of exploring the sensitivity not only to the parameters themselves but also to their particular correlations that can appear in different scenarios for the UV physics.  Finding experimental tests of these potential correlations is undoubtedly of great interest in the high energy physics community.  The main reason is that not only the values of these effective couplings but also their correlations carry relevant information on the particular underlying UV theory which is the final  responsible for generating such low  energy effective couplings.  The different correlations predicted for those parameters $a$ and $b$ (and more concretely for their deviations respect to their SM values given by $\Delta a = 1-a$ and $\Delta b = 1-b$)   in various examples explored in the literature (like 2HDM,   SMEFT,   MHCM,  dilaton models and others) suggest the interest of these kind of tests, like the ones proposed in this work.  
   
We have first explored the relevance of the particular combination ($\kappa_V^2-\kappa_{2V}$) at the subprocess level.  Our investigation of the most relevant subprocess,  given by the $WW \to HH$  scattering, 
indicates that the total cross section has a minimum in the $(\kappa_{V}, \kappa_{2V})$ plane close to the correlation hypotheses given by the line $\kappa_V^2=\kappa_{2V}$ which contains the SM point $(\kappa_V,\kappa_{2V})=(1,1)$.  Our study of the subprocess also indicates that departures from this line,  i.e. for $\kappa_V^2\neq \kappa_{2V}$, are good indicators of BSM  not only in the total cross section but also in the particular distributions with respect to the angular variables of the final state.  We find that these are also good candidates for tests  of the different possible correlations among the $\kappa_V=a$ and $\kappa_{2V}=b$ parameters.  
  Secondly, we have explored the sensitivity to the BSM effective couplings $a$ and $b$ in the $e^+e^- \to HH \nu \bar \nu$ process.   In order to do so, we have performed a scan of the cross section in the $(a,b)$ parameter space  which have been shown in the form of contour line plots.  We have checked that for the particular choice of $a^2 = b$ the contour lines approach a valley region with very small values for the cross section, signaling the existence of a close minimum.  Close to that region $a$ and $b$ will be therefore very difficult to be tested.  The more distant $a$ and $b$ are from this $a^2 = b$ direction the better will be the access to test these parameters at $e^+e^-$ colliders by means of double Higgs production.  Furthermore, thanks to these plots, we have been able to show that the cross section of the process is not only sensible to the absolute value of $\Delta a$ and $\Delta b$, but also to their relative size and sign.  This means that the direction in which we choose to depart from the SM configuration is relevant to the behaviour of the process.  The summary of the predictions for the total cross section in the $(\kappa_{V}, \kappa_{2V})$ plane and their comparison with the different correlation hypotheses defined by $\Delta b = C \Delta a$,  with $C= \pm \frac{1}{4},  \pm \frac{1}{3},  \pm \frac{1}{2},  \pm 1,  \pm 2,  \pm 3,  \pm 4$,  are collected in figure \ref{contour_plots_star}.    From these plots we can conclude  that a given correlation hypotheses for $\Delta a$ versus $\Delta b$ that points in a `perpendicular' direction to such $a^2=b$ line will be more easily tested experimentally, in the sense that it will reach higher sensitivities for small deviations respect to the SM values than other directions.  For instance,  the cases with $C=2,3,4$ seem very difficult to be tested because the corresponding lines lay close to the contourlines with the lowest cross section.  In contrast,  the cases with $C=-1/2, -1/3$ seem to be the easiest ones to be tested.  We have shown in these plots that the behaviour of the contour lines is also sensible to the center-of-mass energy of the process, being the highest energies the most sensitive ones to departures respect to the SM.  CLIC with 3000 GeV  will then offer the best option.  
  
 Regarding our analysis of the correlations by means of the differential cross sections, we conclude that our three proposed ones,   $d\sigma/dM_{HH}$,  $d\sigma/d\eta_H$,  and $d\sigma/d p_H^T$ will provide very good options to disentangle the deviations from BSM predictions with respect to the SM.   Furthermore,  the corresponding results for CLIC,  summarized in Figs. \ref{hist: MHH_corr},  \ref{hist: Eta_corr} and \ref{hist: Pt_corr} respectively,  also demonstrate that the most characteristic departures from the SM predictions are given as enhancements at high $M_{HH}$,  at the central rapidity region close to $\eta_H=0$ and high transverse momentum $p^T_H$, indicating the high transversality of the Higgs bosons (and consequently of their decay products) in the BSM signals.  These differential cross section results also confirm our findings in the total cross sections of the comparative results for the various correlation hypotheses,  showing  that 
 $C=2,3,4$ will be difficult to be tested  whereas  $C=-1/2,-1/3$ will be easy to be tested.

 Finally, in order to quantify the sensitivity to BSM departures from the SM,  we consider the specific Higgs decays into $b \bar b$ pairs that lead to the complete process of our interest given by $e^+e^- \to HH \nu \bar \nu \to b \bar b b \bar b \nu \bar \nu$.  Accordingly,  we explore the final events containing the 4 $b$-jets from the two Higgs decays and the missing transverse energy from the $\nu \bar \nu$ pairs.  For this exploration we compute the ratio $R$ given in Eq. \ref{eqn: R}  in the $(\kappa_{V}, \kappa_{2V})$ plane for the three projected $e^+e^-$ colliders: ILC$(500 \,{\rm GeV}, 4 \,{\rm ab}^{-1} )$, ILC $(1000\, {\rm GeV}, 8\, {\rm ab}^{-1} )$,  and CLIC $(3000 \, { \rm GeV},  5\,{\rm ab}^{-1})$.   The larger the value for $R$ is, the better the sensibility to $\kappa_{V}$ and $\kappa_{2V}$ is reached.  We have defined the accessibility region in Fig. \ref{contourRstar} as $R>3$.   We find the highest accessibility to $(\kappa_{V}, \kappa_{2V})$ at CLIC(3000 GeV) that will definitely improve the present sensitivity at the LHC.  Thus,  for the simple setting of $a = 1$, the accessible values  are found for $b > 1.2$ and $b < 0.95$, and for the simple setting of $b = 1$, the accessible values found are $a < 0.89$ and $a > 1.01$.  The corresponding values for the other collider options are worse.  At ILC(500 GeV),  for $a=1$ the reach in $b$ will be outside the explored interval in this plot, and for $b=1$ the reach in $a$ will be $a>1.19$.  At ILC(1000 GeV),  for $a=1$  the accessibility region will be $b > 1.5$ and $b < 0.91$, and for $b=1$ we find the accessibility  at $a<0.75$ and $a>1.03$.  Regarding the accessibility to reach the different correlations we also find that CLIC will provide the best option.  We have finally determined, in each quadrant of the studied $(\kappa_{V}, \kappa_{2V})$ plane, which ones of the various correlation $C$ factors assumed in the correlation equation $\Delta b = C \Delta a$ will be easier and which ones will be more difficult to be tested at CLIC. 
 In the upper left quadrant ($a < 1$, $b > 1$)  we find that the shortest path to reach the $R = 3$ parabola is provided by $C=-1/2$ and the longest path occurs for $C=-4$.   In the lower left quadrant  ($a < 1$, $b < 1$) the shortest path is for $\Delta b = 4 \Delta a$ and the longest is for $\Delta b = 2 \Delta a$ (which in fact does not even enter the accessible region for the studied range of the parameter space).  In the upper right quadrant ($a > 1$, $b > 1$) the shortest path is for $C=1/4$,  and the longest one is for $C = 4$. Finally,  in the lower right quadrant  ($a > 1$, $b < 1$), there are no remarkable differences in the corresponding distances.
 
  In summary,  the general conclusion that can be drawn from this work is that different values for the LO-HEFT parameters $a$ and $b$ and their different possible correlations, specially if they are not correlated by $a^2=b$, can lead to very different kinematical properties in the final events from Di-Higgs production, as compared to the SM case,  that could be exploited in future $e^+e^-$ colliders in the search for BSM Higgs physics.  This could help us to have a higher understanding of the nature of the scalar sector of the SM, and maybe take a step towards discriminating among the various options for the underlying high energy (UV) theory.

\section*{Acknowledgments}
D.D and M.J.H acknowledge partial financial support by the Spanish Research Agency (Agencia Estatal de Investigación) through the grant  with reference number PID2022-137127NB-I00 
funded by MCIN/AEI/10.13039/501100011033/ FEDER, UE.  We also acknowledge financial support from the  grant IFT Centro de Excelencia Severo Ochoa with reference number CEX2020-001007-S
 funded by MCIN/AEI/10.13039/501100011033, from the previous AEI project PID2019-108892RB-I00 funded by MCIN/AEI/10.13039/501100011033, and from the European Union’s Horizon 2020 research and innovation programme under the Marie Sklodowska-Curie grant agreement No 860881-HIDDeN.  The work of R.M. is supported by CONICET and ANPCyT under projects PICT 2017-2751, PICT 2018-03682 and PICT-2021-I-INVI-00374.
The work of D.D.is also supported by the Spanish Ministry of Science and Innovation via an FPU grant No FPU22/03485. 
\newpage

\section*{Data Availability Statement}

No Data associated in the manuscript.

\section*{Appendices}
\appendix

\section{Diagrams contributing to $e^+ e^- \rightarrow HH\nu \bar{\nu}$}\label{A}
\label{Diagrams}

 Here we display the complete set of Feynman diagrams that contribute to  $e^+ e^- \rightarrow HH\nu \bar{\nu}$.  We are neglecting the electron mass in the generation of this full set of diagrams.  These diagrams are automatically generated by \textsc{MadGraph} and using the UFO file with our HEFT model.  They contain the diagrams with WBF configuration,  diagrams 5, 6, 7 and 8,  which are the ones where the subprocess $W^+ W^- \rightarrow HH$ takes place and the remaining diagrams, 1, 2, 3 and 4 ,  which are mediated by intermediate $Z$ bosons.  Note that \textsc{MadGraph} works in the unitary gauge by default, so it does not take into account the contribution from the GBs.  The coloured dots in these diagrams represent the effective couplings of the Higgs boson with the electroweak gauge bosons,  $HVV$ and $HHVV$ ($V=W, Z$) within the HEFT.  These are given respectively by $a=\kappa_V$ and $b=\kappa_{2V}$ as described in Sec. \ref{effHcoup}. 
 
 \begin{figure}[H]
  \includegraphics[width=\textwidth]{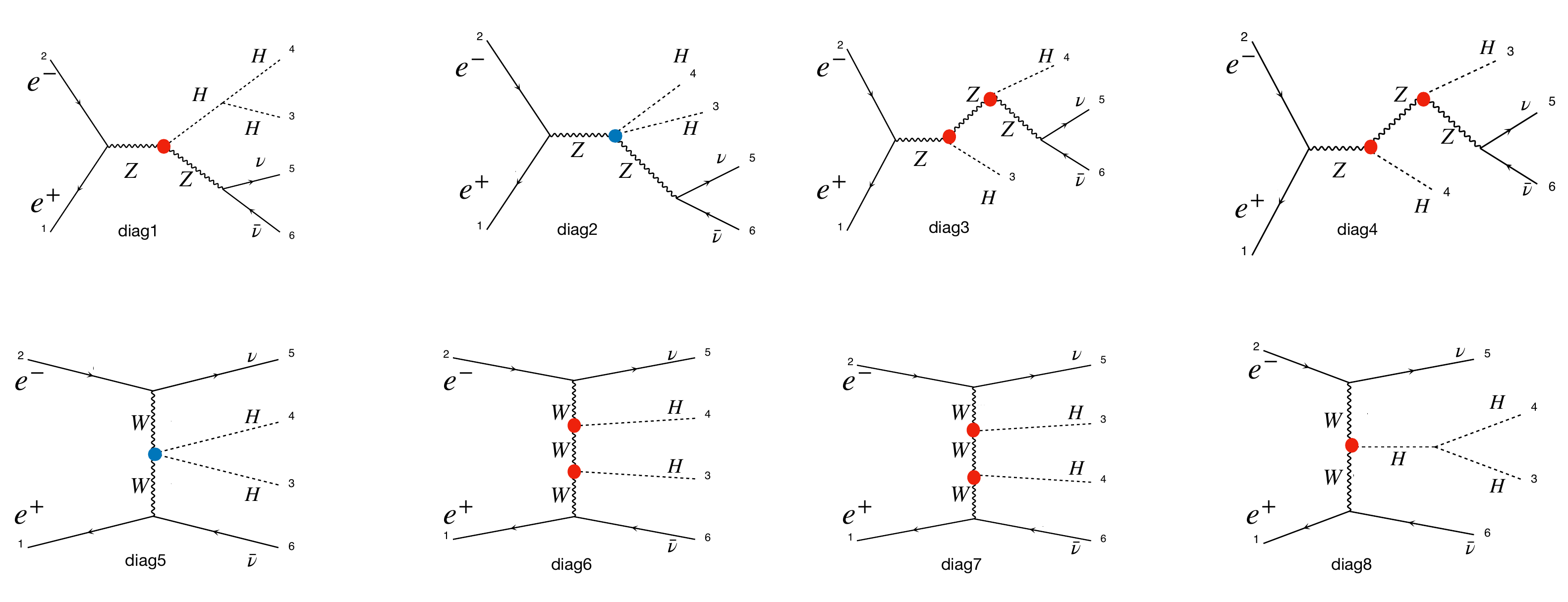}
   \caption{Feynman diagrams for $e^+ e^- \rightarrow HH\nu \bar{\nu}$ generated by MG5 in the unitary gauge.  The coloured dots represent the effective couplings of the Higgs boson with the electroweak gauge bosons, $HVV$ and $HHVV$ ($V=W, Z$) within the HEFT. }
    \label{FDs-epem}
 \end{figure}
    

\newpage

\bibliography{DDHM-EPJC-v3}

\end{document}